\documentclass[reprint,amsmath,amssymb,aps,prx]{revtex4-2}
\usepackage{graphicx}
\usepackage{dcolumn}
\usepackage{bm}
\usepackage{color}
\usepackage{units}
\usepackage{wasysym}
\usepackage{MnSymbol}
\usepackage{comment}
\usepackage{times}
\usepackage{soul}
\usepackage{upgreek}
\usepackage[unicode=true,pdfusetitle,bookmarks=false,colorlinks=true,citecolor=blue,urlcolor=blue,linkcolor=blue]{hyperref}

\begin{document}

\title{Theory of Nonequilibrium Coexistence with Coupled Conserved and Nonconserved Order Parameters}
\author{Daniel Evans}
\author{Ahmad K. Omar}
\email{aomar@berkeley.edu}
\affiliation{Department of Materials Science and Engineering, University of California, Berkeley, California 94720, USA}
\affiliation{Materials Sciences Division, Lawrence Berkeley National Laboratory, Berkeley, California 94720, USA}

\begin{abstract}
Phase separation routinely occurs in both living and synthetic systems.
These phases are often complex and distinguished by features including crystallinity, nematic order, and a host of other nonconserved order parameters. 
For systems at equilibrium, the phase boundaries that characterize these transitions can be straightforwardly determined through the framework of thermodynamics.
The prevalence of phase separation in active and driven systems motivates the need for a genuinely nonequilibrium theory for the coexistence of complex phases.
Here, we develop a dynamical theory of coexistence when both conserved and nonconserved order parameters are present, casting coexistence criteria into the familiar form of equality of state functions.  
Our theory generalizes thermodynamic notions such as the chemical potential and Gibbs-Duhem relation to systems out of equilibrium.
While these notions may not exist for all nonequilibrium systems, we numerically verify their existence for a variety of systems by introducing the phenomenological Active Model C+.
We hope our work aids in the development of a comprehensive theory of high-dimensional nonequilibrium phase diagrams.
\end{abstract}

\maketitle

\section{\label{sec:intro} Introduction}

\begin{figure*}
	\centering
	\includegraphics[width=.95\textwidth]{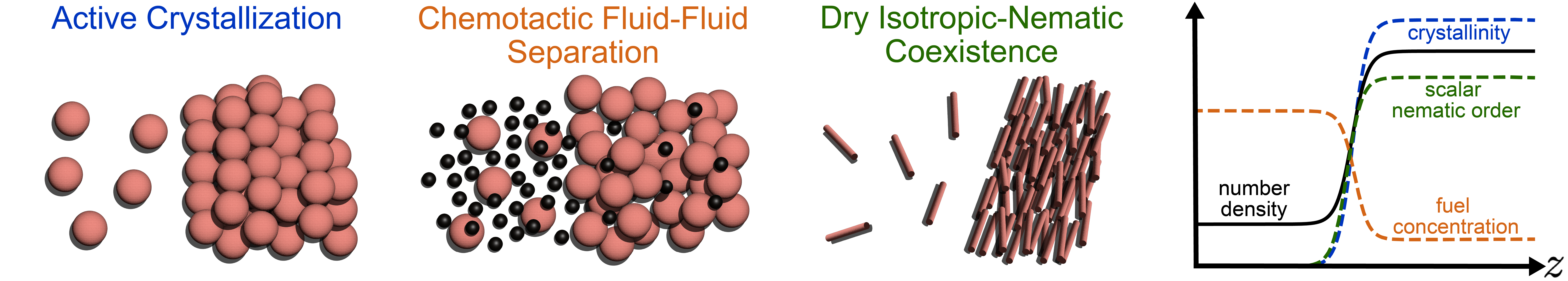}
	\caption{\protect\small{{Three example systems where our theory would apply with $n_N=1$: active crystallization, chemotactic fluid-fluid separation, and active isotropic-nematic coexistence. A schematic of the order parameter profiles is shown, where the nonconserved order parameter represents the local crystallinity, fuel concentration, and nematic order. While chemoattractant fuel is likely to be subject to a constraint (i.e.,~mass conservation) common coarse-grained models treat chemotactic fuel as a wholly unconstrained variable~\cite{Liebchen2018, Zhao2023}.}}}
	\label{fig:schematicapplications}
\end{figure*}

Nonequilibrium systems as diverse as reaction-modulated biomolecular condensates~\cite{Laflamme2020, Kirschbaum2021}, chiral crystals comprised of motile bacteria~\cite{Petroff2015} or starfish embryos~\cite{Tan2022}, and nematically ordered amoeba cells~\cite{Gruler1999} all can display states of phase coexistence.
A hallmark of these nonequilibrium states is the presence of both conserved (e.g.,~number density) and nonconserved (e.g.,~crystallinity, fuel concentration, or nematicity as shown in Fig.~\ref{fig:schematicapplications}) order parameters which couple to generate these intriguing states of phase coexistence.
In many of these systems, the absence of even a local equilibrium precludes the use of the thermodynamic arguments employed to construct equilibrium phase diagrams.
While there is now a growing body of literature on generalizing theories of phase coexistence to nonequilibrium systems with \textit{conserved order parameters}~\cite{Fily2012, Redner2013, Wittkowski2014, Takatori2015, Speck2016, Solon2018, Hermann2019, Hermann2021, Speck2021, Omar2023b, Brauns2024NonreciprocalFields, Greve2024, Saha2024, Langford2024PhaseMatter, Chiu2024TheoryCoexistence}, there are comparatively fewer studies~\cite{Brauns2020, Frohoff2023} for the construction of nonequilibrium phase diagrams when both conserved and nonconserved order parameters are crucial.
A framework for describing these transitions could deepen our understanding of active crystallization~\cite{Bialke2012, Turci2021, Omar2021pd, Caprini2023, Galliano2023, Hermann2023, Shi2023, Ding2024}, isotropic-nematic coexistence~\cite{Putzig2014, Cai2019}, and chemotactic fluid-fluid separation~\cite{Liebchen2018, Zhao2023}.

One can always directly determine phase diagrams by directly measuring the order parameters in heterogeneous states of coexistence in simulation or experiment. 
For systems at equilibrium, thermodynamics allows one to circumvent these measurements and construct \textit{predictive theories} for phase diagrams solely by equating \textit{bulk state functions}.
These coexistence criteria are determined variationally from the system's free energy which is generally ill-defined out of equilibrium.
What, if any, bulk state functions are equal between coexisting phases is thus an open question for driven systems.

In this Article, we explore the questions outlined above by developing an entirely dynamical theory of phase coexistence that is applicable to both equilibrium and nonequilibrium systems with coupled conserved and nonconserved order parameters.
We find coexistence criteria solely \textit{in terms of bulk equations of state}, just as in equilibrium.
While the nonequilibrium coexistence criteria when there is a single conserved order parameter~\cite{Aifantis1983rule, Solon2018, Omar2023b} can always be identified, here we find the criteria to only exist for a subset of nonequilibrium systems.
In particular, we find systems in which: (i) the criteria are exact and expressed as equality of bulk state functions; (ii) the criteria are approximate and expressed as a generalized equal-area Maxwell construction; and (iii) no coexistence criteria exist in terms of bulk state functions.
To validate our approach, we introduce a nonequilibrium field theory: Active Model C+ (AMC+).
We find excellent agreement between the predictions made by our theory and the phase diagrams obtained through numerical simulation of AMC+.
We hope our theory provides a practical procedure to construct first-principles phase diagrams of nonequilibrium systems with coupled conserved and nonconserved order parameters.

\section{Theory of Multiphase Coexistence}

In this Section, we describe a dynamical theory of coexistence between $n_P$ phases described by a \textit{single} scalar conserved field coupled to $n_N$ scalar nonconserved fields that is applicable to both equilibrium and nonequilibrium systems.
Nonconserved fields that are tensorial can be described under our theory by considering each component of the tensor as one of the $n_N$ scalar fields and including the appropriate couplings in the field dynamics.
We first review the thermodynamically-derived equilibrium coexistence criteria in Section~\ref{sec:thermo}.
In Section~\ref{sec:mech1}, we begin with the evolution equations for each of the order parameters and derive equations of state that are equal in coexisting phases.
While not every nonequilibrium system has bulk coexistence criteria, in those that do, we find a pseudopotential and global quantity that are equal in coexisting phases that are analogous to the equilibrium chemical potential and pressure, respectively.

\subsection{\label{sec:thermo}Equilibrium Criteria from Bulk Thermodynamics}
We consider $n_P$ macroscopic coexisting phases with uniform temperature $T$. 
The phases comprise a set $\mathcal{P}$ of length $n_P$. The overall number density is  ${\rho}^{\rm sys}=N/V$, where $N$ is the particle number and ${V}$ is the total system volume.
The density field, $\rho$, can fluctuate in space but is constrained to have a mean of ${\rho}^{\rm sys}$.
The degree of order in the system is characterized by $n_N$ nonconserved intensive order parameters, $\{\psi_i\}$.
The nonconserved order parameters comprise a set $\mathcal{N}$ of length $n_N$.
In contrast to $\rho$, each $\psi_i$ is \textit{unconstrained}.
The system is described by the vector of order parameters, ${\mathbf{X} \equiv \begin{bmatrix} \rho & \psi_1 & \psi_2 & \cdots & \psi_{n_N} \end{bmatrix}^{\rm T}}$, whose indices comprise a set $\mathcal{X}$ of length $n_N + 1$ and are the symbols of the order parameters themselves, e.g.,~$X_{\rho}=\rho$ and $X_{\psi_i} = \psi_i \ \forall i \in \mathcal{N}$.

The bulk (mean-field) free energy density of the system is denoted as ${f^{\rm bulk} \left( \mathbf{X} \right)}$ with the total free energy ${F^{\rm bulk} = V f^{\rm bulk} \left( \mathbf{X} \right)}$.
The absence of a coupling between at least one $\psi_i$ and $\rho$ in the free energy along with the lack of constraints on the nonconserved fields will result in each phase having an identical value of every $\psi_i$: the value that minimizes $f^{\rm bulk}$.
A coupling between a $\psi_i$ and the constrained ${\rho}$ in the free energy can lead to coexistence between dense and dilute phases with different values of at least one $\psi_i$.
More specifically, this nontrivial coupling will be reflected in non-additive contributions of ${\rho}$ and $\psi_i$ to the free energy density, i.e.,~${f^{\rm bulk} \left( \mathbf{X} \right) \neq \sum_{i \in \mathcal{X}} f^{\rm bulk}_i \left( X_i \right)}$ (and thus the mean-field probability cannot be factorized, i.e.,~$P\left( \mathbf{X} \right) \propto \exp[- Vf^{\rm bulk}\left( \mathbf{X} \right) / k_{B} T] \neq \prod_{i \in \mathcal{X}} P_{i} \left( X_i \right)$, where ${k_{B}T}$ is the thermal energy).
A necessary criterion for equilibrium coexistence with coupled conserved and nonconserved order parameters is thus a non-vanishing mixed derivative, ${\partial^2 f^{\rm bulk} / \partial \rho \partial \psi_i}$ for at least one $i \in \mathcal{N}$.

In the scenario of $n_P$ coexisting phases (e.g.,~$n_P=2$ for two-phase coexistence), the total free energy can be expressed as ${F^{\rm bulk} = \sum_{\alpha \in \mathcal{P}} V^{\alpha} f^{\rm bulk} \big( \mathbf{X}^{\alpha} \big)}$ where ${V^{\alpha}}$ and ${\mathbf{X}^{\alpha}}$ are the respective volume and order parameters of the $\alpha$ phase and we have neglected the interfacial free energy (the ratio of the interfacial area to the system volume is negligibly small for macroscopic systems).
Notably, while the phase volumes and number densities are constrained, there are no constraints on each ${\psi_i^{\alpha}}$ (i.e.,~systems prepared at a given density and total volume can take any value of each $\psi_i^{\alpha}$).
Minimizing the free energy with respect to each phase's volume and ${\mathbf{X}^{\alpha}}$, subject to the above constraints, results in the equilibrium coexistence criteria:
\begin{subequations}
    \label{eq:eqmcriteria}
    \begin{align}
        \label{eq:eqmcriteria1}
        & \boldsymbol{\mu}^{\rm bulk} \left( \mathbf{X}^{\alpha} \right) = \boldsymbol{\mu}^{\rm coexist} \ \forall \alpha \in \mathcal{P}, \\
        \label{eq:eqmcriteria4}
        & p^{\rm bulk} \left( \mathbf{X}^{\alpha} \right) = p^{\rm coexist} \ \forall \alpha \in \mathcal{P},
    \end{align}
\end{subequations}
where ${\boldsymbol{\mu}^{\rm bulk} \equiv \partial f^{\rm bulk} / \partial \mathbf{X} \equiv \begin{bmatrix} \mu^{\rm bulk}_{\rho} & \mu^{\rm bulk}_{\psi_1} & \cdots & \mu^{\rm bulk}_{\psi_{n_N}} \end{bmatrix}^{\rm T}}$ (${\mu^{\rm bulk}_{\rho} \equiv \partial f^{\rm bulk} / \partial \rho}$ is the familiar bulk chemical potential and ${\mu^{\rm bulk}_{\psi_i} \equiv \partial f^{\rm bulk} / \partial \psi_i}$) and ${p^{\rm bulk} \equiv \boldsymbol{\mu}^{\rm bulk} \cdot \mathbf{X} - f^{\rm bulk}}$ is the bulk pressure.
Here, ${\mu_{\rho}^{\rm coexist}}$ and $p^{\rm coexist}$ are constants that must be determined and generally depend on $\rho^{\rm sys}$ and $T$.
Contrasting this, ${\mu_{\psi_i}^{\rm coexist} = 0 \ \forall i \in \mathcal{N}}$ as the nonconserved quantities are not subject to any constraint by definition.
While the criteria following from the constrained minimizations of the free energy with respect to ${\rho}$ [the first component of Eq.~\eqref{eq:eqmcriteria1}] and ${V}$ [Eq.~\eqref{eq:eqmcriteria2}] are familiar for any state of equilibrium $n_P$-phase coexistence, the remaining $n_P n_N$ criteria ensure that within each phase, optimum values of each $\psi_i$ are selected for the corresponding ${\rho}$.
Subtracting the number of equations in Eq.~\eqref{eq:eqmcriteria}, $2 n_P - 2 + n_P n_N$, from the number of variables describing the coexisting phases, $n_P (n_N + 1)$, we find the number of degrees of freedom to be $2 - n_P$, the well-known (constant temperature) Gibbs phase rule~\cite{Gibbs1878}.

Equation~\eqref{eq:eqmcriteria4} can be equivalently cast as a Maxwell equal-area construction on the chemical potential vector, resulting in coexistence criteria that do not explicitly contain the thermodynamic pressure:
\begin{subequations}
    \label{eq:eqmcriteria2}
    \begin{align}
        \label{eq:eqmcriteria21}
        & \boldsymbol{\mu}^{\rm bulk} \left( \mathbf{X}^{\alpha} \right) = \boldsymbol{\mu}^{\rm coexist} \ \forall \alpha \in \mathcal{P}, \\
        \label{eq:eqmcriteria24}
        & \int_{\mathbf{X}^{\alpha}}^{\mathbf{X}^{\beta}} \left[ \boldsymbol{\mu}^{\rm bulk} \left( \mathbf{X} \right) - \boldsymbol{\mu}^{\rm coexist} \right] \cdot d \mathbf{X} = 0 \ \forall \alpha, \beta \in \mathcal{P},
    \end{align}
\end{subequations}
which is a direct result of the Gibbs-Duhem relation~\cite{Plischke1994}:
\begin{equation}
\label{eq:GDeqm}
    dp^{\rm bulk} = \mathbf{X} \cdot d\boldsymbol{\mu}^{\rm bulk}.
\end{equation}
We note that for systems with one order parameter, the coexistence criteria can be expressed entirely with \textit{either} the chemical potential \textit{or} pressure as chemical potential equality can be expressed through the appropriate Maxwell construction on the pressure. 
For systems with multiple order parameters, this flexibility is lost: there is no way to express the coexistence criteria without the chemical potential vector. 
However, the path-independence of the integrals in Eq.~\eqref{eq:eqmcriteria24} (its value is $p^{\rm bulk}( \mathbf{X}^{\alpha}) - p^{\rm bulk}( \mathbf{X}^{\beta})$ regardless of the parameterization chosen to evaluate the integrals) allows one to express the criteria in a form that resembles that of one-component systems.
We now consider the integration path $\mathbf{X}^*(\rho)$ such that $\mu_{\psi_i}^{\rm bulk} \left( \mathbf{X}^* \right) = \mu_{\psi_i}^{\rm coexist} = 0 \ \forall i \in \mathcal{N}$.
While there may be multiple possible solutions $\mathbf{X}^*$, one should select the solution that corresponds to the coexistence scenario under consideration (i.e.,~to describe $\alpha$-$\beta$ coexistence, the solution $\mathbf{X}^*(\rho)$ with end points $\mathbf{X}^*(\rho^{\alpha}) = \mathbf{X}^{\alpha}$ and $\mathbf{X}^*(\rho^{\beta}) = \mathbf{X}^{\beta}$ should be chosen).
Along this path, the Maxwell construction resembles that of one-component systems:
\begin{subequations}
    \label{eq:eqmcriteria3}
    \begin{align}
        \label{eq:eqmcriteria31}
        & \mu_{\rho}^{\rm bulk} \left( \mathbf{X}^{\alpha} \right) = \boldsymbol{\mu}^{\rm coexist} \ \forall \alpha \in \mathcal{P}, \\
        \label{eq:eqmcriteria34}
        & \int_{\rho^{\alpha}}^{\rho^{\beta}} \left[ \mu_{\rho}^{\rm bulk} \left( \mathbf{X}^* \right) - \mu_{\rho}^{\rm coexist} \right] d \rho = 0 \ \forall \alpha, \beta \in \mathcal{P}.
    \end{align}
\end{subequations}
Importantly, $\mathbf{X}^*(\rho)$ represents the values of the order parameters that are measurable, as
$\mu_{\rho}^{\rm bulk} \left( \mathbf{X}^* \right)$ [and equivalently $p^{\rm bulk} \left( \mathbf{X}^* \right)$] is what one would measure for a homogeneous system at equilibrium.
While the full dependence of every component of $\boldsymbol{\mu}^{\rm bulk}$ on $\mathbf{X}$ is generally not something one can measure, the path-independence of Eq.~\eqref{eq:eqmcriteria24} renders this inconsequential.

We note that while we have derived these coexistence criteria for systems with coupled conserved and nonconserved fields, the criteria also naturally recover the criteria for coexistence described by a single conserved field (e.g.,~liquid-gas coexistence) with ${\psi_i^{\alpha} = \psi_i^{\beta} \ \forall i \in \mathcal{N}, \ \forall \alpha, \beta \in \mathcal{P}}$ and ${\rho^{\alpha} \neq \rho^{\beta} \ \forall \alpha \neq \beta \in \mathcal{P}}$.
Next, we seek to determine the nonequilibrium analogs of the chemical potentials and Gibbs-Duhem relation in order to express our coexistence criteria solely in terms of measurable bulk equations of state.

\subsection{\label{sec:mech1}Nonequilibrium Criteria from Dynamics}
One may always obtain the phase diagram of a system by determining the full spatial order parameter profiles during coexistence, whether it be through experiment, simulation, or theoretical techniques.
This is often difficult, however, and consequently we aim to develop a framework that allows us to circumvent this task, just as in equilibrium, and determine phase diagrams of nonequilibrium systems by equating state functions across phases.
The approach proposed below is able to reproduce the criteria derived in the previous section for equilibrium systems while also describing a large class of nonequilibrium systems.

We begin by considering the spatial and temporal dynamics of each order parameter.
As we look to describe states of macroscopic phase coexistence, where the radius of curvature of the phases far exceeds any other length scale in the system, we neglect the effects of interfacial curvature and consider a quasi-1D geometry where spatial variations only occur in the $z$-dimension with translational invariance in the other dimensions.
This ensures the $z$-direction will be normal to the interface for phase-separated states.
The conserved order parameter is subject to an integral constraint, $\int_V d \mathbf{r} \rho(\mathbf{r}) = V \rho^{\rm sys}$, where $\rho(\mathbf{r})$ is its local value at position $\mathbf{r}$ and $\rho^{\rm sys}$ is its space-averaged value.
While we explicitly considered a conserved number density in Section~\ref{sec:thermo}, here we allow $\rho$ to be any conserved order parameter subject to an integral constraint of the above form.
Contrasting this, we consider nonconserved order parameters that are not subject to any constraints and therefore each $\int_V d \mathbf{r} \psi_i(\mathbf{r}) \ \forall i \in \mathcal{N}$ can take any value.
The evolution equations for the order parameters satisfy the balance laws:
\begin{equation}
\label{eq:species_continuity_eq}
    \partial_t \mathbf{X} = -\partial_z\mathbf{J} + \mathbf{s},
\end{equation}
where $\partial_t \equiv \partial / \partial t$, $\partial_z \equiv \partial / \partial z$, ${\mathbf{J} \equiv \begin{bmatrix} J_{\rho} & J_{\psi_1} & \cdots & J_{\psi_{n_N}} \end{bmatrix}^{\rm T}}$ is the flux of each order parameter in the $z$-direction, and ${\mathbf{s} \equiv \begin{bmatrix} s_{\rho} & s_{\psi_1} & \cdots & s_{\psi_{n_N}} \end{bmatrix}^{\rm T}}$ is the generation term associated with each order parameter.
As $\rho$ is conserved, $s_{\rho}=0$ by definition.
The fluxes in the directions orthogonal to $z$ are neglected, consistent with our quasi-1D geometry and translational invariance in these directions.

We now require constitutive relations for the fluxes and generation terms appearing in our evolution equations.
With precise definitions of the order parameters, we could formulate exact expressions for these quantities (e.g.,~through an Irving-Kirkwood procedure~\cite{Irving1950}) and precisely identify \textit{the driving forces} behind them.
For example, if the conserved order parameter were the number density, the flux can be directly connected to the linear momentum balance~\cite{Omar2023b}.
Here, to maintain generality in the selection of the order parameters, we instead propose Markovian constitutive relations for $\mathbf{J}$ and $\mathbf{s}$ in terms of ``flux-driving'' and ``generation-driving'' forces, respectively.
Before doing so, to further simplify, we assume the flux of nonconserved fields is negligible with the generation terms controlling the the dynamics of the nonconserved order parameters at each point in space (i.e.,~$|\partial_z J_{\psi_i}| << |s_{\psi_i}| \ \forall i \in \mathcal{N}$). 
While this is consistent with the commonly invoked ``Model A'' dynamics of nonconserved order parameters~\cite{Hohenberg1977} there may be specific cases in which at least one $J_{\psi_i}$ plays a significant role in the dynamics. 
These cases will be the subject of future work.

The remaining  quantities governing the dynamics of the order parameter fields are now just $J_\rho$ and each $s_{\psi_i} \ \forall i \in \mathcal{N}$.
Stationarity of the conserved field ($\partial_t \rho=0$) combined with flux-free boundary conditions results in $J_{\rho}=0$.
Additionally, stationarity of the nonconserved fields ($\partial_t \psi_i = 0 \ \forall i \in \mathcal{N}$) directly implies that $s_{\psi_i} = 0 \ \forall i \in \mathcal{N}$.
We now propose that these steady-state conditions correspond to a ``force'' balance with respect to ``flux-driving'' and ``generation-driving'' forces.
The most general linear model of this form is:
\begin{subequations}
    \label{eq:constitutivedynamics}
    \begin{align}
        & J_{\rho} = L f_{\rho}, \\
        & s_{\psi_i} = \sum_{j \in \mathcal{N}} M_{\psi_i \psi_j} f_{\psi_j} \ \forall i \in \mathcal{N},
    \end{align}
\end{subequations}
where $f_{\rho}$ is the force driving the flux of the conserved field and $f_{\psi_i}$ is the direct force driving the generation of the $i$th nonconserved field.
Here, $f_{\rho}$ and every $f_{\psi_i}$ are generally functionals of the complete spatial profile of the order parameters, $\mathbf{X}(z)$.
$L$ is a transport coefficient which must be positive to ensure the flux is in the same spatial direction as its driving force and generally depends on the local value of $\mathbf{X}$.
Analogously, $M_{\psi_i \psi_j}$ is the $ij$ component of a positive-definite (ensuring nonconserved fields are generated with positive forces and destroyed with negative forces) full-rank matrix of transport coefficients, each generally depending on the local value of $\mathbf{X}$.
Notably, $L$ has no explicit spatial dependence and hence $J_{\rho}$ inherits the spatial parity of $f_{\rho}$, i.e.,~both $J_{\rho}$ and $f_{\rho}$ are both spatially odd [i.e., $f_\rho(z) = -f_\rho(-z)$].
Similarly, each $s_{\psi_i}$ must match the spatially even parity of each $f_{\psi_j} \ \forall j \in \mathcal{N}$ [i.e., $f_{\psi_j}(z) = f_{\psi_j}(-z)$].

For passive systems, the driving forces can be identified through the framework of linear irreversible thermodynamics~\cite{degroot2013}.
In the absence of nonconserved fields, the driving force for the flux can be equivalently expressed using either the chemical potential or pressure, $f_{\rho}=-\partial_z p=-\rho \partial_z \mu_{\rho}$ which are related through the Gibbs-Duhem relation [Eq.~\eqref{eq:GDeqm}].
Note that this ``pressure'' is the mechanical pressure when the conserved field is the number density.
When nonconserved fields are present, the flux of the conserved field is still driven by pressure gradients, however the Gibbs-Duhem relation now contains additional terms related to each $\mu_{\psi_i}$.
Relaxational dynamics with $f_{\psi_i} = - \mu_{\psi_i} \ \forall i \in \mathcal{N}$ ensures that the equilibrium condition of $\mu_{\psi_i} = 0 \ \forall i \in \mathcal{N}$ is reached.

We note that the linearity of the constitutive relations proposed above limits their application to small forces.
For the context of coexistence, the crucial aspect of these relations is that they identify the correct driving forces so that we may describe the stationary state.
If these forces are identified, we can use the steady-state conditions found above to immediately identify that all forces must vanish:
\begin{subequations}
    \label{eq:all_steadystate}
    \begin{align}
        \label{eq:conserved_steadystate}
        & f_{\rho} = 0, \\
        \label{eq:nonconserved_steadystate}
        & f_{\psi_i} = 0 \ \forall i \in \mathcal{N},
    \end{align}
\end{subequations}
which are exact conditions for any system following the constitutive relations in Eq.~\eqref{eq:constitutivedynamics}.

One may microscopically derive Eq.~\eqref{eq:constitutivedynamics} for a given system following an Irving-Kirkwood procedure~\cite{Irving1950} and identify exact microscopic forms for $f_{\rho}$ and each $f_{\psi_i}$.
Alternatively, these expressions can be derived variationally for equilibrium systems using classical density functional theory~\cite{Hansen2013}.
The resulting forms for $f_{\rho}$ and each $f_{\psi_i}$ are often integro-differential equations with nonlocal contributions from long-ranged interparticle interactions and correlations.
We aim to circumvent the often difficult task of solving these equations for the complete spatial steady-state order parameter profiles.

To proceed, we perform a general expansion of $f_{\rho}$ and each $f_{\psi_i}$ with respect to spatial gradients of the order parameters.
We retain spatial gradients up to third-order in our expansion of $f_\rho$ (which has odd parity) and up to second-order in our expansion of each $f_{\psi_i}$ (which have even parity).
It will become apparent that the parity of these expansions play a prominent role in the development of coexistence criteria.
The accuracy of these expansions will depend on the degree to which any non-local terms in these forces can be approximated with local terms.

We first expand the generation-driving forces:
\begin{subequations}
    \label{eq:S_expansion}
    \begin{equation}
        f_{\psi_i} \approx f^{\rm bulk}_{\psi_i} + f^{\rm int}_{\psi_i} \ \forall i \in \mathcal{N},
    \end{equation}
    where:
    \begin{equation}
    f^{\rm int}_{\psi_i} = - \mathbf{f}_{\psi_i}^{(2,1)} : \partial_z\mathbf{X} \partial_z\mathbf{X}
    -  \mathbf{f}_{\psi_i}^{(2,2)} \cdot \partial^2_{zz} \mathbf{X} \ \forall i \in \mathcal{N}.
\end{equation}
\end{subequations}
Here, $f^{\rm bulk}_{\psi_i} \ \forall i \in \mathcal{N}$ are equations of state that we assume have at least one zero such that a homogeneous steady-state is possible.
If interfaces are unfavored, as is expected for states of macroscopic phase-separation, coexistence between phases with differing values of nonconserved fields can only be achieved if at least one $f^{\rm bulk}_{\psi_i}$ depends on $\rho$.
Each $\mathbf{f}_{\psi_i}^{(2,1)}$ and $\mathbf{f}_{\psi_i}^{(2,1)}$ are rank 2 and 1 tensors of equations of state, respectively.
Without loss of generality, we define each $\mathbf{f}_{\psi_i}^{(2,1)}$ to be symmetric as antisymmetric contributions do not impact $f^{\rm int}_{\psi_i}$.
In total, the expanded generation-driving forces contain $n_N + n_N(n_N + 1) + n_N ([n_N+1]^2 + n_N + 1) / 2$ equations of state.

We now expand the flux-driving force: 
\begin{subequations}
\label{eq:body_force_expansion}
\begin{equation}
    f_{\rho}
    \approx f_{\rho}^{(1)} + f_{\rho}^{(3)},
\end{equation}
where:
\begin{align}
    \label{eq:body_force_expansion01}
    f_{\rho}^{(1)} =&  \mathbf{f}_{\rho}^{(1,1)} \cdot\partial_z \mathbf{X}, \\
    \label{eq:body_force_expansion02}
    f_{\rho}^{(3)} =& - \mathbf{f}_{\rho}^{(3,1)} \vdots \partial_z\mathbf{X} \partial_z\mathbf{X} \partial_z\mathbf{X} \nonumber \\
    & - \mathbf{f}_{\rho}^{(3,2)} : \partial^2_{zz}\mathbf{X} \partial_z\mathbf{X} - \mathbf{f}_{\rho}^{(3,3)}\cdot \partial^3_{zzz}\mathbf{X},
\end{align}
\end{subequations}
Every element of $\mathbf{f}_{\rho}^{(1,1)}$, $\mathbf{f}_{\rho}^{(3,1)}$, $\mathbf{f}_{\rho}^{(3,2)}$, and $\mathbf{f}_{\rho}^{(3,3)}$ (rank 1, 3, 2, and 1 tensors, respectively) are generally equations of state that depend on $\mathbf{X}$ but not its spatial gradients.
Equation~\eqref{eq:body_force_expansion02} is unaffected by antisymmetric contributions to $\mathbf{f}_{\rho}^{(3,1)}$ (with respect to permuting a triplet of indices) and thus we define $\mathbf{f}_{\rho}^{(3,1)}$ to be symmetric under permuting each of its three indices.
In total, the flux-driving force contains $2(n_N + 1)+(n_N+1)^2+\frac{1}{6}(n_N+1)(n_N+2)(n_N+3)$ equations of state.

Our goal is to extract bulk coexistence criteria from the steady-state conditions that all forces must vanish.
Broadly, forces with even spatial parity, such as each $f_{\psi_i}$, contain bulk contributions that do not necessarily vanish in the absence of spatial gradients.
It is precisely these bulk contributions that will appear in our coexistence criteria. 
Conversely, forces with odd spatial parity do not contain a bulk contribution and hence one cannot straightforwardly extract coexistence criteria from a spatially uniform odd quantity.
It is for this reason that coexistence criteria for the $f_{\psi_i}=0$ conditions can be immediately obtained, as will be shown, while the coexistence criteria that result from $f_{\rho}=0$ require the conversion of this odd force to a quantity with even spatial parity.

We immediately identify $n_P n_N$ coexistence criteria from the force balance on the nonconserved fields, $f_{\psi_i}=0 \ \forall i \in \mathcal{N}$.
Substituting the absence of spatial variations of the order parameters in the bulk phases into Eq.~\eqref{eq:S_expansion}, we have:
\begin{equation}
    \label{eq:Sbulkzero}
    f^{\rm bulk}_{\psi_i} \left( \mathbf{X}^{\alpha} \right) = 0 \ \forall i \in \mathcal{N}, \ \alpha \in \mathcal{P}.
\end{equation}
This set of criteria apply to any coexistence scenario with nonconserved order parameters and, analogously to that in equilibrium [Eq.~\eqref{eq:eqmcriteria1}], correspond to a generation-free state.

It is now apparent that the conserved field, and consequently $f_{\rho}$, is what makes determining the phase diagrams of these systems a non-trivial task.
$f_{\rho}$ is odd and must be transformed to an even quantity in order to extract bulk coexistence criteria.
We therefore seek a transformation that allows us to equivalently express zero flux-driving force as spatial uniformity of an even potential-like quantity:
\begin{equation}
    \label{eq:transform}
   f_{\rho} = 0 \rightarrow \partial_z u_{\rho} = 0,
\end{equation}
where we have defined $u_{\rho}$ to be a ``pseudopotential''.
If such a transformation can be made, we will find that $u_{\rho}$ is spatially uniform with even spatial parity and therefore coexistence criteria can be extracted from this condition.
Before detailing the transformation between the flux-driving force and $u_{\rho}$, we expand $u_{\rho}$ to second-order in spatial gradients of the order parameters, consistent with our expansions of $f_{\rho}$ and each $f_{\psi_i}$:
\begin{subequations}
    \label{eq:u_expansion}
    \begin{equation}
        u_{\rho} \approx u_{\rho}^{\rm bulk} + u_{\rho}^{\rm int},
    \end{equation}
    where:
    \begin{equation}
    u_{\rho}^{\rm int} = - \mathbf{u}_{\rho}^{(2,1)} : \partial_z\mathbf{X} \partial_z\mathbf{X}
    -  \mathbf{u}_{\rho}^{(2,2)} \cdot \partial^2_{zz} \mathbf{X}.
\end{equation}
\end{subequations}
This expansion has the same form as that of each $f_{\psi_i}$ [Eq.~\eqref{eq:S_expansion}], containing a total of $1 + (n_N + 1) + ([n_N+1]^2 + n_N + 1) / 2$ equations of state in the scalar $u_{\rho}^{\rm bulk}$, vector $\mathbf{u}_{\rho}^{(2,2)}$, and rank 2 tensor $\mathbf{u}_{\rho}^{(2,1)}$.
As was the case for each $\mathbf{f}_{\psi_i}^{(2,1)}$, we define $\mathbf{u}_{\rho}^{(2,1)}$ to be symmetric as any antisymmetric contributions do not impact $u_{\rho}^{\rm int}$.

If we can find a transformation of the form of Eq.~\eqref{eq:transform}, we can simply integrate the steady-state condition $\partial_z u_{\rho}=0$ to determine coexistence criteria. 
By again noting that the order parameters are spatially uniform in the bulk phases, we find the following $n_P-1$ criteria from Eq.~\eqref{eq:u_expansion}:
\begin{equation}
    \label{eq:uequal}
    u_{\rho}^{\rm bulk} \left( \mathbf{X}^{\alpha} \right) = u_{\rho}^{\rm coexist} \ \forall \alpha \in \mathcal{P},
\end{equation}
where $u_{\rho}^{\rm coexist}$ is the coexistence value of $u_{\rho}$ which is a constant that must be determined.
Equation~\eqref{eq:uequal} is the nonequilibrium analog of equality of chemical potentials in coexisting phases [Eq.~\eqref{eq:eqmcriteria1}].

We now look to explicitly determine the form of the transformation in Eq.~\eqref{eq:transform} to extract the coexistence criteria in Eq.~\eqref{eq:uequal}.
This transformation must be bijective such that there is a one-to-one mapping between zero flux-driving force and constant $u_{\rho}$.
To maintain generality, we allow $u_{\rho}$ to contain contributions from generation-driving forces.
Allowing these contributions will prove necessary to ensure one can identify $u_{\rho}=\mu_{\rho}$ in equilibrium when the conserved field is the number density.
The simplest bijective transformation that contains contributions from the generation-driving forces is:
\begin{equation}
    \label{eq:Tansatz}
    \boldsymbol{\mathcal{T}} \cdot \overline{\mathbf{f}} = \partial_z \mathbf{u},
\end{equation}
where $\boldsymbol{\mathcal{T}}$ is the transformation tensor, a full-rank (i.e.,~all $n_N+1$ rows/columns of $\boldsymbol{\mathcal{T}}$ are linearly indepenedent) rank 2 tensor of state functions that we seek to determine.
We have introduced a pseudopotential vector, $\mathbf{u} \equiv \begin{bmatrix} u_{\rho} & u_{\psi_1} & \cdots & u_{\psi_{n_N}} \end{bmatrix}^{\rm T}$, whose elements corresponding to nonconserved fields are simply $u_{\psi_i} = - f_{\psi_i} \ \forall i \in \mathcal{N}$.
Each of the $n_N+1$ components of the effective force vector, $\overline{\mathbf{f}} \equiv \begin{bmatrix} \overline{f}_{\rho} & \overline{f}_{\psi_i} & \cdots & \overline{f}_{\psi_N} \end{bmatrix}^{\rm T}$, must be spatially odd and contain terms up to third-order in spatial gradients such that $\mathbf{u}$ is second-order and even.
We define the effective force on the conserved field to simply be the flux-driving force, $\overline{f}_{\rho} \equiv f_{\rho}$.
Similarly, for simplicity, we take elements of $\overline{\mathbf{f}}$ corresponding to nonconserved fields to only depend on the associated generation-driving force.
The simplest form that satisfies this condition while ensuring that we recover $u_{\psi_i} = - f_{\psi_i} \ \forall i \in \mathcal{N}$ is $\overline{f}_{\psi_i} \propto \partial_z f_{\psi_i} \ \forall i \in \mathcal{N}$. 
This corresponds to the rows of $\boldsymbol{\mathcal{T}}$ that are associated with nonconserved fields being diagonal.
We therefore define $\overline{f}_{\psi_i} \equiv \psi_i \partial_z f_{\psi_i} \ \forall i \in \mathcal{N}$ such that the rows of the transformation tensor corresponding to nonconserved fields can be identified as $\mathcal{T}_{\psi_i \psi_j} = - \delta_{i j} \psi_i^{-1} \ \forall i, j \in \mathcal{N}$ and $\mathcal{T}_{\psi_i \rho} = 0 \ \forall i \in \mathcal{N}$.

The second-order expansion of $\mathbf{u}$ is:
\begin{subequations}
    \label{eq:u_expansion2}
    \begin{equation}
        \mathbf{u} \approx \mathbf{u}^{\rm bulk} + \mathbf{u}^{\rm int},
    \end{equation}
    where:
    \begin{equation}
    \mathbf{u}^{\rm int} = - \mathbf{u}^{(2,1)} : \partial_z\mathbf{X} \partial_z\mathbf{X}
    -  \mathbf{u}^{(2,2)} \cdot \partial^2_{zz} \mathbf{X}.
\end{equation}
\end{subequations}
We again define each $\mathbf{u}^{(2,1)}_{i} \ \forall i \in \mathcal{X}$ to be symmetric as any antisymmetric contributions do not affect $\mathbf{u}^{\rm int}$.
Each of the elements of $\mathbf{u}$ has been expanded previously in this Section [Eq.~\eqref{eq:S_expansion} and \eqref{eq:u_expansion}], however Eq.~\eqref{eq:u_expansion2} allows us to compactly express these equations.

The expanded effective force vector is:
\begin{subequations}
\label{eq:body_force_expansion2}
\begin{equation}
    \overline{\mathbf{f}}
    \approx \overline{\mathbf{f}}^{(1)} + \overline{\mathbf{f}}^{(3)},
\end{equation}
where:
\begin{align}
    \overline{\mathbf{f}}^{(1)} =&  \overline{\mathbf{f}}^{(1,1)} \cdot\partial_z \mathbf{X}, \\
    \overline{\mathbf{f}}^{(3)} =& - \overline{\mathbf{f}}^{(3,1)} \vdots \partial_z\mathbf{X} \partial_z\mathbf{X} \partial_z\mathbf{X} \nonumber \\
    & - \overline{\mathbf{f}}^{(3,2)} : \partial^2_{zz}\mathbf{X} \partial_z\mathbf{X} - \overline{\mathbf{f}}^{(3,3)}\cdot \partial^3_{zzz}\mathbf{X}.
\end{align}
\end{subequations}
We again define each $\overline{\mathbf{f}}^{(3,1)}_i \ \forall i \in \mathcal{X}$ to be symmetric with respect to permuting its three indices.
Every component of $\overline{\mathbf{f}}$ has been previously introduced in this Section, either directly for the conserved field [Eq.~\eqref{eq:body_force_expansion}] or indirectly for the nonconserved fields (found by spatially differentiating Eq.~\eqref{eq:S_expansion} and multiplying by the associated order parameter), however Eq.~\eqref{eq:body_force_expansion2} allows us to compactly express these equations.

From our definition of the effective forces and pseudopotentials of the nonconserved fields, we only need to determine $u_{\rho}$ and one row of the transformation tensor, $\boldsymbol{\mathcal{T}}_{\rho} \equiv \begin{bmatrix} \mathcal{T}_{\rho \rho} & \mathcal{T}_{\rho \psi_1} & \cdots & \mathcal{T}_{\rho \psi_{n_N}} \end{bmatrix}^{\rm T}$, with:
\begin{equation}
    \label{eq:Tansatz_reduced}
    \boldsymbol{\mathcal{T}}_{\rho} \cdot \overline{\mathbf{f}} = \partial_z u_{\rho}.
\end{equation}
In equilibrium and when the conserved field is the number density, $u_{\rho}$ can be identified as either the chemical potential or pressure.
If one identifies $u_{\rho}$ as the pressure, which is straightforward beginning from our force balance, then $\mathcal{T}_{\rho \rho} = -1$ and $\mathcal{T}_{\rho \psi_i} = 0 \ \forall i \in \mathcal{N}$.
Alternatively, if one identifies $u_{\rho}$ as the chemical potential, $\boldsymbol{\mathcal{T}}_{\rho}$ encodes the Gibbs-Duhem relation [Eq.~\eqref{eq:GDeqm}] such that $\mathcal{T}_{\rho \rho} = -1/\rho$ and $\mathcal{T}_{\rho \psi_i} = 1/\rho \ \forall i \in \mathcal{N}$.
Hereafter, we use $u_{\rho}=\mu_{\rho}$ in equilibrium to simplify the following analysis.

Substituting the expansions for $\overline{\mathbf{f}}$ [Eq.~\eqref{eq:body_force_expansion2}] and $u_{\rho}$ [Eq.~\eqref{eq:u_expansion}] into Eq.~\eqref{eq:Tansatz_reduced}, we match terms and find a number of relations that must hold (as detailed in Appendix~\ref{secap:T}):
\begin{subequations}
\label{eq:pseudo_sys_eq}
    \begin{align}
        & \boldsymbol{\mathcal{T}}_{\rho} \cdot \overline{\mathbf{f}}^{(1,1)}  = \partial_{\mathbf{X}} u_{\rho}^{\rm bulk},\label{eq:pseudo_bulk} \\
        & \boldsymbol{\mathcal{T}}_{\rho} \cdot \overline{\mathbf{f}}^{(3,1)} = \left[ \partial_{\mathbf{X}} \mathbf{u}_{\rho}^{(2,1)}\right]^{\rm S},  \\
        & \boldsymbol{\mathcal{T}}_{\rho} \cdot \overline{\mathbf{f}}^{(3,2)} = \left( \partial_{\mathbf{X}} \mathbf{u}_{\rho}^{(2,2)} + 2 \mathbf{u}_{\rho}^{(2,1)} \right), \\
        & \partial_{\mathbf{X}}(\boldsymbol{\mathcal{T}}_{\rho} \cdot \overline{\mathbf{f}}^{(3,3)}) = \partial_{\mathbf{X}} \mathbf{u}_{\rho}^{(2,2)},
    \end{align}
\end{subequations}
where $\partial_{\mathbf{X}} \equiv \partial / \partial \mathbf{X}$ and $[\partial u^{(2, 1)}_{\rho ij} / \partial X_k]^{\rm S} \equiv \frac{1}{6}(\partial u^{(2, 1)}_{\rho ij} / \partial X_k +\partial u^{(2, 1)}_{\rho ik} / \partial X_j +\partial u^{(2, 1)}_{\rho ji} / \partial X_k +\partial u^{(2, 1)}_{\rho jk} / \partial X_i +\partial u^{(2, 1)}_{\rho ki} / \partial X_j +\partial u^{(2, 1)}_{\rho kj} / \partial X_i)$ extracts the symmetric (with respect to exchange of the $i$, $j$, and $k$ indices) portion of the third-rank tensor $\partial_{\mathbf{X}} \mathbf{u}_{\rho}^{(2,1)}$ with components $\partial u^{(2, 1)}_{\rho ij} / \partial X_k$.

Equation~\eqref{eq:pseudo_sys_eq} is an overdetermined system of $1 + n_N +2(n_N + 1)^2+\frac{1}{6}(n_N + 1)(n_N + 2)(n_N + 3)$ linear, first-order, homogeneous PDEs for $3 + 2n_N + n_N + (n_N + 1)(n_N + 2) / 2$ unknown functions (the elements of $\boldsymbol{\mathcal{T}}_{\rho}$ and each coefficient in $u_{\rho}$).
We see that the number of equations exceeds the number of unknown functions when nonserved fields are present ($n_N \geq 1$) and hence \textit{a solution is not guaranteed to exist}.
This makes clear that the coexistence theories developed for a single conserved field~\cite{Aifantis1983rule, Solon2018, Omar2023b} represent a special case where the system of PDEs in Eq.~\eqref{eq:pseudo_sys_eq} becomes an exactly determined (and consequently solveable) system of ODEs.
For this reason, the single-field coexistence theories in Refs.~\cite{Solon2018, Omar2023b} represent an exceptional case where it is guaranteed that $u_{\rho}$ exists.

One approach to solving Eq.~\eqref{eq:pseudo_sys_eq} may be to eliminate the coefficients of $u_{\rho}$ from each equation through a series of linear operations, obtaining a set of equations that solely contain $\boldsymbol{\mathcal{T}}_{\rho}$ and the known force coefficients.
If one does this, for the components of $\boldsymbol{\mathcal{T}}_{\rho}$ and the coefficients of $u_{\rho}$ to be continuous and differentiable functions of $\mathbf{X}$, it must be ensured that partial differentiation of these unknown functions with respect to the order parameters is commutative (this is formally known as involutivity~\cite{Seiler2010}).
A solution that satisfies Eq.~\eqref{eq:pseudo_sys_eq} may not have this property.
We therefore must determine additional conditions that the solution must satisfy to ensure the commutativity of partial differentiation (these additional equations are formally referred to as compatibility conditions~\cite{Seiler2010}).
Determining these conditions for an arbitrary number of order parameters is a difficult task but, nevertheless, Eq.~\eqref{eq:pseudo_sys_eq} serves as the starting point in determining solutions for $\boldsymbol{\mathcal{T}}_{\rho}$ and $u_{\rho}$.
Importantly, a solution to Eq.~\eqref{eq:pseudo_sys_eq} can be subsequently checked to ensure partial differentiation is commutative.
This commutativity is satisfied in passive systems when the pseudopotential of every order parameter is its associated chemical potential, $\mathbf{u}=\boldsymbol{\mu}$, and the transformation tensor has elements $\mathcal{T}_{\psi_i \psi_j}=-\delta_{ij} \psi_i^{-1} \ \forall i, j \in \mathcal{N}$, $\mathcal{T}_{\psi_{i} \rho} = 0 \ \forall i \in \mathcal{N}$, $\mathcal{T}_{\rho \rho} = -\rho^{-1}$, and $\mathcal{T}_{\rho \psi_i} = \rho^{-1} \ \forall i \in \mathcal{N}$.

When a full solution to Eq.~\eqref{eq:pseudo_sys_eq} (one with a full-rank $\boldsymbol{\mathcal{T}}$ that respects the commutativity of partial differentiation) exists, we can identify $u_{\rho}$ as:
\begin{subequations}
\label{eq:u_form}
    \begin{align}
        & u_{\rho}^{\rm bulk} = \int \boldsymbol{\mathcal{T}}_{\rho} \cdot \overline{\mathbf{f}}^{(1,1)} \cdot d\mathbf{X},\\
        & \mathbf{u}^{(2,1)}_{\rho} = \frac{1}{2}\left(\boldsymbol{\mathcal{T}}_{\rho} \cdot \overline{\mathbf{f}}^{(3,2)} - \partial_{\mathbf{X}}\left(\boldsymbol{\mathcal{T}_{\rho}} \cdot \overline{\mathbf{f}}^{(3,3)}\right)\right), \\
        & \mathbf{u}^{(2,2)}_{\rho} = \boldsymbol{\mathcal{T}_{\rho}}\cdot \overline{\mathbf{f}}^{(3,3)}.
    \end{align}
\end{subequations}
Combining this with the solution for the pseudopotentials associated with the nonconserved order parameters, $u_{\psi_i} = - f_{\psi_i} \ \forall i \in \mathcal{N}$, we now have an expression for every component of $\mathbf{u}$.

In systems where a full solution to Eq.~\eqref{eq:pseudo_sys_eq} can be found, we still require an additional set of coexistence criteria, analogous to the equality of pressures in coexisting phases in equilibrium.
To identify the remaining criteria, we introduce a generalized Gibbs-Duhem relation that maps the vector of pseudopotentials, $\mathbf{u}$, to a global quantity with even spatial parity, $\mathcal{G}$:
\begin{equation}
\label{eq:generalized_GD}
    \boldsymbol{\mathcal{E}} \cdot d\mathbf{u} = d\mathcal{G}.
\end{equation}
We have introduced a generalized Maxwell construction vector, $\boldsymbol{\mathcal{E}}$, containing $n_N + 1$ components which are each an equation of state.
The connection between $\boldsymbol{\mathcal{E}}$ and a Maxwell construction will later become apparent, however we can appreciate that the equilibrium Gibbs-Duhem relation [Eq.~\eqref{eq:GDeqm}] allows us to express equality of pressures across phases [Eq.~\eqref{eq:eqmcriteria4}] as the equilibrium Maxwell construction [Eqs.~\eqref{eq:eqmcriteria24} or equivalently \eqref{eq:eqmcriteria34}].

We now expand $\mathcal{G}$ to second-order:
\begin{subequations}
\label{eq:G_expansion}
\begin{equation}
    \mathcal{G} \approx \mathcal{G}^{\rm bulk} + \mathcal{G}^{\rm int},
\end{equation}
where:
\begin{equation}
    \mathcal{G}^{\rm int} = - \boldsymbol{\mathcal{G}}^{(2,1)} : \partial_z\mathbf{X} \partial_z\mathbf{X}
    -  \boldsymbol{\mathcal{G}}^{(2,2)} \cdot \partial^2_{zz}\mathbf{X}.
\end{equation}
\end{subequations}
$\mathcal{G}^{\rm bulk}$ is a scalar equation of state while $\boldsymbol{\mathcal{G}}^{(2,1)}$ and $\boldsymbol{\mathcal{G}}^{(2,2)}$ are tensors of rank 2 and 1, respectively, where each element is an equation of state.
As was the case with $\mathbf{u}_{\rho}^{(2, 1)}$ and each $\mathbf{f}_{\psi_i}^{(2, 1)}$, we define $\boldsymbol{\mathcal{G}}^{(2,1)}$ to be symmetric as antisymmetric contributions to $\boldsymbol{\mathcal{G}}^{(2,1)}$ do not impact $\mathcal{G}^{\rm int}$.
In total, $\mathcal{G}$ contains ${2+n_N+(n_N + 1)(n_N + 2) / 2}$ equations of state that each generally depend on $\mathbf{X}$.

If a global quantity can be found, our remaining coexistence criteria follow from substituting the spatial uniformity of the pseudopotential vector ($d \mathbf{u} = \mathbf{0}$) into the generalized Gibbs-Duhem relation [Eq.~\eqref{eq:generalized_GD}] which informs us that $\mathcal{G}$ is spatially constant.
As the order parameters are homogeneous within the bulk phases, from Eq.~\eqref{eq:G_expansion} we identify that $\mathcal{G}^{\rm bulk}$ is equal in coexisting phases:
\begin{equation}
    \mathcal{G}^{\rm bulk} \left( \mathbf{X}^{\alpha} \right) = \mathcal{G}^{\rm coexist} \ \forall \alpha \in \mathcal{P},
\end{equation}
providing our final $n_P - 1$ coexistence criteria which, in equilibrium, correspond to equality of pressures [Eq.~\eqref{eq:eqmcriteria4}] between phases.

We now look to determine the components of the Maxwell construction vector.
Substituting the expressions for $\mathbf{u}$ [Eq.~\eqref{eq:u_expansion2}] and $\mathcal{G}$ [Eq.~\eqref{eq:G_expansion}] into the generalized Gibbs-Duhem relation [Eq.~\eqref{eq:generalized_GD}], we identify a number of relations that must hold (as detailed in Appendix~\ref{secap:E}):
\begin{subequations}
\label{eq:E_system_eq}
    \begin{align}
        & \boldsymbol{\mathcal{E}} \cdot \partial_{\mathbf{X}} \mathbf{u}^{\rm bulk} = \partial_{\mathbf{X}} \mathcal{G}^{\rm bulk}, \label{eq:G_bulk}\\
        & \partial_{\mathbf{X}}(\boldsymbol{\mathcal{E}} \cdot \mathbf{u}^{(2, 2)}) = \partial_{\mathbf{X}} \boldsymbol{\mathcal{G}}^{(2,2)},  \label{eq:G22}\\
        & \boldsymbol{\mathcal{E}} \cdot \left(2 \mathbf{u}^{(2, 1)}+ \partial_{\mathbf{X}} \mathbf{u}^{(2, 2)}\right) = 2 \boldsymbol{\mathcal{G}}^{(2,1)} + \partial_{\mathbf{X}} \boldsymbol{\mathcal{G}}^{(2,2)}, \label{eq:G22_G21}\\
        & \boldsymbol{\mathcal{E}} \cdot \left[\partial_{\mathbf{X}} \mathbf{u}^{(2, 1)}\right]^{\rm S'}= \left[\partial_{\mathbf{X}} \boldsymbol{\mathcal{G}}^{(2,1)}\right]^{\rm S}, \label{eq:G21}
    \end{align}
\end{subequations}
where $[\partial u^{(2, 1)}_{n ij} / \partial X_k]^{\rm S'} \equiv \frac{1}{6}(\partial u^{(2, 1)}_{n ij} / \partial X_k +\partial u^{(2, 1)}_{n ik} / \partial X_j +\partial u^{(2, 1)}_{n ji} / \partial X_k +\partial u^{(2, 1)}_{n jk} / \partial X_i +\partial u^{(2, 1)}_{n ki} / \partial X_j +\partial u^{(2, 1)}_{n kj} / \partial X_i)$ extracts the symmetric (with respect to exchange of the $i$, $j$, and $k$ indices) portion of the rank 4 tensor $\partial_{\mathbf{X}} \mathbf{u}^{(2,1)}$ with components $\partial u^{(2, 1)}_{n ij} / \partial X_k$. 
Just as was the case with Eq.~\eqref{eq:pseudo_sys_eq}, Eq.~\eqref{eq:E_system_eq} is overdetermined with $1+n_N +2(n_N + 1)^2+\frac{1}{6}(n_N + 1)(n_N+2)(n_N+3)$ linear, first-order, homogeneous PDEs for $3+2n_N+(n_N+1)(n_N+2)/2$ unknown functions (each component of $\boldsymbol{\mathcal{E}}$ and each coefficient of $\mathcal{G}$).
We can again solve Eq.~\eqref{eq:E_system_eq} for $\mathcal{G}$ and $\boldsymbol{\mathcal{E}}$ and subsequently check if partial differentiation with respect to the order parameters is commutative.
In the Supplemental Material (SM)~\footnote{See the SM for supporting equations, numerical methods, and numerical support}, we differentiate Eq.~\eqref{eq:E_system_eq} when $n_N=1$ to find the conditions that ensure partial differentiation is commutative (this is formally known as prolongation~\cite{Seiler2010}), gaining additional PDEs (compatibility conditions) that must be satisfied.
We emphasize that even though we may find the additional equations that ensure partial differentiation is commutative, the system of PDEs is overdetermined and hence one may not be able to find a consistent solution to every equation: \textit{a solution is not guaranteed}.
When a full solution to Eq.~\eqref{eq:E_system_eq} exists, we can identify the form of $\mathcal{G}$:
\begin{subequations}
\label{eq:G_functional_form}
\begin{align}
    & \mathcal{G}^{\rm bulk} = \boldsymbol{\mathcal{E}} \cdot \mathbf{u}^{\rm bulk} - \int \mathbf{u}^{\rm bulk} \cdot d\boldsymbol{\mathcal{E}} , \\
    & \boldsymbol{\mathcal{G}}^{(2,1)} = \left(\boldsymbol{\mathcal{E}} \cdot \mathbf{u}^{(2,1)}  - \frac{1}{2} \mathbf{E}^{\rm T} \cdot \mathbf{u}^{(2,2)} \right),\\
    & \boldsymbol{\mathcal{G}}^{(2,2)} = \boldsymbol{\mathcal{E}}\cdot\mathbf{u}^{(2,2)},
\end{align}
\end{subequations}
where $\mathbf{E} \equiv \partial_{\mathbf{X}} \boldsymbol{\mathcal{E}}$.

In equilibrium, the generalized Gibbs-Duhem relation [Eq.~\eqref{eq:generalized_GD}] reduces to its equilibrium form [Eq.~\eqref{eq:GDeqm}], where the pseudopotential vector is simply the chemical potential vector, $\mathbf{u} = \boldsymbol{\mu}$, the global quantity is the pressure, $\mathcal{G}=p$, and the Maxwell construction vector is the order parameter vector, $\boldsymbol{\mathcal{E}} = \mathbf{X}$.
Notably, the global quantity appears to only have a clear mechanical interpretation in equilibrium.

Our final nonequilibrium coexistence criteria are:
\begin{subequations}
\label{eq:final_coex_criteria}
\begin{align}
    & \mathbf{u}^{\rm bulk} (\mathbf{X}^\alpha) = \mathbf{u}^{\rm coexist} \ \forall \alpha \in \mathcal{P}, \label{eq:u_coex_criteria} \\
    & \mathcal{G}^{\rm bulk} (\mathbf{X}^\alpha) = \mathcal{G}^{\rm coexist} \ \forall \alpha \in \mathcal{P}, \label{eq:G_coex_criteria}
\end{align}
\end{subequations}
where $u_{\psi_i}^{\rm coexist}=0 \ \forall i \in \mathcal{N}$ from Eq.~\eqref{eq:Sbulkzero} while $u_{\rho}^{\rm coexist}$ and $\mathcal{G}^{\rm coexist}$ are generally non-vanishing and must be determined by equating $u^{\rm bulk}_{\rho}$ and $\mathcal{G}^{\rm bulk}$ across phases, respectively.
Subtracting the number of equations in Eq.~\eqref{eq:final_coex_criteria}, $2 n_P - 2 + n_P n_N$, from the number of variables describing the coexisting phases, $n_P(n_N + 1)$, we find the number of degrees of freedom to be $2-n_P$, recovering the equilibrium Gibbs phase rule as expected~\cite{Chiu2024TheoryCoexistence}.

In an analogous manner to the equilibrium Maxwell construction [Eq.~\eqref{eq:eqmcriteria24} or equivalently Eq.~\eqref{eq:eqmcriteria34}], we can use the generalized Gibbs-Duhem relation [Eq.~\eqref{eq:generalized_GD}] to re-express equality of $\mathcal{G}^{\rm bulk}$ across phases as a generalized equal-area Maxwell construction:
\begin{subequations}
\label{eq:allG_maxwell_dE_form}
    \begin{equation}
    \label{eq:G_maxwell_dE_form}
    \int_{\boldsymbol{\mathcal{E}}^\alpha}^{\boldsymbol{\mathcal{E}}^\beta} \left[\mathbf{u}^{\rm bulk}(\mathbf{X}) - \mathbf{u}^{\rm coexist} \right]
    \cdot d\boldsymbol{\mathcal{E}} = 0 \ \forall \alpha, \beta \in \mathcal{P},
\end{equation}
where the integration is path-independent as it is equivalent to an integral over $-d \mathcal{G}^{\rm bulk}$ between the $\alpha$ and $\beta$ phases. 
This path-independence allows us to select the most convenient path to integrate, which is often $\mathbf{X}^*(\rho)$ satisfying $u_{i}(\mathbf{X}^*) = 0 \ \forall i \in \mathcal{N}$ for all $\rho$ considered. 
This path selection reduces Eq.~\eqref{eq:G_maxwell_dE_form} to a one-dimensional integral:
    \begin{equation}
    \label{eq:G_maxwell_dE_form2}
    \int_{\mathcal{E}_\rho^{\alpha}}^{\mathcal{E}_\rho^{\beta}} \left[u_{\rho}^{\rm bulk}(\mathbf{X}^*) - u_{\rho}^{\rm coexist} \right]
    d \mathcal{E}_{\rho} = 0 \ \forall \alpha, \beta \in \mathcal{P}.
\end{equation}
\end{subequations}
Equation~\eqref{eq:allG_maxwell_dE_form} is the nonequilibrium generalization of Eqs.~\eqref{eq:eqmcriteria24} and \eqref{eq:eqmcriteria34}.

If the components of $\boldsymbol{\mathcal{E}}$ are not strictly monotonic functions of $\mathbf{X}$, there is ambiguity in evaluating the generalized Maxwell constructions in Eq.~\eqref{eq:allG_maxwell_dE_form}.
This ambiguity is eliminated by expressing these criteria as ``weighted-area'' constructions, substituting $d \boldsymbol{\mathcal{E}} = \mathbf{E} \cdot d \mathbf{X}$ into Eq.~\eqref{eq:G_maxwell_dE_form}:
\begin{subequations}
    \begin{equation}
    \label{eq:G_maxwell_drho_form}
    \int_{\boldsymbol{\mathcal{E}}^\alpha}^{\boldsymbol{\mathcal{E}}^\beta} \left[\mathbf{u}^{\rm bulk}(\mathbf{X}) - \mathbf{u}^{\rm coexist} \right]
    \cdot \mathbf{E} \cdot  d\mathbf{X} = 0 \ \forall \alpha, \beta \in \mathcal{P},
\end{equation}
which, along the path $\mathbf{X}^*(\rho)$, reduces to:
    \begin{equation}
    \label{eq:G_maxwell_drho_form2}
    \int_{\rho^{\alpha}}^{\rho^{\beta}} \left[u_{\rho}^{\rm bulk}(\mathbf{X}^*) - u_{\rho}^{\rm coexist} \right]
    \mathbf{E}_{\rho} \cdot d \mathbf{X}^* = 0 \ \forall \alpha, \beta \in \mathcal{P},
\end{equation}
\end{subequations}
where $\mathbf{E}_{\rho} \equiv \partial_{\mathbf{X}} \mathcal{E}_{\rho}$ is the relevant row of $\mathbf{E}$, i.e.,~its components are $E_{\rho i} \ \forall i \in \mathcal{X}$.

Equation~\eqref{eq:final_coex_criteria} can be straightforwardly recast as a generalized common tangent construction on a suitably defined effective bulk free energy, $w^{\rm bulk} \equiv \int \mathbf{u}^{\rm bulk} \cdot d\boldsymbol{\mathcal{E}}$, with respect to $\boldsymbol{\mathcal{E}}$:
\begin{subequations}
    \label{eq:common_tangent}
    \begin{align}
        &\frac{\partial w^{\rm bulk}}{\partial \boldsymbol{\mathcal{E}}} \bigg |_{\mathbf{X}=\mathbf{X}^{\alpha}} = \mathbf{u}^{\rm coexist} \ \forall \alpha \in \mathcal{P}, \label{eq:common_tangent_u}\\
        &\mathbf{u}^{\rm coexist} \cdot \boldsymbol{\mathcal{E}}(\mathbf{X}^{\alpha}) - w^{\rm bulk}(\mathbf{X}^{\alpha}) = \mathcal{G}^{\rm coexist} \ \forall \alpha \in \mathcal{P}. \label{eq:common_tangent_G}
    \end{align}
\end{subequations}
The definition of $w^{\rm bulk}$ requires its Hessian to be symmetric, which is guaranteed if the Hessian of $\mathcal{G}^{\rm bulk}$ is symmetric.
It is then clear that when the generalized Gibbs-Duhem relation is satisfied and every element of $\boldsymbol{\mathcal{E}}$ is non-constant, the coexistence criteria can be expressed as a common tangent construction on $w^{\rm bulk}$ with respect to $\boldsymbol{\mathcal{E}}$.
While Eq.~\eqref{eq:common_tangent} is an appealing form of the criteria, only in equilibrium can it be derived from an extremization of $w^{\rm bulk}$ subject to the appropriate physical constraints.
In Appendix~\ref{secap:freeenergy}, we demonstrate that the conditions for defining an effective free energy functional, $\mathcal{W}[\mathbf{X}(z)]$, such that $\mathbf{u} = \delta \mathcal{W}[\mathbf{X}(z)] / \delta \boldsymbol{\mathcal{E}}$, are more restrictive than the conditions necessary for $\mathcal{G}$ to exist [Eq.~\eqref{eq:E_system_eq}]. 
This contrasts the case in which there is a single (conserved) order parameter, where Solon \textit{et al}.~\cite{Solon2018} demonstrated that a free energy functional can be readily defined.

While the criteria described above constitute, to the best of our knowledge, the first set of genuinely nonequilibrium coexistence criteria with coupled conserved and nonconserved order parameters, exact solutions for $\mathbf{u}$ and $\mathcal{G}$ may not always be possible. 
For systems in which formally exact solutions cannot be obtained but macroscopic phase separation remains a steady-state, we look to develop \textit{approximate} coexistence criteria that one can use to estimate the phase boundaries.
To do so, we consider systems where $u_{\rho}$ and associated transformation tensor, $\boldsymbol{\mathcal{T}}_{\rho}$, can be found (and thus a solution to Eq.~\eqref{eq:pseudo_sys_eq} exists).
In these systems, the final set of coexistence criteria corresponds to equality of $\mathcal{G}^{\rm bulk}$ across phases and relies on a solution to Eq.~\eqref{eq:E_system_eq}, which does not necessarily exist.
In systems where we cannot find a full solution to Eq.~\eqref{eq:E_system_eq}, we propose partitioning the generalized Gibbs-Duhem relation into bulk and interfacial contributions (as detailed in Appendix~\ref{secap:partial}):
\begin{subequations}
\label{eq:GD_split}
\begin{align}
    & \boldsymbol{\mathcal{E}}^{\rm bulk} \cdot d\mathbf{u}^{\rm bulk} = d\mathcal{G}^{\rm bulk}, \label{eq:E_bulk}\\
    & \boldsymbol{\mathcal{E}}^{\rm int} \cdot d\mathbf{u}^{\rm int} = d\mathcal{G}^{\rm int}. \label{eq:E_int}
\end{align}
\end{subequations}
such that $\boldsymbol{\mathcal{E}}^{\rm bulk}$ satisfies Eq.~\eqref{eq:G_bulk} and $\boldsymbol{\mathcal{E}}^{\rm int}$ satisfies Eqs.~\eqref{eq:G22}-\eqref{eq:G21}.
Only when $\boldsymbol{\mathcal{E}} = \boldsymbol{\mathcal{E}}^{\rm bulk} = \boldsymbol{\mathcal{E}}^{\rm int}$ is the global quantity rigorously defined with Eq.~\eqref{eq:E_system_eq}.
When this is the case, the generalized Gibbs-Duhem relation [Eq.~\eqref{eq:generalized_GD}] holds and $\mathcal{G}^{\rm bulk} \equiv \boldsymbol{\mathcal{E}}^{\rm bulk} \cdot \mathbf{u}^{\rm bulk} - \int \mathbf{u}^{\rm bulk} \cdot d \boldsymbol{\mathcal{E}}^{\rm bulk}$ is equal in coexisting phases.
Otherwise, when $\boldsymbol{\mathcal{E}}^{\rm bulk} \neq \boldsymbol{\mathcal{E}}^{\rm int}$, $\mathcal{G}^{\rm bulk}$ generally takes a different value in each of the coexisting phases.

As shown in Appendix~\ref{secap:weightedarea}, the weighted-area construction [Eq.~\eqref{eq:G_maxwell_drho_form}] using $\mathbf{E} = \mathbf{E}^{\rm int}\equiv \partial_{\mathbf{X}} \boldsymbol{\mathcal{E}}^{\rm int}$ is guaranteed to vanish when evaluated along the spatial coexistence profile, $\mathbf{X}^{\rm c} (z)$, if a solution for $\boldsymbol{\mathcal{E}}^{\rm int}$ [Eqs.~\eqref{eq:G22}-\eqref{eq:G21}] can be found:
\begin{multline}
    \label{eq:G_maxwell_drho_form_int}
    \int_{z^\alpha}^{z^\beta} \bigg[\mathbf{u}^{\rm bulk}\left (\mathbf{X}^{\rm c}\right) \\ - \mathbf{u}^{\rm coexist} \bigg]
    \cdot \mathbf{E}^{\rm int} \left (\mathbf{X}^{\rm c}\right) \cdot  \partial_z \mathbf{X}^{\rm c} dz = 0 \ \forall \alpha, \beta \in \mathcal{P},
\end{multline}
where $z^{\alpha}$ and $z^{\beta}$ are positions deep in the homogeneous $\alpha$ and $\beta$ phases, respectively.
The value of this integral is now path-dependent, as it is no longer the difference in $\mathcal{G}^{\rm bulk}$ between phases.
Importantly, $\mathbf{X}^{\rm c} (z)$ is the full spatial coexistence profile that we set out to avoid determining.
We can then gain approximate coexistence criteria by performing the weighted-area construction with $\mathbf{E} = \mathbf{E}^{\rm int}$, evaluating the integral along a path other than the generally unknown $\mathbf{X}^{\rm c} (z)$ such as $\mathbf{X}^{*}(\rho)$ in Eq.~\eqref{eq:G_maxwell_drho_form2}.
However, the weighted-area construction is not guaranteed to vanish along these other paths, where its value quantifies the departure of the criteria from exactness.
Importantly, the approximation gains accuracy as the selected integration path approaches the actual coexistence profiles.

The approximate criteria in Eq.~\eqref{eq:G_maxwell_drho_form_int} makes use of $\boldsymbol{\mathcal{E}}^{\rm int}$, however we can alternatively approximate the final set of criteria by setting $\mathcal{G}^{\rm bulk}$ equal across phases, i.e.,~using the weighted-area construction [Eq.~\eqref{eq:G_maxwell_drho_form}] with $\mathbf{E}=\mathbf{E}^{\rm bulk} \equiv \partial_{\mathbf{X}} \boldsymbol{\mathcal{E}}^{\rm bulk}$.
When using $\mathbf{E}=\mathbf{E}^{\rm bulk}$, the weighted-area construction is path-independent as it is equal to the difference in $\mathcal{G}^{\rm bulk}$ between phases, however this difference is generally finite when $\boldsymbol{\mathcal{E}}^{\rm bulk} \neq \boldsymbol{\mathcal{E}}^{\rm int}$.
This approach, in contrast to Eq.~\eqref{eq:G_maxwell_drho_form_int}, does not inherently contain a degree of freedom (such as the integration path) to change the quality of the approximation. 
In the event that one can access the coexistence profile to check the approximation, the difference in $\mathcal{G}^{\rm bulk}$ between $\alpha$ and $\beta$ phases can be determined (as shown in Appendix~\ref{secap:bulkapprox}) with:
\begin{multline}
    \label{eq:deltaG}
    \Delta^{\alpha \beta} \mathcal{G}^{\rm bulk} \equiv \mathcal{G}^{\rm bulk} \left( \mathbf{X}^{\alpha}\right) - \mathcal{G}^{\rm bulk} \left( \mathbf{X}^{\beta}\right) \\ = \int_{z^{\alpha}}^{z^{\beta}} \bigg[ \mathbf{u}^{\rm bulk}\left(\mathbf{X}^{\rm c} \right) - \mathbf{u}^{\rm coexist} \bigg] \cdot \bigg[ \mathbf{E}^{\rm bulk}\left(\mathbf{X}^{\rm c} \right) \\ - \mathbf{E}^{\rm int}\left(\mathbf{X}^{\rm c} \right) \bigg] \cdot \partial_z \mathbf{X}^{\rm c} dz.
\end{multline}
When $\Delta^{\alpha \beta} \mathcal{G}^{\rm bulk} \approx 0$ for every pair of coexisting phases, equating $\mathcal{G}^{\rm bulk}$ across phases is a good approximation for the final set of coexistence criteria.

\section{\label{sec:models} Application to Nonequilibrium Models}

In this Section, we introduce phenomenological models of nonequilibrium field dynamics and numerically obtain the phase diagrams of these systems by determining the complete spatial order parameter profiles. 
This allows us to directly evaluate our theory of nonequilibrium coexistence which enables the construction of phase diagrams solely from bulk equations of state.

A number of phenomenological nonequilibrium field theories have been proposed to describe systems with a single nonconserved field, such as Active Model A (AMA)~\cite{Cates2019, Caballero2020}, a single conserved field, such as Active Model B+ (AMB+)~\cite{Wittkowski2014, Tjhung2018, Cates2019}, and coupled conserved fields, such as the Nonreciprocal Cahn-Hilliard (NRCH) model~\cite{Saha2020,You2020}.
Here, we introduce Active Model C+ (AMC+) which couples a conserved field that undergoes dynamics similar to those of AMB+ with $n_N$ nonconserved fields that undergo dynamics similar to AMA.
In particular, AMC+ can be seen as a form of the recently introduced Multicomponent Active Model B+ (MAMB+)~\cite{Chiu2024TheoryCoexistence} where $n_N$ of the fields are nonconserved.
It was demonstrated in Ref.~\cite{Chiu2024TheoryCoexistence} that MAMB+ preserves clear limits of gradient and passive dynamics while capturing all possible first and third-order contributions to flux-driving forces and all bulk and second-order contributions to the pseudopotentials.
AMC+ retains this structure with the simplification that the pseudopotentials of the nonconserved fields can be immediately identified.
In one spatial dimension (this can be straightforwardly generalized), AMC+ can be mapped to the dynamics in Eq.~\eqref{eq:constitutivedynamics} with the following forms for the flux-driving and generation-driving forces:
\begin{subequations}
\label{eq:AMC+_definition}
\begin{align}
    & f_{\rho} = -\rho \bigg( \partial_z u_{\rho}^0 - \sum_{i \in \mathcal{N}} \frac{\psi_i}{\rho} \partial_z f_{\psi_i} + \boldsymbol{\eta}_{\rho}^{\rm N} \cdot \partial_z\mathbf{X} \nonumber \\ & - \boldsymbol{\xi}_{\rho}^{{\rm A}} : \partial^2_{zz}\mathbf{X} \partial_z \mathbf{X}  - \boldsymbol{\theta}_{\rho} \vdots \partial_z\mathbf{X} \partial_z\mathbf{X} \partial_z\mathbf{X} \bigg), \label{eq:amc+_frho}\\
    & u_{\rho}^0 = \frac{\delta \mathcal{F}}{\delta \rho} + \tau_{\rho}^{\rm N} - \boldsymbol{\lambda}_{\rho} : \partial_z\mathbf{X} \partial_z\mathbf{X} - \boldsymbol{\pi}_{\rho}^{{\rm A'}} \cdot \partial^2_{zz} \mathbf{X}, \label{eq:amc+_urho} \\
    & f_{\psi_i} = -\frac{\delta \mathcal{F}}{\delta \psi_i} - \tau_{\psi_i}^{\rm N} + \boldsymbol{\lambda}_{\psi_i} : \partial_z\mathbf{X} \partial_z\mathbf{X} + \boldsymbol{\pi}_{\psi_i}^{{\rm A'}} \cdot \partial^2_{zz} \mathbf{X} \ \forall i \in \mathcal{N}, \label{eq:amc+_upsi} \\
    & \mathcal{F} \left[ \mathbf{X}(z) \right] = \int_V d \mathbf{r} \left( f^{\rm bulk} + \frac{1}{2} \boldsymbol{\kappa} : \partial_z\mathbf{X} \partial_z\mathbf{X} \right) \label{eq:amc+_f}.
\end{align}
\end{subequations}
There are four distinct contributions to $f_{\rho}$.
The first is driven by spatial gradients of $u_{\rho}^0$ and each $f_{\psi_i}$, resembling passive dynamics where gradients in chemical potentials drive fluxes.
Here, $u_{\rho}^0$ would be $u_{\rho}$ if every element of the ``non-gradient'' tensors $\boldsymbol{\eta}_{\rho}^{\rm N}$, $\boldsymbol{\xi}_{\rho}^{\rm A}$, and $\boldsymbol{\theta}_{\rho}$ were zero.
First-order contributions to $f_{\rho}$ that cannot be integrated into the bulk contribution to $u_{\rho}^0$ are captured by $\boldsymbol{\eta}_{\rho}^{\rm N}$, a rank 1 tensor of state functions.
The next two terms capture third-order non-integrable contributions to $f_{\rho}$, where $\boldsymbol{\xi}_{\rho}^{\rm A}$ is an antisymmetric rank 2 tensor of state functions and $\boldsymbol{\theta}_{\rho}$ is a symmetric (with respect to three-fold permutation of indices) rank 3 tensor of state functions.

When $f_{\rho}$ is fully integrable, $u_{\rho}=u_{\rho}^0$ can be immediately identified, whereas the pseudopotential of each nonconserved field is always known, $u_{\psi_i}=-f_{\psi_i} \ \forall i \in \mathcal{N}$.
Each pseudopotential, for both conserved and nonconserved fields, contains four contributions.
The first is variational in that it is derived from a free energy functional, $\mathcal{F}$, which would be the system free energy in the absence of the other terms.
The next three contributions are non-variational, the first of which is the rank 1 tensor of state functions $\boldsymbol{\tau}_{\gamma}^{\rm N} \ \forall \gamma \in \mathcal{X}$ that captures contributions to $u_{\gamma}^{\rm bulk}$ that cannot be integrated into the bulk portion of the would-be free energy, $f^{\rm bulk}$.
The last two non-variational contributions are second-order in spatial gradients.
The coefficient $\boldsymbol{\lambda}_{\gamma} \ \forall \gamma \in \mathcal{X}$ is a symmetric rank 2 tensor of state functions that breaks the relationship between interfacial terms in the associated $u_{\gamma}^{\rm int}$ that results from the functional derivative.
Finally, the coefficient $\boldsymbol{\pi}_{\gamma}^{{\rm A'}} \ \forall \gamma \in \mathcal{X}$ is a rank 1 tensor of state functions capturing contributions that cannot result from a functional derivative due to the symmetries of $\boldsymbol{\kappa}$, a positive-definite rank 2 tensor of state functions that ensures gradients in the order parameters are penalized in the free energy.
In particular, $\boldsymbol{\kappa}$ is symmetric and is also symmetric under permuting indices after differentiating, i.e.,~exchanging any of the $\nu$, $\gamma$, and $\zeta$ indices does not alter $\partial_{\nu} \kappa_{\gamma \zeta}$~\cite{Hansen2013}.

\begin{figure}
	\centering
	\includegraphics[width=.475\textwidth]{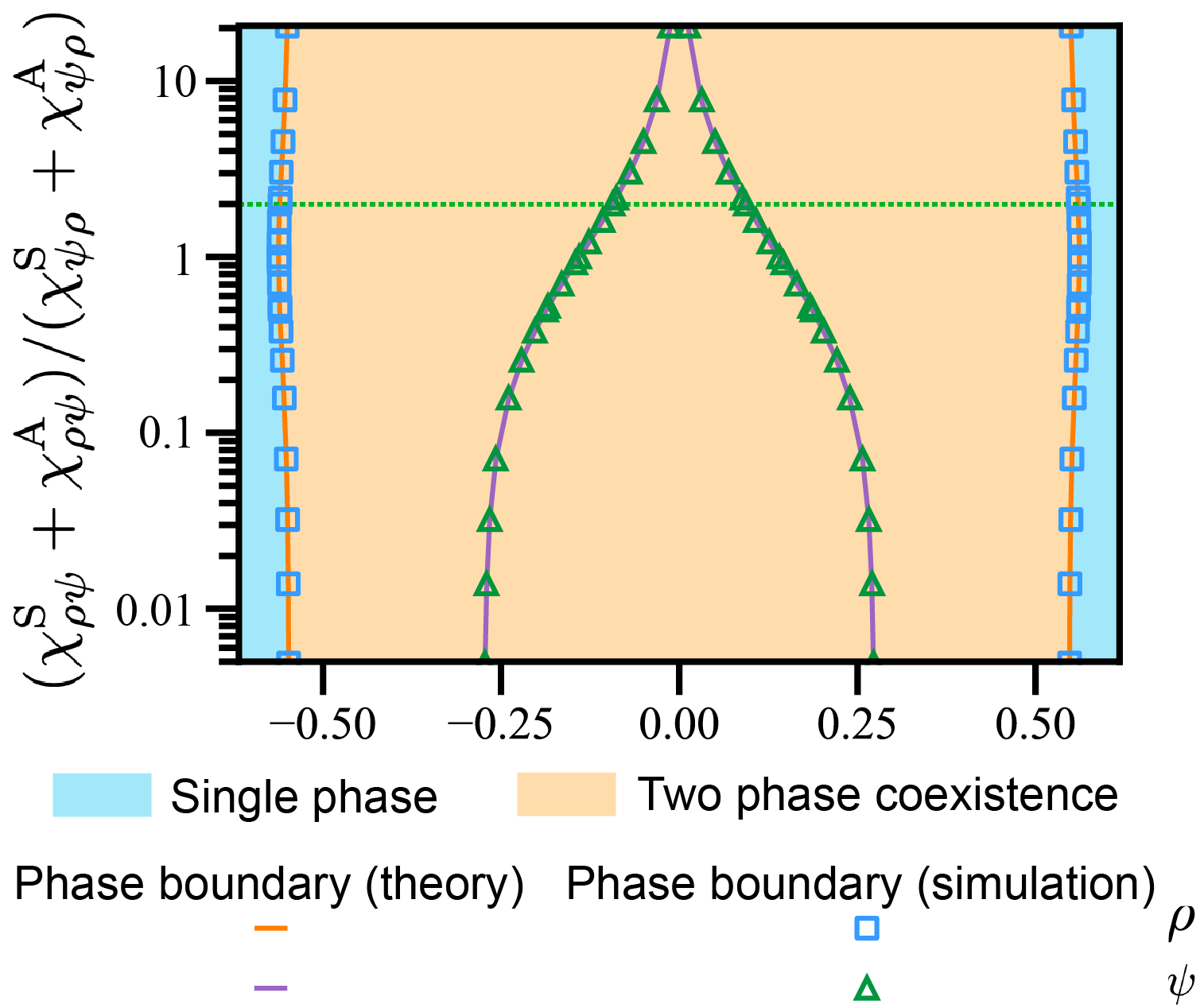}
	\caption{\protect\small{{Phase diagram of AMC with $n_N=1$, displaying the coexistence values of $\rho$/$\psi$ as a function of the ratio $(\chi + \alpha) / (\chi - \alpha)$. We set $\overline{\kappa}_{\rho \rho} = \overline{\kappa}_{\psi \psi} = \overline{\kappa}_{\rho \psi} = 0.01$, $\overline{\kappa}_{\psi \rho}=0.005$, and $\chi = 1/2$. The dotted green line indicates the limit where $\overline{\kappa}_{\rho \psi} (\chi - \alpha) = \overline{\kappa}_{\psi \rho} (\chi + \alpha)=2$ and the coexistence criteria are exact.}}}
	\label{fig:nrpd}
\end{figure}

Our model reduces to AMB+~\cite{Tjhung2018, Cates2019} in the absence of any nonconserved fields ($n_N=0$).
In AMB+, there are two nonequilibrium parameters: one analogous to our $\xi$ that breaks the gradient structure of $f_{\rho}$ and one analogous to our $\lambda$ that breaks the variational structure of $u_{\rho}^0$.
In one spatial dimension when $\xi$ is a constant, $\lambda$ can be redefined to include the effects of $\xi$ such that the dynamics are in gradient form.
Introducing nonconserved fields ($n_N>0$) allows both the gradient and variational structure of passive dynamics to be broken in more ways.
Counting the number of unique, nonzero elements of $\boldsymbol{\eta}_{\rho}^{\rm N}$, $\boldsymbol{\xi}_{\rho}^{\rm A}$, and $\boldsymbol{\theta}_{\rho}$ we see that there are $\frac{1}{6} n_N^3 + \frac{3}{2}n_N^2 + \frac{13}{6} n_N + 1$ terms that can break the gradient structure of $f_{\rho}$.
Performing the same counting for each $\boldsymbol{\tau}_{\gamma}^{\rm N}$, $\boldsymbol{\lambda}_{\gamma}$, and $\boldsymbol{\pi}_{\gamma}^{{\rm A'}}$ we find $\frac{1}{2}n_N^3 + \frac{5}{2} n_N^2 + 4 n_N + 1$ terms that can break the variational structure of $u_{\rho}^0$ and each $f_{\psi_i}$.
In contrast to AMB+, the asymmetry of $\boldsymbol{\xi}_{\rho}^{\rm A}$ means $\boldsymbol{\lambda}_{\rho}$ cannot be redefined to include the effects of a constant $\boldsymbol{\xi}_{\rho}^{\rm A}$, even in one spatial dimension.

We define Active Model C (AMC) to be a simplified version of AMC+ where $\boldsymbol{\eta}_{\rho}^{\rm N} = \mathbf{0}$, $\boldsymbol{\xi}_{\rho}^{\rm A} = \mathbf{0}$, and $\boldsymbol{\theta}_{\rho} = \mathbf{0}$ such that the conserved field experiences gradient dynamics with the passive form of the transformation tensor ($\mathcal{T}_{\psi_i \psi_j}=-\delta_{ij} \psi_i^{-1} \ \forall i, j \in \mathcal{N}$, $\mathcal{T}_{\psi_{i} \rho} = 0 \ \forall i \in \mathcal{N}$, $\mathcal{T}_{\rho \rho} = -\rho^{-1}$, and $\mathcal{T}_{\rho \psi_i} = \rho^{-1} \ \forall i \in \mathcal{N}$) and $u_{\rho}=u_{\rho}^0$.
To achieve the variational Model C~\cite{Hohenberg1977} (MC) dynamics of passive systems, every element of every $\tau_{\gamma}^{\rm N}$, $\boldsymbol{\lambda}_{\gamma}$, and $\boldsymbol{\pi}_{\gamma}^{\rm A'}$ must additionally be zero.
A nonzero component of any of these tensors breaks the passive structure of the dynamics.

\begin{figure*}
	\centering
	\includegraphics[width=.95\textwidth]{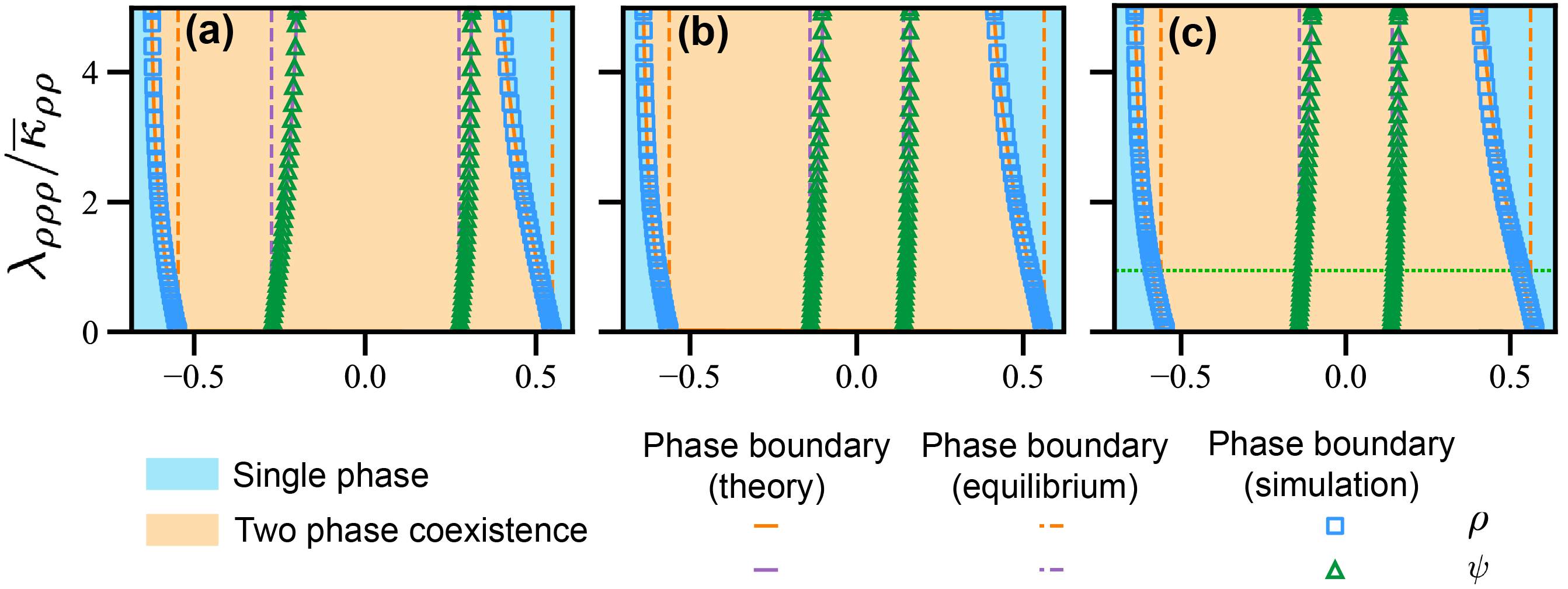}
	\caption{\protect\small{{Phase diagrams of AMC with $n_N=1$ as a function of $\rho$/$\psi$ and $\lambda_{\rho \rho \rho} / \overline{\kappa}_{\rho \rho}$. We set $\overline{\kappa}_{\psi \rho}=0$ and $\overline{\kappa}_{\rho \rho} = \overline{\kappa}_{\psi \psi} = 0.01$ in every case. We consider systems with (a) exact coexistence criteria by setting $\chi=-\alpha=1/4$ and $\overline{\kappa}_{\rho \psi} = \lambda_{\rho \psi \rho} = \lambda_{\rho \rho \psi} = \lambda_{\rho \psi \psi} = 0$, (b) approximate coexistence criteria by setting $\chi=1/2$, $\alpha=0$, and $\overline{\kappa}_{\rho \psi} = \lambda_{\rho \psi \rho} = \lambda_{\rho \rho \psi} = \lambda_{\rho \psi \psi} = 0$, and (c) no coexistence criteria within our theory by setting $\chi=1/2$, $\alpha=0$, and $\overline{\kappa}_{\rho \psi} = \lambda_{\rho \psi \rho} = \lambda_{\rho \rho \psi} = \lambda_{\rho \psi \psi} = 0.01$. The dotted green line in (c) indicates where $\lambda_{\rho \rho \rho} = \lambda_{\rho \psi \psi}$ and well-defined approximate coexistence criteria can be formulated. When this is the case, the interfacial Maxwell construction vector takes the form ${\boldsymbol{\mathcal{E}}^{\rm int} = \begin{bmatrix} \exp \left( 2 \lambda_{\rho \rho \rho} \rho / \overline{\kappa}_{\rho \rho} + 2 \lambda_{\rho \rho \psi} \psi / \overline{\kappa}_{\rho \psi} \right) & 0 \end{bmatrix}^T}$. When $\lambda_{\rho \rho \rho} \neq \lambda_{\rho \psi \psi}$, a solution to Eqs.~\eqref{eq:G22}-\eqref{eq:G21} does not exist. In this case, we continue to use the weighted-area construction with $\mathbf{E}^{\rm int}$ to predict the phase diagram in (c) even though it is not a well-defined approximation here.}}}
	\label{fig:ambApd}
\end{figure*}

We now consider systems with dynamics corresponding to AMC, where nonequilibrium effects can come from nonzero $\tau_{\gamma}^{\rm N}$, $\boldsymbol{\lambda}_{\gamma}$, and $\boldsymbol{\pi}_{\gamma}^{\rm A'}$ for any order parameter $\gamma \in \mathcal{X}$.
We consider the bulk contributions to the pseudopotentials to result from:
\begin{subequations}
    \begin{align}
        & f^{\rm bulk} = \sum_{i \in \mathcal{X}} \left( \frac{a_i}{2}X_i^2+\frac{b_i}{4}X_i^4 \right) + \frac{1}{2}\boldsymbol{\chi}^{\rm S} : \mathbf{X} \mathbf{X}, \\
        & \boldsymbol{\tau}^{\rm N} = \boldsymbol{\chi}^{\rm A} \cdot \mathbf{X},
    \end{align}
\end{subequations}
where $f^{\rm bulk}$ appears in the free energy functional [see Eq.~\eqref{eq:amc+_f}], $a_i$ and $b_i$ are constants, and $\boldsymbol{\chi}^{\rm S}$ and $\boldsymbol{\chi}^{\rm A}$ are symmetric and antisymmetric rank 2 tensors of constants, respectively.
We further assume that $\boldsymbol{\kappa}$ and every $\boldsymbol{\lambda}_{\gamma}$ and $\boldsymbol{\pi}_{\gamma}^{\rm A'}$ are constant. 
For convenience, we define $\overline{\kappa}_{ij} \equiv \kappa_{ij} + \pi_{ij}^{\rm A'} \ \forall i, j \in \mathcal{X}$.
We now focus on a single nonconserved field ($n_N=1$), $\psi$, and set $a_{\rho}=-1.2$, $a_{\psi}=1$, $b_{\rho}=4$, $b_{\psi}=0$, $\chi_{\rho \rho}^{\rm S} = \chi_{\psi \psi}^{\rm S} = 0$, $\chi_{\rho \psi}^{\rm S} = \chi_{\psi \rho}^{\rm S} = \chi$, $\chi_{\rho \psi}^{\rm A} = -\chi_{\psi \rho}^{\rm A} = \alpha$, and $\boldsymbol{\lambda}_{\psi}=\mathbf{0}$.

We first consider systems where $\boldsymbol{\lambda}_{\rho} = \mathbf{0}$ and therefore nonequilibrium effects are solely due to $\alpha \neq 0$ and $\overline{\kappa}_{\rho \psi} \neq \overline{\kappa}_{\psi \rho}$.
These dynamics resemble the two-component NRCH model~\cite{Saha2020, You2020} except one of the components is nonconserved.
When the parameters of this model are constrained such that $\overline{\kappa}_{\rho \psi} (\chi - \alpha) = \overline{\kappa}_{\psi \rho} (\chi + \alpha)$, it has been shown that the equilibrium coexistence criteria [Eq.~\eqref{eq:eqmcriteria}] are exact~\cite{Brauns2024NonreciprocalFields, Saha2024, Greve2024} despite the nonequilibrium nature of the system dynamics.
This is due to the fact that these systems have ``spurious'' gradient dynamics~\cite{Frohoff2023}.
We aim to recover this result using our theory while extending the coexistence criteria to systems where $\overline{\kappa}_{\rho \psi} (\chi - \alpha) \neq \overline{\kappa}_{\psi \rho} (\chi + \alpha)$.

We now look to solve Eq.~\eqref{eq:E_system_eq} for $\boldsymbol{\mathcal{E}}$ and $\mathcal{G}$, augmented with the compatibility conditions provided in the SM~\cite{Note1}.
For arbitrary values of the parameters, these equations do not have a general solution.
However, solving for $\boldsymbol{\mathcal{E}}^{\rm bulk}$ and $\boldsymbol{\mathcal{E}}^{\rm int}$ independently (as described in Appendix~\ref{secap:partial}), we find:
\begin{subequations}
    \begin{align}
    & \boldsymbol{\mathcal{E}}^{\rm bulk} = \begin{bmatrix}
        \rho & \frac{\chi+\alpha}{\chi-\alpha}\psi
    \end{bmatrix}^{\rm T},\\
    & \boldsymbol{\mathcal{E}}^{\rm int} = \begin{bmatrix}
        \rho & \frac{\overline{\kappa}_{\rho \psi}}{\overline{\kappa}_{\psi \rho}}\psi
    \end{bmatrix}^{\rm T}.
\end{align}
\end{subequations}
If $\overline{\kappa}_{\rho \psi} (\chi - \alpha) = \overline{\kappa}_{\psi \rho} (\chi + \alpha)$, it is clear that $\boldsymbol{\mathcal{E}}^{\rm bulk}= \boldsymbol{\mathcal{E}}^{\rm int}$ and Eq.~\eqref{eq:final_coex_criteria} is the exact coexistence criteria.
This is the limit where the system experiences spurious gradient dynamics~\cite{Frohoff2023, Greve2024} and where our coexistence criteria are equivalent to those in Refs.~\cite{Brauns2024NonreciprocalFields, Saha2024,Greve2024}.
If $\overline{\kappa}_{\rho \psi} (\chi - \alpha) \neq \overline{\kappa}_{\psi \rho} (\chi + \alpha)$, we approximate the criterion corresponding to equality of $\mathcal{G}^{\rm bulk}$ ($n_P=2$ such that this is a single criterion) [Eq.~\eqref{eq:G_coex_criteria}] by defining $\mathcal{G}^{\rm bulk} \equiv \boldsymbol{\mathcal{E}}^{\rm bulk} \cdot \mathbf{u}^{\rm bulk} - \int \mathbf{u}^{\rm bulk} \cdot d \boldsymbol{\mathcal{E}}^{\rm bulk}$ and equating it across phases.

Using our exact and approximate coexistence criterion, we predict the phase diagram of this system and compare it to that obtained by numerically determining the full coexistence profiles, $\rho^c(z)$ and $\psi^c(z)$, following Refs.~\cite{Uecker2014, Thiele2019, Holl2020EfficientModel, Note1}.
Figure~\ref{fig:nrpd} shows the coexistence values of the order parameters as a function of the ratio $(\chi + \alpha) / (\chi - \alpha)$, demonstrating that our exact and approximate criterion both provide highly accurate predictions.
To understand the accuracy of the approximate criterion, in the SM~\cite{Note1} we compute $\Delta^{\alpha \beta} \mathcal{G}^{\rm bulk}$ [Eq.~\eqref{eq:deltaG}]:
\begin{equation}
    \Delta^{\alpha \beta} \mathcal{G}^{\rm bulk}  = \left( \frac{\chi + \alpha}{\chi - \alpha} - \frac{\overline{\kappa}_{\rho \psi}}{\overline{\kappa}_{\psi \rho}} \right) \int_{z^{\alpha}}^{z^{\beta}} u^{\rm bulk}_\psi (\mathbf{X}^{\rm c}) \partial_z \psi^{\rm c} dz.
\end{equation}
We find $\Delta^{\alpha \beta} \mathcal{G}^{\rm bulk}$ to be approximately zero, justifying equating $\mathcal{G}^{\rm bulk}$ across phases as an approximate criterion when $\overline{\kappa}_{\rho \psi} (\chi - \alpha) \neq \overline{\kappa}_{\psi \rho} (\chi + \alpha)$.

While the system in Fig.~\ref{fig:nrpd} provided a minimal platform to explore nonequilibrium coexistence criteria, both exact and approximate, each element of the Maxwell construction vector(s) is linear in the order parameters, just as with passive systems.
We seek a system where the Maxwell construction vector is a nonlinear function of the order parameters to further validate our theory.
This is the case when $\overline{\kappa}_{\rho \psi} = \overline{\kappa}_{\psi \rho} = 0$ and $\lambda_{\rho \rho \rho}$ is the only nonzero element of $\boldsymbol{\lambda}_{\rho}$:
\begin{subequations}
\label{eq:EintEbulkAMC}
    \begin{align}
        & \boldsymbol{\mathcal{E}}^{\rm bulk} = \begin{bmatrix} \exp\left(\frac{2\lambda_{\rho \rho \rho}}{\overline{\kappa}_{\rho \rho}}\rho\right) & \frac{\chi + \alpha}{\chi - \alpha} \exp\left(\frac{2\lambda_{\rho \rho \rho}}{\overline{\kappa}_{\rho \rho}}\rho\right) \end{bmatrix}^{\rm T}, \\
        & \boldsymbol{\mathcal{E}}^{\rm int} = \begin{bmatrix} \exp\left(\frac{2\lambda_{\rho \rho \rho}}{\overline{\kappa}_{\rho \rho}}\rho\right) & 0 \end{bmatrix}^{\rm T}.
    \end{align}
\end{subequations}
We see that $\boldsymbol{\mathcal{E}}^{\rm bulk}=\boldsymbol{\mathcal{E}}^{\rm int}$ when $\alpha=-\chi$ and hence the coexistence criteria are exact.
We predict the phase diagram using our criteria when $\boldsymbol{\mathcal{E}}^{\rm bulk}=\boldsymbol{\mathcal{E}}^{\rm int}$ in Fig.~\ref{fig:ambApd}(a), finding excellent agreement compared to numerical results.
In the SM~\cite{Note1}, we offer further support that our coexistence criteria are exact in this system by keeping the ratio $\lambda_{\rho \rho \rho} / \overline{\kappa}_{\rho \rho}$ fixed while tuning the raw value of these interfacial parameters, finding that the shape of the interface changes but the binodals do not.

We require an approximate criterion when $\alpha\neq-\chi$ (correspondingly $\boldsymbol{\mathcal{E}}^{\rm bulk}\neq\boldsymbol{\mathcal{E}}^{\rm int}$), for which we may use $\boldsymbol{\mathcal{E}}^{\rm int}$ in a weighted-area construction [Eq.~\eqref{eq:G_maxwell_drho_form_int}] along the path $\mathbf{X}^*(\rho)$~\footnote{Due to the forms of $\boldsymbol{\mathcal{E}}^{\rm bulk}$ and $\boldsymbol{\mathcal{E}}^{\rm int}$ in Eq.~\eqref{eq:EintEbulkAMC}, the weighted area construction along the path $\mathbf{X}^*(\rho)$ is identical whether one uses $\mathbf{E}^{\rm bulk}$ or $\mathbf{E}^{\rm int}$.}.
We use this as the approximate coexistence criterion, again finding excellent agreement with the numerically determined phase diagram as shown in Fig.~\ref{fig:ambApd}(b).
In the SM~\cite{Note1}, we provide numerical support for using this approximate criterion by demonstrating that the weighted-area construction using $\mathbf{E}^{\rm int}$ is similar in value when evaluated along $\mathbf{X}^*$ and $\mathbf{X}^{\rm c}$.

When $\overline{\kappa}_{\rho \psi} \neq 0$ or elements of $\boldsymbol{\lambda}_{\rho}$ other than $\lambda_{\rho \rho \rho}$ are nonzero, $\mathbf{E}^{\rm int}$ cannot be determined.
There is no well-defined approximate final criterion in this case, however the system retains coexistence as a steady-state.
Strikingly, using a form of $\mathbf{E}^{\rm int}$ that only solves a subset of the required equations [i.e.,~not every equation in Eqs.~\eqref{eq:G22}-\eqref{eq:G21} is satisfied] in the weighted-area construction still provides quantitatively accurate predictions as shown in Fig.~\ref{fig:ambApd}(c).
In the SM~\cite{Note1}, we plot the relative error of our predicted binodals in Fig.~\ref{fig:ambApd} compared to the numerically determined binodals.
We find the relative error between the numerical results and our predictions to meet our expectation that the error increases as one moves from the exact criterion in Fig.~\ref{fig:ambApd}(a) to the poorly-defined approximate criterion in Fig.~\ref{fig:ambApd}(c).

The phase diagram in Fig.~\ref{fig:nrpd} is symmetric about the vertical line $\rho=0$, or equivalently $\psi=0$.
This is due to the symmetry of the model when $\boldsymbol{\lambda}_{\rho} = \mathbf{0}$, where inverting the order parameter values, $\mathbf{X} \rightarrow-\mathbf{X}$, always results in $\mathbf{u} \rightarrow -\mathbf{u}$.
When $\boldsymbol{\lambda}_{\rho} \neq \mathbf{0}$, the inversion symmetry of $\mathbf{u}$ is broken and $\mathbf{X} \rightarrow-\mathbf{X}$ no longer implies $\mathbf{u} \rightarrow -\mathbf{u}$.
Consequently, the phase diagrams in Fig.~\ref{fig:ambApd} are not symmetric about $\rho=0$ (or equivalently $\psi=0$).

\section{Discussion and Conclusions} 

We have outlined a procedure to determine the phase diagram of systems described by a conserved field coupled to any number of nonconserved fields, both in and out of equilibrium.
Beginning from a force-balance representation of steady-state conditions, our theory allows one to derive a pseudopotential for the conserved field that must be equal in coexisting phases which, in equilibrium, is the chemical potential.
With this pseudopotential, one can introduce a generalized Gibbs-Duhem relation to derive the final set of coexistence criteria, the equality of a global quantity across phases.
This global quantity becomes the pressure in equilibrium.
These coexistence criteria are equivalent to a generalized common tangent construction on an effective free energy.
When a solution to the generalized Gibbs-Duhem relation does not exist, we introduce and justify two approximate forms of the criteria that one may use to estimate the phase diagram.
In equilibrium systems, the coexistence criteria are always equality of the chemical potential of the conserved field across phases, zero chemical potential of every nonconserved field in each phase, and equality of pressures across phases.
Out of equilibrium, however, the functional form of the analogous state functions follow from system-specific steady-state force balances.
We validate our theory by introducing Active Model C+ (AMC+) and predicting the phase diagrams of various systems, which we find to be in excellent agreement with the phase diagrams obtained by numerically determining the full spatial order parameter profiles during coexistence.

To use our theory, a number of equations of state in the flux-driving force and every generation-driving force must be determined.
In equilibrium, there are a number of techniques, both experimental and computational, one can use to determine these equations of state.
For example, there are structural expressions for both the bulk and interfacial contributions to the free energy~\cite{Hansen2013, Lowen1990}, allowing one to determine the passive forces given the microscopic structure of a system.
While determining the functional dependence of the generation-driving forces of some nonconserved fields may prove challenging (e.g.~a crystallinity field), $\mu_{\rho}(\mathbf{X}^*)$ and $p(\mathbf{X}^*)$ are more easily measured.
Due to the path-independence of the equilibrium Maxwell construction, only one of either $\mu_{\rho}(\mathbf{X}^*)$ or $p(\mathbf{X}^*)$, along with the parameterization $\mathbf{X}^*(\rho)$, is required to determine the phase diagram of a system and thus the difficulty in determining generation-driving forces is not an obstacle in equilibrium.
It is clear in a number of contexts that theoretical, computational, and experimental approaches are consistent in equilibrium.
Out of equilibrium, a similar understanding is still absent.

If the order parameters are known, there are theoretical techniques (such as an Irving-Kirkwood procedure~\cite{Irving1950}) to formally determine the coefficients in our forces.
The coefficients resulting from these techniques generally depend on the microscopic degrees of freedom in the system and therefore can be determined from particle-based simulations.
Experimentally, microscopic information is not always accessible and thus we desire a continuum-level procedure to determine the coefficients in our forces.
One potential route towards this is to make dynamical measurements that can be compared to our constitutive relations in Eq.~\eqref{eq:constitutivedynamics}.
As was suggested in Ref.\cite{Chiu2024TheoryCoexistence}, measuring the flux of the conserved quantity may allow the form of the flux-driving force to be determined.
The dynamics of the nonconserved fields are generally less straightforward to measure than that of the conserved field, however there is growing work to determine the coefficients in our forces using machine learning approaches~\cite{Colen2021, Joshi2022, Supekar2023}.

There are macroscopically phase-separating systems that cannot be described by our theory, such as the system in Fig.~\ref{fig:ambApd}(c).
It remains an open question as to whether there is a coexistence theory that can describe these systems.
Moreover, elucidating the structure of AMC+ and how the coexistence criteria change by introducing nonzero elements of the various nonequilibrium terms would facilitate a continuum-level understanding of how to tune a given nonequilibrium system such that a desired phase diagram is achieved.

This work focused on systems described by a single conserved field coupled to an arbitrary number of strictly nonconserved fields whose dynamics can solely be described by a generation term.
We aim to extend this work to systems where fluxes of nonconserved order parameters (not considered here) play an important role in the dynamics. 
Furthermore, many nonconserved order parameters are subject to aggregate constraints (e.g.,~constraints on linear combinations of the order parameters), such as mass conservation in reaction-diffusion systems, which we will also look to incorporate in future work.
By combining the present work with Ref.~\cite{Chiu2024TheoryCoexistence} (where we develop a mechanical coexistence theory for systems with multiple conserved fields), it is our hope that we can begin to determine the phase diagram of complex nonequilibrium systems, including multicomponent crystallization~\cite{Chiu2023}, nematic mixtures~\cite{Skamrahl2023,Bhattacharyya2024}, and reaction-diffusion systems with multiple conservation laws~\cite{Robinson2024}.

\begin{acknowledgments}
We thank Yizhi Shen, Dimitrios Fraggedakis, Yu-Jen Chiu, and Luke Langford for helpful discussions and feedback on this manuscript. 
We acknowledge support from the Laboratory Directed Research and Development Program of Lawrence Berkeley National Laboratory under U.S. Department of Energy Contract No. DE-AC02-05CH11231, the UC Berkeley College of Engineering, and the U.S. Department of Defense through the National Defense Science and Engineering Graduate Fellowship Program.
This research used the Savio computational cluster resource provided by the Berkeley Research Computing program. 
\end{acknowledgments}

\appendix

\section{\label{secap:theory_exact}Derivation of Exact Coexistence Criteria}
\subsection{Transformation Tensor}
\label{secap:T}
We first look to derive the transformation tensor, $\boldsymbol{\mathcal{T}}$, that allows us to perform the transformation $\overline{\mathbf{f}}=\mathbf{0}\rightarrow\partial_z \mathbf{u}=\mathbf{0}$.
Explicitly writing the terms in each component of $\overline{\mathbf{f}}$ [Eq.~\eqref{eq:body_force_expansion2}] we have:
\begin{subequations}
\label{eqap:body_force_expansion}
\begin{equation}
    \overline{f}_i \left( z\right)
    \approx \overline{f}^{(1)}_i + \overline{f}^{(3)}_i \ \forall i \in \mathcal{X},
\end{equation}
where:
\begin{equation}
    \overline{f}^{(1)}_i = \sum_{j \in \mathcal{X}} \overline{f}_{ij}^{(1,1)} \partial_z X_j \ \forall i \in \mathcal{X},
\end{equation}
and:
\begin{multline}
    \overline{f}^{(3)}_i = - \sum_{j \in \mathcal{X}}\sum_{k \in \mathcal{X}}\sum_{l \in \mathcal{X}} \overline{f}_{ijkl}^{(3,1)}\partial_z X_j \partial_z X_k \partial_z X_l  \\ - \sum_{j \in \mathcal{X}} \sum_{k \in \mathcal{X}} \overline{f}_{ijk}^{(3,2)} \partial^2_{zz} X_j \partial_z X_k \\ - \sum_{j \in \mathcal{X}} \overline{f}_{ij}^{(3,3)}\partial^3_{zzz} X_j \ \forall i \in \mathcal{X}.
\end{multline}
\end{subequations}
We define each $\overline{f}_{ijkl}^{(3,1)}$ to be symmetric with respect to permuting the $j$, $k$, and $l$ indices as any asymmetric contributions do not affect the value of $\overline{f}^{(3)}_i$.
Similarly writing the terms of each component of $\mathbf{u}$ [Eq.~\eqref{eq:u_expansion2}] we have:
\begin{subequations}
\label{eqap:pseudo_pot_expansion}
\begin{equation}
    u_i \approx u_i^{\rm bulk} + u_i^{\rm int} \ \forall i \in \mathcal{X},
\end{equation}
where:
\begin{equation}
    u_i^{\rm int} = -\sum_{j \in \mathcal{X}} \sum_{k \in \mathcal{X}} u_{ijk}^{(2,1)} \partial_z X_j \partial_z X_k 
    - \sum_{j \in \mathcal{X}} u_{ij}^{(2,2)} \partial^2_{zz} X_j \ \forall i \in \mathcal{X}.
\end{equation}
\end{subequations}
Antisymmetric (with respect to exchanging $j$ and $k$) contributions to $u_{ijk}^{(2,1)}$ do not impact the value of $u_i^{\rm int}$ and hence we define each $u_{ijk}^{(2,1)}$ to be symmetric.

The transformation tensor is defined to linearly convert $\overline{\mathbf{f}}$ into $\partial_z \mathbf{u}$:
\begin{subequations}
    \begin{equation}
\label{eqap:ansatmech}
    \sum_{j \in \mathcal{X}} \mathcal{T}_{ij} \overline{f}_j= \partial_z u_i \ \forall i \in \mathcal{X}.
\end{equation}
From our definition of $u_{\psi_i} = -f_{\psi_i} = -\int d z \overline{f}_{\psi_i} / \psi_i \ \forall i \in \mathcal{N}$, we find $\mathcal{T}_{\psi_i \psi_j}=-\delta_{ij} \psi_i^{-1} \ \forall i, j \in \mathcal{N}$ and $\mathcal{T}_{\psi_{i} \rho} = 0 \ \forall i \in \mathcal{N}$.
We then only need to determine one row of the transformation tensor, $\mathcal{T}_{\rho j} \ \forall j \in \mathcal{X}$, and $u_{\rho}$:
\begin{equation}
\label{eqap:ansatmechsub}
    \sum_{j \in \mathcal{X}} \mathcal{T}_{\rho j} \overline{f}_j= \partial_z u_{\rho}.
\end{equation}
\end{subequations}
It is clear that introducing a global multiplicative constant into each $\mathcal{T}_{\rho j}$ simply rescales $u_{\rho}$, i.e.,~$\mathcal{T}_{\rho j} \rightarrow c \mathcal{T}_{\rho j} \implies u_{\rho} \rightarrow c u_{\rho}$, and thus the condition of uniformity of $u_{\rho}$ is unaffected.
Introducing an additive constant, $\mathcal{T}_{\rho j} \rightarrow \mathcal{T}_{\rho j} + c$, \textit{does} nontrivially affect $u_{\rho}$, however.

Differentiating $u_{\rho}$ in the $z$-direction we have:
\begin{multline}
    \label{eqap:psi_zderive}
    \partial_z u_{\rho} = \sum_{j \in \mathcal{X}}\frac{\partial u_{\rho}^{\rm bulk}}{\partial X_j} \partial_z X_j - \sum_{j \in \mathcal{X}} u_{\rho j}^{(2,2)} \partial^3_{zzz} X_j \\ - \sum_{j \in \mathcal{X}}\sum_{k \in \mathcal{X}} \left( \frac{\partial u_{\rho j}^{(2,2)}}{\partial X_k} + 2 u_{\rho jk}^{(2,1)} \right) \partial^2_{zz} X_j \partial_z X_k \\ - \sum_{j \in \mathcal{X}}\sum_{k \in \mathcal{X}}\sum_{l \in \mathcal{X}} \frac{\partial u_{\rho jk}^{(2,1)}}{\partial X_l} \partial_z X_j \partial_z X_k \partial_z X_l,
\end{multline}
where antisymmetric (with respect to exchanging $j$, $k$ and $l$) contributions to $\partial u_{\rho jk}^{(2,1)} / \partial X_l$ do not affect the value of $\partial_z u_{\rho}$.
Substituting this into Eq.~\eqref{eqap:ansatmechsub} along with Eq.~\eqref{eqap:body_force_expansion}, matching terms, and differentiating until every algebraic equation is first-order, we find an overdetermined system of $n_N + 1 + 2(n_N+1)^2 + (n_N + 1) (n_N + 2) (n_N + 3) / 6$ linear, first-order, homogeneous, PDEs for $1 + 2(n_N + 1) + (n_N + 1)(n_N + 2) / 2$ unknown functions (each $\mathcal{T}_{\rho j}$ and every coefficient in $u_{\rho}$):
\begin{subequations}
\label{eqap:T_system_relation}
    \begin{align}
        & \sum_{i \in \mathcal{X}} \mathcal{T}_{\rho i} \overline{f}_{ij}^{(1,1)} = \frac{\partial u_n^{\rm bulk}}{\partial X_j} \ \forall j \in \mathcal{X}, \label{eqap:t_bulk_relation}\\
        & \sum_{i \in \mathcal{X}} \mathcal{T}_{\rho i} \overline{f}_{ijkl} ^{(3,1)} = \left[ \frac{\partial u_{\rho jk}^{(2,1)}}{\partial X_l}\right]^{{\rm S}_{jkl}} \ \forall j, k, l \in \mathcal{X}, \label{eqap:t_int_relation_1} \\
        & \sum_{i \in \mathcal{X}} \mathcal{T}_{\rho i} \overline{f}_{ijk}^{(3,2)} = \frac{\partial u_{\rho j}^{(2,2)}}{\partial X_k} + 2 u_{\rho jk}^{(2,1)} \ \forall j, k \in \mathcal{X}, \label{eqap:t_int_relation_2}\\
        & \sum_{i \in \mathcal{X}} \frac{\partial}{\partial X_k} \left(\mathcal{T}_{\rho i}  \overline{f}_{ij}^{(3,3)} \right)= \frac{\partial u_{\rho j}^{(2,2)}}{\partial X_k} \ \forall j, k \in \mathcal{X},\label{eqap:t_int_relation_3}
    \end{align}
\end{subequations}
where the superscript ${\rm S}_{jkl}$ denotes an object that is symmetric with respect to permuting the $j$, $k$, and $l$ indices, e.g.,~$[A_{ijkl}]^{{\rm S}_{jkl}} \equiv \frac{1}{6} (A_{ijkl} + A_{ijlk} + A_{ikjl} + A_{iklj} + A_{iljk} + A_{ilkj})$.

To make further progress, we look to eliminate the coefficients of $u_{\rho}$ from Eq.~\eqref{eqap:T_system_relation} such that we obtain a system of equations for each $\mathcal{T}_{\rho j}$ solely in terms of the known force coefficients.
If one does this, care must be taken to ensure the commutativity of partial differentiation of each of the unknown functions with respect to the order parameters.
This procedure to obtain a system of PDEs where commutativity is ensured is known as ``completion to involution''~\cite{Seiler2010}.
Here, we demonstrate how to eliminate the coefficients of $u_{\rho}$ from Eq.~\eqref{eqap:T_system_relation}, noting that a solution to the resulting equations can be checked subsequently to ensure the commutativity of partial differentiation is indeed respected.

The first relation we find follows from Eq.~\eqref{eqap:t_bulk_relation} and identifying that $\partial^2  u_{\rho}^{\rm bulk} / \partial X_j \partial X_k$ must be symmetric with respect to exchanging $j$ and $k$: 
\begin{multline}
\label{eqap:bulkTFrelationship}
    \sum_{i \in \mathcal{X}} \Bigg[ \frac{\partial \mathcal{T}_{\rho i}}{\partial X_k} \overline{f}_{ij}^{(1,1)} - \frac{\partial \mathcal{T}_{\rho i}}{\partial X_j} \overline{f}_{ik}^{(1,1)}\\ 
    + \mathcal{T}_{\rho i} \frac{\partial \overline{f}_{ij}^{(1,1)}}{\partial X_k} - \mathcal{T}_{\rho i} \frac{\partial \overline{f}_{ik}^{(1,1)}}{\partial X_j} \Bigg]
    = 0 \ \forall j, k \in \mathcal{X}.
\end{multline}
Next, we note $u_{\rho jk}^{(2, 1)}$ must also be symmetric with respect to exchanging $j$ and $k$ and therefore:
\begin{multline}
\label{eqap:intsymTFrelationship}
    \sum_{i \in \mathcal{X}} \bigg[ \mathcal{T}_{\rho i} \overline{f}_{ijk}^{(3,2)} - \mathcal{T}_{\rho i} \overline{f}_{ikj}^{(3,2)} - \frac{\partial \mathcal{T}_{\rho i}}{\partial X_k} \overline{f}_{ij}^{(3, 3)} + \frac{\partial \mathcal{T}_{\rho i}}{\partial X_j} \overline{f}_{ik}^{(3, 3)} \\ - \mathcal{T}_{\rho i} \frac{\partial \overline{f}_{ij}^{(3, 3)}}{\partial X_k} + \mathcal{T}_{\rho i} \frac{\partial \overline{f}_{ik}^{(3, 3)}}{\partial X_j} \bigg] = 0 \ \forall j, k \in \mathcal{X}.
\end{multline}
Our last relation ensures $u_{\rho jk}^{(2, 1)}$ and $u_{\rho j}^{(2, 2)}$ appear consistently in Eqs.~\eqref{eqap:t_int_relation_1}-\eqref{eqap:t_int_relation_3}.
Differentiating Eqs.~\eqref{eqap:t_int_relation_2} and \eqref{eqap:t_int_relation_3} and substituting the result into Eq.~\eqref{eqap:t_int_relation_1} we find:
\begin{multline}
\label{eqap:intTFrelationship}
    \sum_{i \in \mathcal{X}} \Bigg[ 6\mathcal{T}_{\rho i} \overline{f}_{ijkl}^{(3,1)} - \frac{\partial}{\partial X_l}(\mathcal{T}_{\rho i} \overline{f}_{ijk}^{(3,2)}) - \frac{\partial}{\partial X_k}(\mathcal{T}_{\rho i} \overline{f}_{ijl}^{(3,2)}) \\
    - \frac{\partial}{\partial X_j}(\mathcal{T}_{\rho i} \overline{f}_{ilk}^{(3,2)}) + 2\frac{\partial^2 }{\partial X_k\partial X_l}(\mathcal{T}_{\rho i} \overline{f}_{ij}^{(3,3)}) \\ + \frac{\partial^2 }{\partial X_k\partial X_j}(\mathcal{T}_{\rho i} \overline{f}_{i l}^{(3,3)}) \Bigg] = 0 \ \forall j, k, l \in \mathcal{X}.
\end{multline}
Importantly, the system of PDEs in Eqs.~\eqref{eqap:bulkTFrelationship}, \eqref{eqap:intsymTFrelationship}, and \eqref{eqap:intTFrelationship} does not contain the compatibility conditions that ensure partial differentiation is commutative.
To demonstrate how to ensure this commutivity, we perform the procedure for the system of PDEs that result from the Gibbs-Duhem relation (derived in Appendix~\ref{secap:E}) when $n_N=1$ in the SM~\cite{Note1}.

\subsection{Maxwell Construction Vector}
\label{secap:E}
We now derive the Maxwell construction vector, $\boldsymbol{\mathcal{E}}$, that defines the relationship between the global quantity, $\mathcal{G}$, and pseudopotential vector, $\mathbf{u}$.
Explicitly writing each term in $\mathcal{G}$ we have:
\begin{subequations}
\label{eqap:G_expansion}
\begin{equation}
    \mathcal{G} \approx \mathcal{G}^{\rm bulk} + \mathcal{G}^{\rm int},
\end{equation}
where:
\begin{equation}
    \mathcal{G}^{\rm int} = -\sum_{j \in \mathcal{X}} \sum_{k \in \mathcal{X}} \mathcal{G}_{jk}^{(2,1)}\partial_z X_j \partial_z X_k
    - \sum_{j \in \mathcal{X}} \mathcal{G}_{j}^{(2,2)} \partial^2_{zz} X_j.
\end{equation}
\end{subequations}
Again, antisymmetric (with respect to exchanging $j$ and $k$) contributions to $\mathcal{G}_{jk}^{(2,1)}$ do not impact the value of $\mathcal{G}^{\rm int}$.

The generalized Gibbs-Duhem relation is:
\begin{equation}
\label{eqap:generalized_GD}
    \sum_{i \in \mathcal{X}} \mathcal{E}_i du_i = d\mathcal{G},
\end{equation}
which defines $\boldsymbol{\mathcal{E}}$ and $\mathcal{G}$.
Here, the expansion of $\mathbf{u}$ in Eq.~\eqref{eqap:pseudo_pot_expansion} is used.
We can immediately appreciate that introducing a global multiplicative constant into the Maxwell construction vector simply rescales $\mathcal{G}$, $\boldsymbol{\mathcal{E}} \rightarrow c \boldsymbol{\mathcal{E}} \implies \mathcal{G} \rightarrow c \mathcal{G}$, and hence does not alter the spatial uniformity of $\mathcal{G}$.
Furthermore, this uniformity is also unaffected by introducing a vector of additive constants into the Maxwell construction vector, $\boldsymbol{\mathcal{E}} \rightarrow \boldsymbol{\mathcal{E}} + \mathbf{c}$, as this simply adds a linear combination of the pseudopotentials to the global quantity, $\mathcal{G} \rightarrow \mathcal{G} + \mathbf{c} \cdot \mathbf{u}$.
As $\mathbf{u}$ is spatially uniform, $\mathbf{c} \cdot \mathbf{u}$ trivially is as well.

Differentiating $\mathcal{G}$ in the $z$-direction we have:
\begin{multline}
    \label{eqap:G_zderive}
    \partial_z \mathcal{G} = \sum_{j \in \mathcal{X}}\frac{\partial \mathcal{G}^{\rm bulk}}{\partial X_j} \partial_z X_j - \sum_{j \in \mathcal{X}} \mathcal{G}_{j}^{(2,2)} \partial^3_{zzz} X_j \\ - \sum_{j \in \mathcal{X}}\sum_{k \in \mathcal{X}} \left( \frac{\partial \mathcal{G}_{j}^{(2,2)}}{\partial X_k} + 2 \mathcal{G}_{jk}^{(2,1)} \right) \partial^2_{zz} X_j \partial_z X_k \\
    - \sum_{j \in \mathcal{X}}\sum_{k \in \mathcal{X}}\sum_{l \in \mathcal{X}} \frac{\partial \mathcal{G}_{jk}^{(2,1)}}{\partial X_l} \partial_z X_j \partial_z X_k \partial_z X_l,
\end{multline}
where again antisymmetric (with respect to exchanging $j$, $k$, and $l$) contributions to $\partial \mathcal{G}_{jk}^{(2,1)}/\partial X_l$ do not affect the value of $\partial_z \mathcal{G}$.
Substituting $\partial_z u_i$ [Eq.~\eqref{eqap:psi_zderive}] and $\partial_z \mathcal{G}$ [Eq.~\eqref{eqap:G_zderive}] into Eq.~\eqref{eqap:generalized_GD} and matching terms, we find an overdetermined system of $1+n_N +2(n_N + 1)^2+\frac{1}{6}(n_N + 1)(n_N+2)(n_N+3)$ linear, first-order, homogeneous PDEs for $3+2n_N+(n_N+1)(n_N+2)/2$ unknown functions (each component of $\boldsymbol{\mathcal{E}}$ and each coefficient in $\mathcal{G}$):
\begin{subequations}
\label{eqap:E_relations_appendix}
    \begin{align}
        &\sum_{i \in \mathcal{X}} \mathcal{E}_i \frac{\partial u_i^{\rm bulk}}{\partial X_j} = \frac{\partial \mathcal{G}^{\rm bulk}}{\partial X_j} \ \forall j \in \mathcal{X}, \label{eqap:E_bulk_relation} \\
        &\sum_{i \in \mathcal{X}}\frac{\partial}{\partial X_k}(\mathcal{E}_i u_{ij}^{(2, 2)}) = \frac{\partial \mathcal{G}_{j}^{(2,2)}}{\partial X_k} \ \forall j, k \in \mathcal{X}, \label{eqap:E_int_relation1} \\
        &\sum_{i \in \mathcal{X}} \mathcal{E}_i \left( 2 u_{ijk}^{(2, 1)} +\frac{\partial u_{ij}^{(2, 2)}}{\partial X_k}\right) = 2 \mathcal{G}_{jk}^{(2,1)} + \frac{\partial \mathcal{G}_{j}^{(2,2)}}{\partial X_k} \ \forall j, k \in \mathcal{X}, \label{eqap:E_int_relation2} \\
        &\sum_{i \in \mathcal{X}} \mathcal{E}_i \left[\frac{\partial u_{ijk}^{(2, 1)}}{\partial X_l} \right]^{{\rm S}_{jkl}}= \left[\frac{\partial \mathcal{G}_{jk}^{(2,1)}}{\partial X_l}\right]^{{\rm S}_{jkl}} \ \forall j, k, l \in \mathcal{X}. \label{eqap:E_int_relation3}
    \end{align}
\end{subequations}

To proceed, we seek equations for the components of $\boldsymbol{\mathcal{E}}$ solely in terms of coefficients in $\mathbf{u}$, i.e.,~we look to eliminate the coefficients of $\mathcal{G}$ from Eq.~\eqref{eqap:E_relations_appendix}.
Just as was the case for the transformation tensor, one generally must find compatibility conditions that ensure partial differentiation is commutative.
Here, we eliminate the coefficients of $\mathcal{G}$ from Eq.~\eqref{eqap:E_relations_appendix}, finding equations one can solve and subsequently check if partial differentiation is commutative.
In the SM~\cite{Note1}, we follow a formal procedure (completion to involution~\cite{Seiler2010}) when $n_N=1$ to find the compatibility conditions one needs to augment Eq.~\eqref{eqap:E_relations_appendix} with to ensure partial differentiation is commutative.

The first set of relationships we find follow from Eq.~\eqref{eqap:E_bulk_relation}, noting $\partial^2 \mathcal{G}^{\rm bulk} / \partial X_j \partial X_k$ must be symmetric upon exchanging $j$ and $k$:
\begin{equation}
\label{eq:E_bulk_appendix}
    \sum_{i \in \mathcal{X}}\Bigg[E_{ij} \frac{\partial u_i^{\rm bulk}}{\partial X_k} - E_{ik} \frac{\partial u_i^{\rm bulk}}{\partial X_j} \Bigg]= 0 \ \forall j, k \in \mathcal{X},
\end{equation}
where $E_{ij} \equiv \partial \mathcal{E}_i / \partial X_j$.
Additionally, $\mathcal{G}_{jk}^{(2, 1)}$ must be symmetric with respect to exchanging $j$ and $k$:
\begin{equation}
\label{eq:E_int_appendix1}
    \sum_{i \in \mathcal{X}} \left[ u_{ij}^{(2,2)}E_{ik} - u_{ik}^{(2,2)}E_{ij} \right] = 0 \ \forall j, k \in \mathcal{X}.
\end{equation}
Our last set of equations ensures $\mathcal{G}_{jk}^{(2,1)}$ and $\mathcal{G}_{j}^{(2,2)}$ appear consistently in Eqs.~\eqref{eqap:E_int_relation1}-\eqref{eqap:E_int_relation3}.
Differentiating Eqs.~\eqref{eqap:E_int_relation1} and \eqref{eqap:E_int_relation2} and substituting the result into Eq.~\eqref{eqap:E_int_relation3}, we find:
\begin{multline}
    \label{eq:E_int_appendix3}
    \sum_{i \in \mathcal{X}} \bigg[ 2 E_{il} u_{ijk}^{(2,1)} + 2 E_{ik} u_{ijl}^{(2,1)} + 2 E_{ij} u_{ilk}^{(2,1)} - \frac{\partial}{\partial X_l}(E_{ik} u_{ij}^{(2,2)}) \\ -  \frac{\partial}{\partial X_k}(E_{il} u_{ij}^{(2,2)}) - \frac{\partial}{\partial X_j}(E_{ik} u_{il}^{(2,2)}) \bigg] = 0 \ \forall j, k, l \in \mathcal{X}.
\end{multline}

\section{\label{secap:theory_approx}Derivation of Approximate Coexistence Criteria}
We now consider systems where $\mathcal{T}_{\rho j} \ \forall j \in \mathcal{N}$ and $u_{\rho}$ can be determined and thus the existence of exact coexistence criteria relies on a solution to the generalized Gibbs-Duhem relation [Eq.~\eqref{eqap:generalized_GD} and consequently Eq.~\eqref{eqap:E_relations_appendix}] for $\boldsymbol{\mathcal{E}}$ and $\mathcal{G}$.
We develop two approximate forms of the final set of coexistence criteria, equality of $\mathcal{G}^{\rm bulk}$ across phases, when a full solution to the generalized Gibbs-Duhem relation does not exist.

\subsection{Partial Solutions to the Generalized Gibbs-Duhem Relation}
\label{secap:partial}
In this Appendix, we consider systems where a full solution to the generalized Gibbs-Duhem relation [Eq.~\eqref{eqap:generalized_GD}] does not exist.
We propose splitting the generalized Gibbs-Duhem relation into bulk and interfacial portions:
\begin{subequations}
    \label{eqap:generalized_GD_split}
    \begin{align}
        & \boldsymbol{\mathcal{E}}^{\rm bulk} \cdot \mathbf{u}^{\rm bulk} = d \mathcal{G}^{\rm bulk}, \label{eqap:generalized_GD_bulk} \\
        & \boldsymbol{\mathcal{E}}^{\rm int} \cdot \mathbf{u}^{\rm int} = d \mathcal{G}^{\rm int}. \label{eqap:generalized_GD_int}
    \end{align}
\end{subequations}
Here, $\boldsymbol{\mathcal{E}}^{\rm bulk}$ is defined to satisfy Eq.~\eqref{eqap:E_bulk_relation}:
\begin{equation}
    \sum_{i \in \mathcal{X}} \mathcal{E}_i^{\rm bulk} \frac{\partial u_i^{\rm bulk}}{\partial X_j} = \frac{\partial \mathcal{G}^{\rm bulk}}{\partial X_j} \ \forall j \in \mathcal{X}, \label{eqap:E_bulk_relation2}
\end{equation}
while $\boldsymbol{\mathcal{E}}^{\rm int}$ is defined to satisfy Eqs.~\eqref{eqap:E_int_relation1}-\eqref{eqap:E_int_relation3}:
\begin{subequations}
    \label{eqap:E_int_relation2s}
    \begin{align}
        &\sum_{i \in \mathcal{X}}\frac{\partial}{\partial X_k}(\mathcal{E}_i u_{ij}^{(2, 2)}) = \frac{\partial \mathcal{G}_{j}^{(2,2)}}{\partial X_k} \ \forall j, k \in \mathcal{X}, \label{eqap:E_int_relation12} \\
        &\sum_{i \in \mathcal{X}} \mathcal{E}_i \left( 2 u_{ijk}^{(2, 1)} +\frac{\partial u_{ij}^{(2, 2)}}{\partial X_k}\right) = 2 \mathcal{G}_{jk}^{(2,1)} + \frac{\partial \mathcal{G}_{j}^{(2,2)}}{\partial X_k} \ \forall j, k \in \mathcal{X}, \label{eqap:E_int_relation22} \\
        &\sum_{i \in \mathcal{X}} \mathcal{E}_i \left[\frac{\partial u_{ijk}^{(2, 1)}}{\partial X_l} \right]^{{\rm S}_{jkl}}= \left[\frac{\partial \mathcal{G}_{jk}^{(2,1)}}{\partial X_l}\right]^{{\rm S}_{jkl}} \ \forall j, k, l \in \mathcal{X}. \label{eqap:E_int_relation32}
    \end{align}
\end{subequations}
When these equations can be solved, $\mathcal{G}$ can be defined, however it is not necessarily spatially uniform in a steady-state.

\subsection{Approximate Weighted-Area Construction}
\label{secap:weightedarea}
We now derive an approximate set of coexistence criteria when the interfacial generalized Gibbs-Duhem relation [Eq.~\eqref{eqap:generalized_GD_int} and correspondingly Eq.~\eqref{eqap:E_int_relation2s}] can be solved for $\boldsymbol{\mathcal{E}}^{\rm int}$.
The following result is independent of a solution to the bulk generalized Gibbs-Duhem relation [Eq.~\eqref{eqap:generalized_GD_bulk} and correspondingly Eq.~\eqref{eqap:E_bulk_relation2}].

The interfacial generalized Gibbs-Duhem relation [Eq.~\eqref{eqap:generalized_GD_int}] can be expressed as $\partial_z \mathcal{G}^{\rm int} = \partial_z ( \boldsymbol{\mathcal{E}}^{\rm int} \cdot \mathbf{u}^{\rm int}) - \mathbf{u}^{\rm int} \cdot \mathbf{E}^{\rm int} \cdot \partial_z \mathbf{X}$ and therefore $\mathbf{u}^{\rm int} \cdot \mathbf{E}^{\rm int} \cdot \partial_z \mathbf{X}$ must be the $z$-derivative of a contribution to $\mathcal{G}^{\rm int}$.
Importantly, every contribution to $\mathcal{G}^{\rm int}$ vanishes in the homogeneous phases by definition and thus the spatial integral of $\mathbf{u}^{\rm int} \cdot \mathbf{E}^{\rm int} \cdot \partial_z \mathbf{X}$ over the interfacial profile is guaranteed to vanish.
In other words, the weighted-area construction [Eq.~\eqref{eq:G_maxwell_drho_form}] with $\mathbf{E}=\mathbf{E}^{\rm int}$ vanishes when evaluated along $\mathbf{X}^{\rm c}(z)$:
\begin{multline}
    \label{eqap:approx1}
    \sum_{i \in \mathcal{X}} \sum_{j \in \mathcal{X}} \int_{z^{\alpha}}^{z^{\beta}} \bigg( u_i^{\rm bulk} (\mathbf{X}^{\rm c}) \\ - u_i^{\rm coexist} \bigg) E_{i j}^{\rm int} (\mathbf{X}^{\rm c}) \partial_z X_j^{\rm c} dz = 0 \ \forall \alpha, \beta \in \mathcal{P},
\end{multline}
where the value of this integral is \textit{path-dependent} as it is not equal to the difference in $\mathcal{G}^{\rm bulk}$ between phases if $\boldsymbol{\mathcal{E}}^{\rm bulk} \neq \boldsymbol{\mathcal{E}}^{\rm int}$.
The full spatial coexistence profile, $\mathbf{X}^{\rm c}(z)$, is what we aim to avoid determining.
We then see that as an approximate set of coexistence criteria, we may evaluate the integral in Eq.~\eqref{eqap:approx1} along a different path, such as $\mathbf{X}^*(\rho)$. This approximation gains accuracy as the selected integration path approaches the coexistence profile.

\subsection{Approximate Equality of the Bulk Global Quantity}
\label{secap:bulkapprox}
We now derive a second approximate form of the final set of coexistence criteria, applicable to systems where both the interfacial generalized Gibbs-Duhem relation [Eq.~\eqref{eqap:generalized_GD_int} and correspondingly Eq.~\eqref{eqap:E_int_relation2s}] and bulk generalized Gibbs-Duhem relation [Eq.~\eqref{eqap:generalized_GD_bulk} and correspondingly Eq.~\eqref{eqap:E_bulk_relation}] have solutions.
Rearranging Eq.~\eqref{eqap:approx1} and identifying $\mathcal{G}^{\rm bulk} \equiv \boldsymbol{\mathcal{E}}^{\rm bulk} \cdot \mathbf{u}^{\rm bulk} - \int \mathbf{u}^{\rm bulk} \cdot d\boldsymbol{\mathcal{E}}^{\rm bulk}$ [from Eq.~\eqref{eqap:E_bulk_relation2}], we find:
\begin{multline}
    \Delta^{\alpha \beta} \mathcal{G}^{\rm bulk} \equiv \mathcal{G}^{\rm bulk} \left( \mathbf{X}^{\alpha} \right) - \mathcal{G}^{\rm bulk} \left( \mathbf{X}^{\beta} \right) \\ = \sum_{i \in \mathcal{X}} \sum_{j \in \mathcal{X}} \int_{z^{\alpha}}^{z^{\beta}} \bigg( u_i^{\rm bulk} (\mathbf{X}^{\rm c}) - u_i^{\rm coexist} \bigg) \bigg( E_{i j}^{\rm bulk} (\mathbf{X}^{\rm c}) \\- E_{i j}^{\rm int} (\mathbf{X}^{\rm c}) \bigg) \partial_z X_j^{\rm c} dz \ \forall \alpha, \beta \in \mathcal{P}.
\end{multline}
We then see that one may equate $\mathcal{G}^{\rm bulk}$ across coexisting phases as an approximate set of coexistence criteria, with the approximation gaining accuracy as the weighted-area construction with $\mathbf{E}=\mathbf{E}^{\rm bulk} - \mathbf{E}^{\rm int}$ [evaluated along $\mathbf{X}^{\rm c}(z)$] gets closer to zero.

\section{\label{secap:freeenergy} Existence of Effective Free Energy}
In this Appendix, we derive the conditions for steady-states of a system to be governed by an effective free energy functional:
\begin{subequations}
    \label{eqap:effectivefreeenergy}
    \begin{equation}
    \label{eqap:Wdef}
        \mathcal{W} [\mathbf{X}(z)] = \int_V d \mathbf{r} \left( w^{\rm bulk} + \frac{1}{2} \sum_{k \in \mathcal{X}} \sum_{l \in \mathcal{X}} w_{k l}^{(2, 1)} \partial_z X_{k} \partial_z X_{l} \right),
    \end{equation}
    where $w^{\rm bulk}$ is the bulk effective free energy and $w_{k l}^{(2, 1)}$ is the $k l$ component of the symmetric matrix parameterizing interfacial contributions to the effective free energy.
    We seek the conditions for when the pseudopotential of each order parameter can be expressed as the functional derivative:
    \begin{equation}
        \label{eqap:funcderivansatz2}
        u_i =  \frac{\delta \mathcal{W}}{\delta \mathcal{E}_i} \ \forall i \in \mathcal{X}.
    \end{equation}
    Given that each component of $\boldsymbol{\mathcal{E}}$ is an entirely local state function of $\mathbf{X}$, the chain rule for functional differentiation yields a simple, equivalent form of Eq.~\eqref{eqap:funcderivansatz2}:
    \begin{equation}
        \label{eqap:funcderivansatz}
        u_i =  \sum_{j \in \mathcal{X}} \frac{\delta \mathcal{W}}{\delta X_j} E_{j i}^{-1} \ \forall i \in \mathcal{X},
    \end{equation}
    which will be the more straightforward form to work with.
    Both Eqs.~\eqref{eqap:funcderivansatz2} and \eqref{eqap:funcderivansatz} require $\boldsymbol{\mathcal{E}}$ to be exist [and hence a solution to Eq.~\eqref{eqap:E_relations_appendix}] and every component to be nonconstant.
    Moreover, Eq.~\eqref{eqap:funcderivansatz} requires $\mathbf{E}$ to be invertible which is slightly more restrictive than every component of $\boldsymbol{\mathcal{E}}$ being nonconstant.
\end{subequations}

To proceed, we contract Eq.~\eqref{eqap:funcderivansatz}  with $\mathbf{E}$ to obtain $\sum_{i \in \mathcal{X}} u_i E_{i j} = \delta \mathcal{W} / \delta X_j$.
Evaluating the functional derivative of $\mathcal{W}$:
\begin{multline}
    \sum_{i \in \mathcal{X}} u_i E_{i j} = \frac{\partial w^{\rm bulk}}{\partial X_j}  - \sum_{k \in \mathcal{X}} w_{j k}^{(2, 1)} \partial^2_{zz} X_k \\ - \sum_{k \in \mathcal{X}} \sum_{l \in \mathcal{X}} \left[ \frac{\partial w_{j k}^{(2, 1)}}{\partial X_{l}} - \frac{1}{2} \frac{\partial w_{k l}^{(2, 1)}}{\partial X_{j}} \right]^{{\rm S}_{k l}} \partial_z X_{k} \partial_z X_{l} \ \forall j \in \mathcal{X},
\end{multline}
where the ${\rm S}_{k l}$ superscript extracts the symmetric (with respect to exchanging $k$ and $l$) portion of a matrix.
Substituting the expansion for each $u_i$ [Eq.~\eqref{eqap:pseudo_pot_expansion}] and matching terms, we find:
\begin{subequations}
    \begin{align}
        &\sum_{i \in \mathcal{X}} u_i^{\rm bulk} E_{i j} = \frac{\partial w^{\rm bulk}}{\partial X_j} \label{eqap:freenergyrelations1}, \\
        & \sum_{i \in \mathcal{X}} u_{ik}^{(2, 2)} E_{i j} = w_{j k}^{(2, 1)} \label{eqap:freenergyrelations2}, \\
        & 2 \sum_{i \in \mathcal{X}} u_{ikl}^{(2, 1)} E_{i j} = \frac{\partial w_{j k}^{(2, 1)}}{\partial X_{l}} + \frac{\partial w_{j l}^{(2, 1)}}{\partial X_{k}} - \frac{\partial w_{k l}^{(2, 1)}}{\partial X_{j}} \label{eqap:freenergyrelations3}.
    \end{align}
\end{subequations}
Substituting these definitions into Eq.~\eqref{eqap:Wdef} and mandating translational invariance of the effective free energy functional (i.e.,~$\delta \mathcal{W}=0$ for the transformation $z \rightarrow z + \delta z$, where $\delta z$ is a constant), we recover the generalized Gibbs-Duhem relation, just as one can recover the equilibrium Gibbs-Duhem relation by mandating translational invariance of an equilibrium free energy functional~\cite{Julicher2018}.
It is then apparent that the generalized Gibbs-Duhem holds if the effective free energy exists.
However, it is not clear if the converse is true (with the condition that $\mathbf{E}$ must be invertible).

We first consider the bulk equation [Eq.~\eqref{eqap:freenergyrelations1}], finding $u_{i}^{\rm bulk} = \partial w^{\rm bulk} / \partial \mathcal{E}_i$.
For the Hessian of $w^{\rm bulk}$ to be symmetric, we must have:
\begin{equation}
\label{eq:E_bulk_appendix2}
    \sum_{i \in \mathcal{X}}\Bigg[E_{ij} \frac{\partial u_i^{\rm bulk}}{\partial X_k} - E_{ik} \frac{\partial u_i^{\rm bulk}}{\partial X_j} \Bigg]= 0 \ \forall j, k \in \mathcal{X},
\end{equation}
which is identical to Eq.~\eqref{eq:E_bulk_appendix}.
As $\partial_{\mathbf{X}} \mathcal{G}^{\rm bulk} = \boldsymbol{\mathcal{E}} \cdot \partial_{\mathbf{X}} \mathbf{u}^{\rm bulk}$ from the generalized Gibbs-Duhem relation [Eq.~\eqref{eqap:generalized_GD} and consequently Eq.~\eqref{eqap:E_bulk_relation}], Eq.~\eqref{eqap:freenergyrelations1} implies $\mathcal{G}^{\rm bulk} = \boldsymbol{\mathcal{E}} \cdot \mathbf{u}^{\rm bulk} - w^{\rm bulk}$.
This confirms that our coexistence criteria are indeed consistent with a common tangent construction on $w^{\rm bulk}$ with respect to $\boldsymbol{\mathcal{E}}$ [Eq.~\eqref{eq:common_tangent}].

We next consider the interfacial equations [Eqs.~\eqref{eqap:freenergyrelations2} and \eqref{eqap:freenergyrelations3}].
Noting $w_{j k}^{(2, 1)}=w_{kj}^{(2, 1)}$, we find a relationship identical to Eq.~\eqref{eq:E_int_appendix1}
\begin{equation}
\label{eq:E_int_appendix12}
    \sum_{i \in \mathcal{X}} \left[ u_{ij}^{(2,2)}E_{ik} - u_{ik}^{(2,2)}E_{ij} \right] = 0 \ \forall j, k \in \mathcal{X}.
\end{equation}
To find our last relationship, we substitute Eq.~\eqref{eqap:freenergyrelations2} into Eq.~\eqref{eqap:freenergyrelations3}:
\begin{multline}
    \label{eq:E_int_appendix32pre}
    2 \sum_{i \in \mathcal{X}} u_{ikl}^{(2, 1)} E_{i j} = \frac{\partial }{\partial X_{l}} \left( u_{ik}^{(2,2)}E_{ij} \right) + \frac{\partial}{\partial X_{k}} \left( u_{il}^{(2,2)}E_{ij} \right) \\ - \frac{\partial}{\partial X_{j}} \left( u_{il}^{(2,2)}E_{ik} \right).
\end{multline}
As $u_{ikl}^{(2, 1)} = u_{ilk}^{(2, 1)} \ \forall i, k, l, \in \mathcal{X}$, we have $(n_N + 1)^2(n_N + 2)/2$ equations in Eq.~\eqref{eq:E_int_appendix32pre}.
Notably, summing over unique triplets of $j$, $k$, and $l$ we find $(n_N+1)(n_N+2)(n_N+3)/6$ equations identical to Eq.~\eqref{eq:E_int_appendix3}:
\begin{multline}
    \label{eq:E_int_appendix32}
    \sum_{i \in \mathcal{X}} \bigg[ 2 E_{il} u_{ijk}^{(2,1)} + 2 E_{ik} u_{ijl}^{(2,1)} + 2 E_{ij} u_{ilk}^{(2,1)} - \frac{\partial}{\partial X_l}(E_{ik} u_{ij}^{(2,2)}) \\ -  \frac{\partial}{\partial X_k}(E_{il} u_{ij}^{(2,2)}) - \frac{\partial}{\partial X_j}(E_{ik} u_{il}^{(2,2)}) \bigg] = 0 \ \forall j, k, l \in \mathcal{X}.
\end{multline}
We again recover that the existence of an effective free energy implies the generalized Gibbs-Duhem, however we see that the converse is not true.
Equation~\eqref{eq:E_int_appendix32pre} contains $n_N^3/3 + n_N^2 + 2 n_N / 3$ more equations than Eq.~\eqref{eq:E_int_appendix32} and thus, when there are nonconserved fields present ($n_N>0$), an effective free energy [Eq.~\eqref{eqap:effectivefreeenergy}] exists in less systems than the generalized Gibbs-Duhem relation [Eq.~\eqref{eqap:generalized_GD}] is well-defined for.

\end{document}


\title{Supplemental Material -- Theory of Nonequilibrium Coexistence with Coupled Conserved and Nonconserved Order Parameters}
\author{Daniel Evans}
\author{Ahmad K. Omar}
\email{aomar@berkeley.edu}
\affiliation{Department of Materials Science and Engineering, University of California, Berkeley, California 94720, USA}
\affiliation{Materials Sciences Division, Lawrence Berkeley National Laboratory, Berkeley, California 94720, USA}

\maketitle
\tableofcontents

\section{Completion of System of PDEs to Involution}
We now complete the system of PDEs resulting from the generalized Gibbs-Duhem relation, $d \mathcal{G}=\sum_{i \in \mathcal{X}} \mathcal{E}_i d u_i$, to involution~\cite{Seiler2010} when $n_N=1$.
First considering an arbitrary number of nonconserved order parameters, the system of linear PDEs derived to solve for $\boldsymbol{\mathcal{E}}$ in the main text is:
\begin{subequations}
    \label{eq:Esys}
    \begin{align}
    \label{eq:Esys1}
        & \partial_j \mathcal{G}^{\rm bulk} = \sum_{i \in \mathcal{X}} \mathcal{E}_i \partial_j u_i^{\rm bulk}, \\
        \label{eq:Esys2}
        & \partial_k \mathcal{G}_j^{(2, 2)} = \sum_{i \in \mathcal{X}} \partial_k \left( \mathcal{E}_i u_{i j}^{(2,2)} \right), \\
        \label{eq:Esys3}
        & 2 \mathcal{G}_{j k}^{(2, 1)} + \partial_k \mathcal{G}_j^{(2, 2)} = \sum_{i \in \mathcal{X}} \mathcal{E}_i \left( 2 u_{i j k}^{(2,1)} +  \partial_k u_{i j}^{(2,2)} \right), \\
        \label{eq:Esys4}
        & \bigg[\partial_n \mathcal{G}_{j k}^{(2, 1)}\bigg]^{{\rm S}_{n j k}} = \sum_{i \in \mathcal{X}} \mathcal{E}_i \left[  \partial_n u_{i j k}^{(2,1)} \right]^{{\rm S}_{n j k}},
    \end{align}
\end{subequations}
where $\partial_i \equiv \partial / \partial X_i$. 
Here, each component of the vector $\mathbf{X}$ are the independent variables while the vector $\mathbf{u}^{\rm bulk}(\mathbf{X})$ and each of the matrices $\mathbf{u}^{(2,1)}(\mathbf{X})$ and $\mathbf{u}^{(2,2)}(\mathbf{X})$ are presumed to be be known.
The dependent variables we must solve for are the scalar $\mathcal{G}^{\rm bulk}$, each component of the vectors $\boldsymbol{\mathcal{E}}$ and $\boldsymbol{\mathcal{G}}^{(2, 2)}$, and each component of the matrix $\boldsymbol{\mathcal{G}}^{(2, 1)}$.
The superscript ${\rm S}_{n j k}$ indicates a matrix that is symmetric with respect to exchanging the $n$, $j$, and $k$ indices.

Inserting the definition of ${\rm S}_{n j k}$ (see main text) we have:
\begin{subequations}
    \label{eq:Esysn}
    \begin{align}
        \label{eq:Esys1n}
        & \partial_j \mathcal{G}^{\rm bulk} = \sum_{i \in \mathcal{X}} \mathcal{E}_i \partial_j u_i^{\rm bulk}, \\
        \label{eq:Esys2n}
        & \partial_k \mathcal{G}_j^{(2, 2)} = \sum_{i \in \mathcal{X}} \partial_k \left( \mathcal{E}_i u_{i j}^{(2,2)} \right), \\
        \label{eq:Esys3n}
        & 2 \mathcal{G}_{j k}^{(2, 1)} + \partial_k \mathcal{G}_j^{(2, 2)} = \sum_{i \in \mathcal{X}} \mathcal{E}_i \left( 2 u_{i j k}^{(2,1)} +  \partial_k u_{i j}^{(2,2)} \right), \\
        \label{eq:Esys4n}
        & \partial_n \mathcal{G}_{j k}^{(2, 1)} + \partial_j \mathcal{G}_{n k}^{(2, 1)} + \partial_k \mathcal{G}_{j n}^{(2, 1)} = \sum_{i \in \mathcal{X}} \mathcal{E}_i\left(  \partial_n u_{i j k}^{(2,1)} + \partial_j u_{i n k}^{(2,1)} + \partial_k u_{i j n}^{(2,1)}\right).
    \end{align}
\end{subequations}
We have $n_N + 1$ independent variables (the number of components of $\mathbf{X}$), and have $n_N + 1$ dependent variables each in $\boldsymbol{\mathcal{E}}$ and $\boldsymbol{\mathcal{G}}^{(2, 2)}$, one in $\mathcal{G}^{\rm bulk}$, and $(n_N+1)(n_N + n_1)/2$ in $\boldsymbol{\mathcal{G}}^{(2, 1)}$.
We therefore have $m=n_N^2/2 + 7n_N/2 + 4$ dependent variables in total.
Equation~\eqref{eq:Esys1n} contains $n_N + 1$ equations, Eqs.~\eqref{eq:Esys2n} and \eqref{eq:Esys3n} each contain $(n_N+1)^2$ more, and Eq.~\eqref{eq:Esys4n} contains $(n_N+1)(n_N+2)(n_N+3)/6$ equations.
Before prolongation, we thus have a system of $t= n_N^3 / 6 + 3 n_N^2 + 41 n_N / 6 + 4$ equations.

It proves convenient to express all of the $t$ equations appearing Eq.~\eqref{eq:Esysn} in vector notation with $\boldsymbol{\Phi} = \mathbf{0}$, where $\boldsymbol{\Phi}$ is defined as:
\begin{equation}
    \boldsymbol{\Phi} = \begin{bmatrix}
        \partial_1 \mathcal{G}^{\rm bulk} - \sum_{i \in \mathcal{X}} \mathcal{E}_i \partial_1 u^{\rm bulk}_i \\
        \vdots \\
        \partial_{n_X} \mathcal{G}^{\rm bulk} - \sum_{i \in \mathcal{X}} \mathcal{E}_i \partial_{n_X} u^{\rm bulk}_i \\
        \partial_1 \mathcal{G}_1^{(2, 2)} - \sum_{i \in \mathcal{X}} \left( \mathcal{E}_i \partial_1 u^{(2,2)}_{i 1} + \partial_1 \mathcal{E}_i u^{(2,2)}_{i 1} \right) \\
        \vdots \\
        \partial_{n_X} \mathcal{G}_{n_X}^{(2, 2)} - \sum_{i \in \mathcal{X}} \left( \mathcal{E}_i \partial_{n_X} u^{(2,2)}_{i n_X} + \partial_n \mathcal{E}_i u^{(2,2)}_{i n_X} \right) \\
        2 \mathcal{G}_{11}^{(2, 1)} + \partial_1 \mathcal{G}_1^{(2, 2)} - \sum_{i \in \mathcal{X}} \mathcal{E}_i \left( 2 u^{(2,1)}_{i 11} + \partial_1 u^{(2,2)}_{i 1} \right) \\
        \vdots \\
        2 \mathcal{G}_{nn}^{(2, 1)} + \partial_{n_X} \mathcal{G}_{n_X}^{(2, 2)} - \sum_{i \in \mathcal{X}} \mathcal{E}_i \left( 2 u^{(2,1)}_{i n_X n_X} + \partial_{n_X} u^{(2,2)}_{i n_X} \right) \\
        \partial_{1} \mathcal{G}_{11}^{(2, 1)} - 2 \mathcal{E}_i \partial_{1} u^{(2,1)}_{i 11} \\
        \vdots \\
        \partial_{n_X} \mathcal{G}_{n_X n_X}^{(2, 1)} - 2 \sum_{i \in \mathcal{X}} \mathcal{E}_i \partial_{n_X} u^{(2,1)}_{i n_X n_X},
    \end{bmatrix}.
\end{equation}

We now consider a system described by two independent variables, $\rho$ and $\psi$, and hence $n_N=1$, $m=8$, and $t=14$.
Explicitly, our dependent variables are $\mathcal{E}_{\rho}$, $\mathcal{E}_{\psi}$, $\mathcal{G}^{\rm bulk}$, $\mathcal{G}^{(2, 2)}_{\rho}$, $\mathcal{G}^{(2, 2)}_{\psi}$, $\mathcal{G}^{(2, 1)}_{\rho \rho}$, $\mathcal{G}^{(2, 1)}_{\rho \psi}$, and $\mathcal{G}^{(2, 1)}_{\psi \psi}$.
Writing out each term in $\boldsymbol{\Phi}$ and simplifying we have:
\begin{equation}
    \label{eq:phifull}
    \boldsymbol{\Phi} = \begin{pmatrix}
        \begin{smallmatrix}
        \partial_{\rho} \mathcal{G}^{\rm bulk} - \mathcal{E}_{\rho} \partial_{\rho} u^{\rm bulk}_{\rho} - \mathcal{E}_{\psi} \partial_{\rho} u^{\rm bulk}_{\psi} 
        \end{smallmatrix} \\
        \begin{smallmatrix}
        \partial_{\psi} \mathcal{G}^{\rm bulk} - \mathcal{E}_{\rho} \partial_{\psi} u^{\rm bulk}_{\rho} - \mathcal{E}_{\psi} \partial_{\psi} u^{\rm bulk}_{\psi} 
        \end{smallmatrix} \\
        \begin{smallmatrix}
        \partial_{\rho} \mathcal{G}_{\rho}^{(2, 2)} - \mathcal{E}_{\rho} \partial_{\rho} u^{(2,2)}_{\rho \rho} - \partial_{\rho} \mathcal{E}_{\rho} u^{(2,2)}_{\rho \rho} - \mathcal{E}_{\psi} \partial_{\rho} u^{(2,2)}_{\psi \rho} - \partial_{\rho} \mathcal{E}_{\psi} u^{(2,2)}_{\psi \rho}
        \end{smallmatrix} \\
        \begin{smallmatrix}
        \partial_{\psi} \mathcal{G}_{\rho}^{(2, 2)} - \mathcal{E}_{\rho} \partial_{\psi} u^{(2,2)}_{\rho \rho} - \partial_{\psi} \mathcal{E}_{\rho} u^{(2,2)}_{\rho \rho} - \mathcal{E}_{\psi} \partial_{\psi} u^{(2,2)}_{\psi \rho} - \partial_{\psi} \mathcal{E}_{\psi} u^{(2,2)}_{\psi \rho}
        \end{smallmatrix} \\
        \begin{smallmatrix}
        \partial_{\rho} \mathcal{G}_{\psi}^{(2, 2)} - \mathcal{E}_{\rho} \partial_{\rho} u^{(2,2)}_{\rho \psi} - \partial_{\rho} \mathcal{E}_{\rho} u^{(2,2)}_{\rho \psi} - \mathcal{E}_{\psi} \partial_{\rho} u^{(2,2)}_{\psi \psi} - \partial_{\rho} \mathcal{E}_{\psi} u^{(2,2)}_{\psi \psi}
        \end{smallmatrix} \\
        \begin{smallmatrix}
        \partial_{\psi} \mathcal{G}_{\psi}^{(2, 2)} - \mathcal{E}_{\rho} \partial_{\psi} u^{(2,2)}_{\rho \psi} - \partial_{\psi} \mathcal{E}_{\rho} u^{(2,2)}_{\rho \psi} - \mathcal{E}_{\psi} \partial_{\psi} u^{(2,2)}_{\psi \psi} - \partial_{\psi} \mathcal{E}_{\psi} u^{(2,2)}_{\psi \psi}
        \end{smallmatrix} \\
        \begin{smallmatrix}
        2 \mathcal{G}_{\rho \rho}^{(2, 1)} + \partial_{\rho} \mathcal{G}_{\rho}^{(2, 2)} - \mathcal{E}_{\rho} \left( 2 u^{(2,1)}_{\rho \rho \rho} + \partial_{\rho} u^{(2,2)}_{\rho \rho} \right) - \mathcal{E}_{\psi} \left( 2 u^{(2,1)}_{\psi \rho \rho} + \partial_{\rho} u^{(2,2)}_{\psi \rho} \right)
        \end{smallmatrix} \\
        \begin{smallmatrix}
        2 \mathcal{G}_{\rho \psi}^{(2, 1)} + \partial_{\psi} \mathcal{G}_{\rho}^{(2, 2)} - \mathcal{E}_{\rho} \left( 2 u^{(2,1)}_{\rho \rho \psi} + \partial_{\psi} u^{(2,2)}_{\rho \rho} \right) - \mathcal{E}_{\psi} \left( 2 u^{(2,1)}_{\psi \rho \psi} + \partial_{\psi} u^{(2,2)}_{\psi \rho} \right)
        \end{smallmatrix} \\
        \begin{smallmatrix}
        2 \mathcal{G}_{\rho \psi}^{(2, 1)} + \partial_{\rho} \mathcal{G}_{\psi}^{(2, 2)} - \mathcal{E}_{\rho} \left( 2 u^{(2,1)}_{\rho \psi \rho} + \partial_{\rho} u^{(2,2)}_{\rho \psi} \right) - \mathcal{E}_{\psi} \left( 2 u^{(2,1)}_{\psi \psi \rho} + \partial_{\rho} u^{(2,2)}_{\psi \psi} \right)
        \end{smallmatrix} \\
        \begin{smallmatrix}
        2 \mathcal{G}_{\psi \psi}^{(2, 1)} + \partial_{\psi} \mathcal{G}_{\psi}^{(2, 2)} - \mathcal{E}_{\rho} \left( 2 u^{(2,1)}_{\rho \psi \psi} + \partial_{\psi} u^{(2,2)}_{\rho \psi} \right) - \mathcal{E}_{\psi} \left( 2 u^{(2,1)}_{\psi \psi \psi} + \partial_{\psi} u^{(2,2)}_{\psi \psi} \right) \\
        \end{smallmatrix} \\
        \begin{smallmatrix}
        \partial_{\rho} \mathcal{G}_{\rho \rho}^{(2, 1)} - \mathcal{E}_{\rho} \partial_{\rho} u^{(2,1)}_{\rho \rho \rho} - \mathcal{E}_{\psi} \partial_{\rho} u^{(2,1)}_{\psi \rho \rho}
        \end{smallmatrix} \\
        \begin{smallmatrix}
        2 \partial_{\rho} \mathcal{G}_{\rho \psi}^{(2, 1)} + \partial_{\psi} \mathcal{G}_{\rho \rho}^{(2, 1)} - \mathcal{E}_{\rho} \left( 2 \partial_{\rho} u^{(2,1)}_{\rho \rho \psi} + \partial_{\psi} u^{(2,1)}_{\rho \rho \rho} \right) - \mathcal{E}_{\psi} \left( 2 \partial_{\rho} u^{(2,1)}_{\psi \rho \psi} + \partial_{\psi} u^{(2,1)}_{\psi \rho \rho} \right)
        \end{smallmatrix} \\
        \begin{smallmatrix}
        2 \partial_{\psi} \mathcal{G}_{\rho \psi}^{(2, 1)} + \partial_{\rho} \mathcal{G}_{\psi \psi}^{(2, 1)} - \mathcal{E}_{\rho} \left( 2 \partial_{\psi} u^{(2,1)}_{\rho \rho \psi} + \partial_{\rho} u^{(2,1)}_{\rho \psi \psi} \right) - \mathcal{E}_{\psi} \left( 2 \partial_{\psi} u^{(2,1)}_{\psi \rho \psi} + \partial_{\rho} u^{(2,1)}_{\psi \psi \psi} \right)
        \end{smallmatrix} \\
        \begin{smallmatrix}
        \partial_{\psi} \mathcal{G}_{\psi \psi}^{(2, 1)} - \mathcal{E}_{\rho} \partial_{\psi} u^{(2,1)}_{\rho \psi \psi} - \mathcal{E}_{\psi} \partial_{\psi} u^{(2,1)}_{\psi \psi \psi}
        \end{smallmatrix}
    \end{pmatrix}.
\end{equation}
where we have used the $n \leftrightarrow j \leftrightarrow k$ symmetry of Eq.~\eqref{eq:Esys4}.

We now look to complete the system of equations in Eq.~\eqref{eq:phifull} to involution following Ref.~\cite{Seiler2010}.
We consider the ordering ${\rho} \succ {\psi}$, noting that the reverse may be chosen as well if the following analysis is repeated swapping $\rho \leftrightarrow \psi$.
We define the order of differentiation of a term to be $q=q_{\rho} + q_{\psi}$ where $q_{\rho}$ derivatives are with respect to $\rho$ and $q_{\psi}$ are with respect to ${\psi}$.
Due to our ordering ${\rho} \succ {\psi}$, $\psi$ is a ``multiplicative'' variable for every PDE considered.
$\rho$ is multiplicative for PDEs of order $q$ if any order $q$ term in the equation has $q_{\rho} > q_{\psi}$.
Otherwise, $\rho$ is ``non-multiplicative'' for that PDE.
A system of PDEs, $\boldsymbol{\Phi}=\mathbf{0}$, that is locally involutive has the property that the prolongations (i.e.~differentiations) of each equation with respect to its non-multiplicative variables can be expressed as a linear combination of the equations themselves and the prolongations with respect to the multiplicative variables.
If a non-multiplicative prolongation is linearly independent of $\boldsymbol{\Phi}$ and the multiplicative prolongations of $\boldsymbol{\Phi}$, then this non-multiplicative prolongation is a \textit{compatibility condition}: an additional PDE that a solution to $\boldsymbol{\Phi}=\mathbf{0}$ must also satisfy.
Prolonging Eq.~\eqref{eq:phifull} once, we find four second-order compatibility conditions.
An additional prolongation results in a third-order compatibility condition.
A third prolongation yields no compatibility conditions, meaning our system of PDEs is involutively complete.
The five necessary compatability conditions to complete Eq.~\eqref{eq:phifull} to involution are:
\begin{equation}
    \boldsymbol{\Phi}^{\rm compat} = 
    \resizebox{.9\hsize}{!}{ $
    \begin{pmatrix}
        \begin{smallmatrix}
        \partial^2_{\rho \psi} \mathcal{G}^{\rm bulk} - \mathcal{E}_{\rho} \partial^2_{\rho \psi} u^{\rm bulk}_{\rho} - \partial_{\rho} \mathcal{E}_{\rho} \partial_{\psi} u^{\rm bulk}_{\rho} - \mathcal{E}_{\psi} \partial^2_{\rho \psi} u^{\rm bulk}_{\psi} - \partial_{\rho} \mathcal{E}_{\psi} \partial_{\psi} u^{\rm bulk}_{\psi}
        \end{smallmatrix} \\
        \begin{smallmatrix}
            \partial^2_{\rho \psi} \mathcal{G}_{\rho}^{(2, 2)} + 2 \partial_{\rho} \mathcal{G}_{\rho \psi}^{(2, 1)} - \mathcal{E}_{\rho} \left(2 \partial_{\rho} u^{(2, 1)}_{\rho \rho \psi} + \partial^2_{\rho \psi} u^{(2, 2)}_{\rho \rho} \right) - \partial_{\rho} \mathcal{E}_{\rho} \left(2 u^{(2, 1)}_{\rho \rho \psi} + \partial_{\psi} u^{(2, 2)}_{\rho \rho} \right) - \mathcal{E}_{\psi} \left(2 \partial_{\rho} u^{(2, 1)}_{\psi \rho \psi} + \partial^2_{\rho \psi} u^{(2, 2)}_{\psi \rho} \right) - \partial_{\rho} \mathcal{E}_{\psi} \left(2 u^{(2, 1)}_{\psi \rho \psi} + \partial_{\psi} u^{(2, 2)}_{\psi \rho} \right)
        \end{smallmatrix} \\
        \begin{smallmatrix}
            \partial^2_{\rho \psi} \mathcal{G}_{\psi}^{(2, 2)} + 2 \partial_{\rho} \mathcal{G}_{\psi \psi}^{(2, 1)}  - \mathcal{E}_{\rho} \left(2 \partial_{\rho} u^{(2, 1)}_{\rho \psi \psi} + \partial^2_{\rho \psi} u^{(2, 2)}_{\rho \psi} \right) - \partial_{\rho} \mathcal{E}_{\rho} \left(2 u^{(2, 1)}_{\rho \psi \psi} + \partial_{\psi} u^{(2, 2)}_{\rho \psi} \right) - \mathcal{E}_{\psi} \left(2 \partial_{\rho} u^{(2, 1)}_{\psi \psi \psi} + \partial^2_{\rho \psi} u^{(2, 2)}_{\psi \psi} \right) - \partial_{\rho} \mathcal{E}_{\psi} \left(2 u^{(2, 1)}_{\psi \psi \psi} + \partial_{\psi} u^{(2, 2)}_{\psi \psi} \right)
        \end{smallmatrix} \\
        \begin{smallmatrix}
        \partial^2_{\rho \psi} \mathcal{G}_{\psi \psi}^{(2, 1)} - \mathcal{E}_{\rho} \partial^2_{\rho \psi} u^{(2, 1)}_{\rho \psi \psi} - \partial_{\rho} \mathcal{E}_{\rho} \partial_{\psi} u^{(2, 1)}_{\rho \psi \psi} - \mathcal{E}_{\psi} \partial^2_{\rho \psi} u^{(2, 1)}_{\psi \psi \psi} - \partial_{\rho} \mathcal{E}_{\psi} \partial_{\psi} u^{(2, 1)}_{\psi \psi \psi}
        \end{smallmatrix} \\
        \begin{smallmatrix}
        \partial^3_{\rho \rho \psi} \mathcal{G}_{\psi \psi}^{(2, 1)} - 2 \partial_{\rho} \mathcal{E}_{\rho} \partial^2_{\rho \psi} u^{(2, 1)}_{\rho \psi \psi} - \mathcal{E}_{\rho} \partial^3_{\rho \rho \psi} u^{(2, 1)}_{\rho \psi \psi} - \partial^2_{\rho \rho} \mathcal{E}_{\rho} \partial_{\psi} u^{(2, 1)}_{\rho \psi \psi} - 2 \partial_{\rho} \mathcal{E}_{\psi} \partial^2_{\rho \psi} u^{(2, 1)}_{\psi \psi \psi} - \mathcal{E}_{\psi} \partial^3_{\rho \rho \psi} u^{(2, 1)}_{\psi \psi \psi} - \partial^2_{\rho \rho} \mathcal{E}_{\psi} \partial_{\psi} u^{(2, 1)}_{\psi \psi \psi}
        \end{smallmatrix}
    \end{pmatrix}. $ }
\end{equation}
Therefore, we have our involutive system of equations $\tilde{\boldsymbol{\Phi}} = \begin{bmatrix}
    \Phi_1 & \cdots & \Phi_{14} & \Phi^{\rm compat}_1 & \cdots & \Phi^{\rm compat}_5
\end{bmatrix}$:
\begin{equation}
    \label{eq:phicomplete}
    \tilde{\boldsymbol{\Phi}} = \resizebox{.9\hsize}{!}{$\begin{pmatrix}
    \begin{smallmatrix}
        \partial_{\rho} \mathcal{G}^{\rm bulk} - \mathcal{E}_{\rho} \partial_{\rho} u^{\rm bulk}_{\rho} - \mathcal{E}_{\psi} \partial_{\rho} u^{\rm bulk}_{\psi} 
        \end{smallmatrix} \\
        \begin{smallmatrix}
        \partial_{\psi} \mathcal{G}^{\rm bulk} - \mathcal{E}_{\rho} \partial_{\psi} u^{\rm bulk}_{\rho} - \mathcal{E}_{\psi} \partial_{\psi} u^{\rm bulk}_{\psi} 
        \end{smallmatrix} \\
        \begin{smallmatrix}
        \partial_{\rho} \mathcal{G}_{\rho}^{(2, 2)} - \mathcal{E}_{\rho} \partial_{\rho} u^{(2,2)}_{\rho \rho} - \partial_{\rho} \mathcal{E}_{\rho} u^{(2,2)}_{\rho \rho} - \mathcal{E}_{\psi} \partial_{\rho} u^{(2,2)}_{\psi \rho} - \partial_{\rho} \mathcal{E}_{\psi} u^{(2,2)}_{\psi \rho}
        \end{smallmatrix} \\
        \begin{smallmatrix}
        \partial_{\psi} \mathcal{G}_{\rho}^{(2, 2)} - \mathcal{E}_{\rho} \partial_{\psi} u^{(2,2)}_{\rho \rho} - \partial_{\psi} \mathcal{E}_{\rho} u^{(2,2)}_{\rho \rho} - \mathcal{E}_{\psi} \partial_{\psi} u^{(2,2)}_{\psi \rho} - \partial_{\psi} \mathcal{E}_{\psi} u^{(2,2)}_{\psi \rho}
        \end{smallmatrix} \\
        \begin{smallmatrix}
        \partial_{\rho} \mathcal{G}_{\psi}^{(2, 2)} - \mathcal{E}_{\rho} \partial_{\rho} u^{(2,2)}_{\rho \psi} - \partial_{\rho} \mathcal{E}_{\rho} u^{(2,2)}_{\rho \psi} - \mathcal{E}_{\psi} \partial_{\rho} u^{(2,2)}_{\psi \psi} - \partial_{\rho} \mathcal{E}_{\psi} u^{(2,2)}_{\psi \psi}
        \end{smallmatrix} \\
        \begin{smallmatrix}
        \partial_{\psi} \mathcal{G}_{\psi}^{(2, 2)} - \mathcal{E}_{\rho} \partial_{\psi} u^{(2,2)}_{\rho \psi} - \partial_{\psi} \mathcal{E}_{\rho} u^{(2,2)}_{\rho \psi} - \mathcal{E}_{\psi} \partial_{\psi} u^{(2,2)}_{\psi \psi} - \partial_{\psi} \mathcal{E}_{\psi} u^{(2,2)}_{\psi \psi}
        \end{smallmatrix} \\
        \begin{smallmatrix}
        2 \mathcal{G}_{\rho \rho}^{(2, 1)} + \partial_{\rho} \mathcal{G}_{\rho}^{(2, 2)} - \mathcal{E}_{\rho} \left( 2 u^{(2,1)}_{\rho \rho \rho} + \partial_{\rho} u^{(2,2)}_{\rho \rho} \right) - \mathcal{E}_{\psi} \left( 2 u^{(2,1)}_{\psi \rho \rho} + \partial_{\rho} u^{(2,2)}_{\psi \rho} \right)
        \end{smallmatrix} \\
        \begin{smallmatrix}
        2 \mathcal{G}_{\rho \psi}^{(2, 1)} + \partial_{\psi} \mathcal{G}_{\rho}^{(2, 2)} - \mathcal{E}_{\rho} \left( 2 u^{(2,1)}_{\rho \rho \psi} + \partial_{\psi} u^{(2,2)}_{\rho \rho} \right) - \mathcal{E}_{\psi} \left( 2 u^{(2,1)}_{\psi \rho \psi} + \partial_{\psi} u^{(2,2)}_{\psi \rho} \right)
        \end{smallmatrix} \\
        \begin{smallmatrix}
        2 \mathcal{G}_{\rho \psi}^{(2, 1)} + \partial_{\rho} \mathcal{G}_{\psi}^{(2, 2)} - \mathcal{E}_{\rho} \left( 2 u^{(2,1)}_{\rho \psi \rho} + \partial_{\rho} u^{(2,2)}_{\rho \psi} \right) - \mathcal{E}_{\psi} \left( 2 u^{(2,1)}_{\psi \psi \rho} + \partial_{\rho} u^{(2,2)}_{\psi \psi} \right)
        \end{smallmatrix} \\
        \begin{smallmatrix}
        2 \mathcal{G}_{\psi \psi}^{(2, 1)} + \partial_{\psi} \mathcal{G}_{\psi}^{(2, 2)} - \mathcal{E}_{\rho} \left( 2 u^{(2,1)}_{\rho \psi \psi} + \partial_{\psi} u^{(2,2)}_{\rho \psi} \right) - \mathcal{E}_{\psi} \left( 2 u^{(2,1)}_{\psi \psi \psi} + \partial_{\psi} u^{(2,2)}_{\psi \psi} \right) \\
        \end{smallmatrix} \\
        \begin{smallmatrix}
        \partial_{\rho} \mathcal{G}_{\rho \rho}^{(2, 1)} - \mathcal{E}_{\rho} \partial_{\rho} u^{(2,1)}_{\rho \rho \rho} - \mathcal{E}_{\psi} \partial_{\rho} u^{(2,1)}_{\psi \rho \rho}
        \end{smallmatrix} \\
        \begin{smallmatrix}
        2 \partial_{\rho} \mathcal{G}_{\rho \psi}^{(2, 1)} + \partial_{\psi} \mathcal{G}_{\rho \rho}^{(2, 1)} - \mathcal{E}_{\rho} \left( 2 \partial_{\rho} u^{(2,1)}_{\rho \rho \psi} + \partial_{\psi} u^{(2,1)}_{\rho \rho \rho} \right) - \mathcal{E}_{\psi} \left( 2 \partial_{\rho} u^{(2,1)}_{\psi \rho \psi} + \partial_{\psi} u^{(2,1)}_{\psi \rho \rho} \right)
        \end{smallmatrix} \\
        \begin{smallmatrix}
        2 \partial_{\psi} \mathcal{G}_{\rho \psi}^{(2, 1)} + \partial_{\rho} \mathcal{G}_{\psi \psi}^{(2, 1)} - \mathcal{E}_{\rho} \left( 2 \partial_{\psi} u^{(2,1)}_{\rho \rho \psi} + \partial_{\rho} u^{(2,1)}_{\rho \psi \psi} \right) - \mathcal{E}_{\psi} \left( 2 \partial_{\psi} u^{(2,1)}_{\psi \rho \psi} + \partial_{\rho} u^{(2,1)}_{\psi \psi \psi} \right)
        \end{smallmatrix} \\
        \begin{smallmatrix}
        \partial_{\psi} \mathcal{G}_{\psi \psi}^{(2, 1)} - \mathcal{E}_{\rho} \partial_{\psi} u^{(2,1)}_{\rho \psi \psi} - \mathcal{E}_{\psi} \partial_{\psi} u^{(2,1)}_{\psi \psi \psi}
        \end{smallmatrix} \\
                \begin{smallmatrix}
            \partial^2_{\rho \psi} \mathcal{G}^{\rm bulk} - \mathcal{E}_{\rho} \partial^2_{\rho \psi} u^{\rm bulk}_{\rho} - \partial_{\rho} \mathcal{E}_{\rho} \partial_{\psi} u^{\rm bulk}_{\rho} - \mathcal{E}_{\psi} \partial^2_{\rho \psi} u^{\rm bulk}_{\psi} - \partial_{\rho} \mathcal{E}_{\psi} \partial_{\psi} u^{\rm bulk}_{\psi}
        \end{smallmatrix} \\
        \begin{smallmatrix}
            \partial^2_{\rho \psi} \mathcal{G}_{\rho}^{(2, 2)} + 2 \partial_{\rho} \mathcal{G}_{\rho \psi}^{(2, 1)} - \mathcal{E}_{\rho} \left(2 \partial_{\rho} u^{(2, 1)}_{\rho \rho \psi} + \partial^2_{\rho \psi} u^{(2, 2)}_{\rho \rho} \right) - \partial_{\rho} \mathcal{E}_{\rho} \left(2 u^{(2, 1)}_{\rho \rho \psi} + \partial_{\psi} u^{(2, 2)}_{\rho \rho} \right) - \mathcal{E}_{\psi} \left(2 \partial_{\rho} u^{(2, 1)}_{\psi \rho \psi} + \partial^2_{\rho \psi} u^{(2, 2)}_{\psi \rho} \right) - \partial_{\rho} \mathcal{E}_{\psi} \left(2 u^{(2, 1)}_{\psi \rho \psi} + \partial_{\psi} u^{(2, 2)}_{\psi \rho} \right)
        \end{smallmatrix} \\
        \begin{smallmatrix}
            \partial^2_{\rho \psi} \mathcal{G}_{\psi}^{(2, 2)} + 2 \partial_{\rho} \mathcal{G}_{\psi \psi}^{(2, 1)}  - \mathcal{E}_{\rho} \left(2 \partial_{\rho} u^{(2, 1)}_{\rho \psi \psi} + \partial^2_{\rho \psi} u^{(2, 2)}_{\rho \psi} \right) - \partial_{\rho} \mathcal{E}_{\rho} \left(2 u^{(2, 1)}_{\rho \psi \psi} + \partial_{\psi} u^{(2, 2)}_{\rho \psi} \right) - \mathcal{E}_{\psi} \left(2 \partial_{\rho} u^{(2, 1)}_{\psi \psi \psi} + \partial^2_{\rho \psi} u^{(2, 2)}_{\psi \psi} \right) - \partial_{\rho} \mathcal{E}_{\psi} \left(2 u^{(2, 1)}_{\psi \psi \psi} + \partial_{\psi} u^{(2, 2)}_{\psi \psi} \right)
        \end{smallmatrix} \\
        \begin{smallmatrix}
        \partial^2_{\rho \psi} \mathcal{G}_{\psi \psi}^{(2, 1)} - \mathcal{E}_{\rho} \partial^2_{\rho \psi} u^{(2, 1)}_{\rho \psi \psi} - \partial_{\rho} \mathcal{E}_{\rho} \partial_{\psi} u^{(2, 1)}_{\rho \psi \psi} - \mathcal{E}_{\psi} \partial^2_{\rho \psi} u^{(2, 1)}_{\psi \psi \psi} - \partial_{\rho} \mathcal{E}_{\psi} \partial_{\psi} u^{(2, 1)}_{\psi \psi \psi}
        \end{smallmatrix} \\
        \begin{smallmatrix}
        \partial^3_{\rho \rho \psi} \mathcal{G}_{\psi \psi}^{(2, 1)} - 2 \partial_{\rho} \mathcal{E}_{\rho} \partial^2_{\rho \psi} u^{(2, 1)}_{\rho \psi \psi} - \mathcal{E}_{\rho} \partial^3_{\rho \rho \psi} u^{(2, 1)}_{\rho \psi \psi} - \partial^2_{\rho \rho} \mathcal{E}_{\rho} \partial_{\psi} u^{(2, 1)}_{\rho \psi \psi} - 2 \partial_{\rho} \mathcal{E}_{\psi} \partial^2_{\rho \psi} u^{(2, 1)}_{\psi \psi \psi} - \mathcal{E}_{\psi} \partial^3_{\rho \rho \psi} u^{(2, 1)}_{\psi \psi \psi} - \partial^2_{\rho \rho} \mathcal{E}_{\psi} \partial_{\psi} u^{(2, 1)}_{\psi \psi \psi}
    \end{smallmatrix}
    \end{pmatrix}.$}
\end{equation}
The importance of completing the system of PDEs to involution can now be appreciated, as it yields five additional compatibility conditions (when $n_N=1$) that must be satisfied for a well-defined solution to the generalized Gibbs-Duhem relation ($d \mathcal{G} = \mathcal{E}_{\rho} d u_{\rho} + \mathcal{E}_{\psi} d u_{\psi}$) to exist.

\section{\label{secSI:numerics}Numerical Simulation Details}
The numerical phase diagrams in the main text were determined using the MATLAB package \texttt{PDE2PATH}~\cite{Uecker2014, Thiele2019, Holl2020EfficientModel} to solve for the steady-state order parameter profiles of AMC.
All profiles were determined in a box of length $L_z=2$ with flux-free boundary conditions using the expressions for $u_{\rho}^{\rm bulk}$ and $u_{\psi}^{\rm bulk}$ provided in the main text.
Macroscopically phase-separated steady-states were found from primary bifurcations from the homogeneous steady-state at $\int_{L_z} d z \rho(z) = 0$, using $\alpha$ and $\lambda_{\rho \rho \rho}$ as the respective tuning parameters in main text Figs.~2 and 3.

\section{\label{secSI:support}Numerical Support for Coexistence Criteria}
\subsection{\label{secSI:exactsupport}Further Demonstration of Exact Criteria}
\begin{figure}
    \centering
    \includegraphics[width=0.85\linewidth]{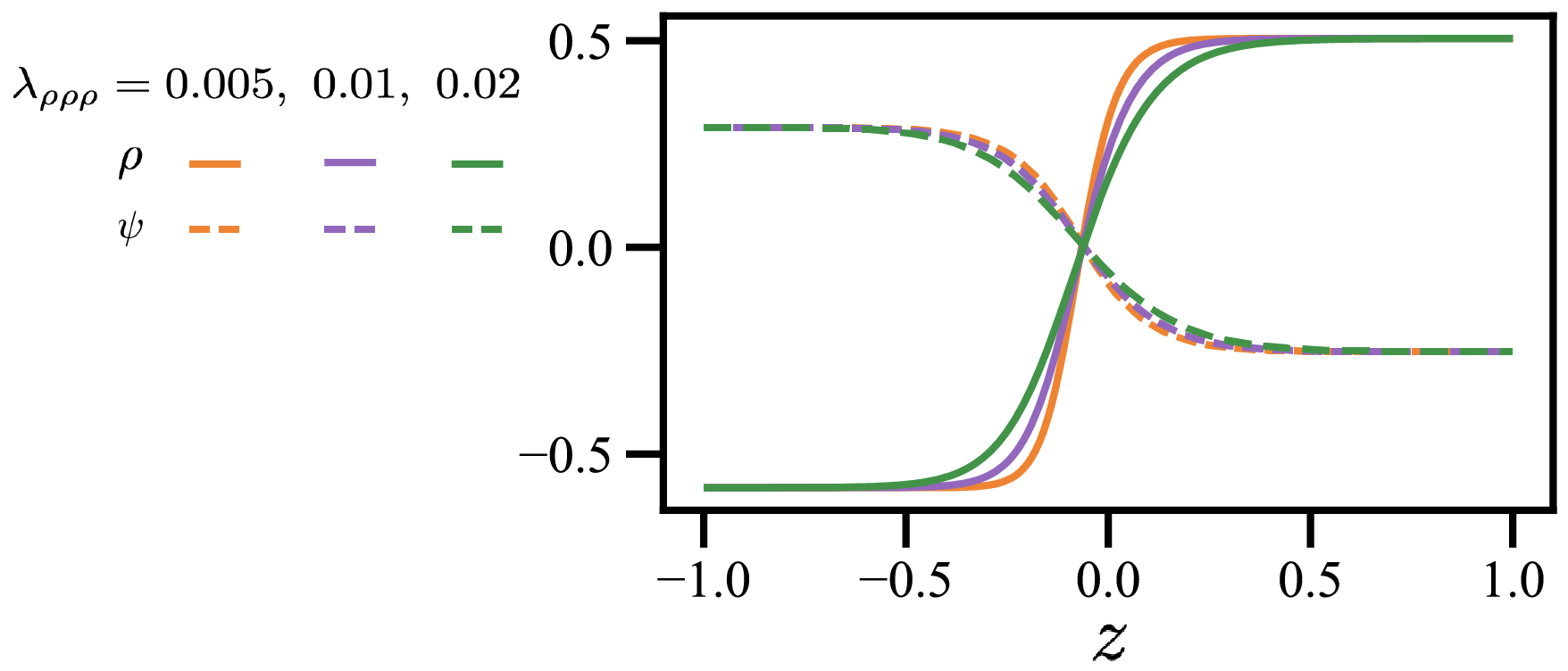}
    \caption{Coexistence profiles of systems with exact coexistence criteria, the phase diagram of which are shown in Fig.~3(a) in the main text. We set $\chi=-\alpha=1/4$, $\overline{\kappa}_{\psi \rho}=\overline{\kappa}_{\rho \psi} = \lambda_{\rho \psi \rho} = \lambda_{\rho \rho \psi} = \lambda_{\rho \psi \psi} = 0$, and $\overline{\kappa}_{\psi \psi} = 0.01$ in every case. We determine three profiles with $\overline{\kappa}_{\rho \rho} = \lambda_{\rho \rho \rho}$ while varying $\lambda_{\rho \rho \rho}$.}
    \label{fig:amcprof}
\end{figure}

We now offer further support for the coexistence criteria presented in the main text when they are in their exact form.
Our theory predicts that the binodals in Fig.~3(a) are a function of the ratio $\lambda_{\rho \rho \rho} / \overline{\kappa}_{\rho \rho}$.
In Fig.~\ref{fig:amcprof}, we numerically determine the coexistence profiles for this system, tuning $\lambda_{\rho \rho \rho}$ and $\overline{\kappa}_{\rho \rho}$ while keeping the ratio between the two fixed.
As expected, we find the shape of the interface changes with the raw values of $\lambda_{\rho \rho \rho}$ and $\overline{\kappa}_{\rho \rho}$, however the values of the order parameters in the bulk phases remain unchanged for fixed $\lambda_{\rho \rho \rho} / \overline{\kappa}_{\rho \rho}$.
This demonstrates our theory correctly identifies the relevant parameters that control the phase diagram of a system.

\subsection{Percent Error of Binodal Predictions}

\begin{figure}
	\centering
	\includegraphics[width=.475\textwidth]{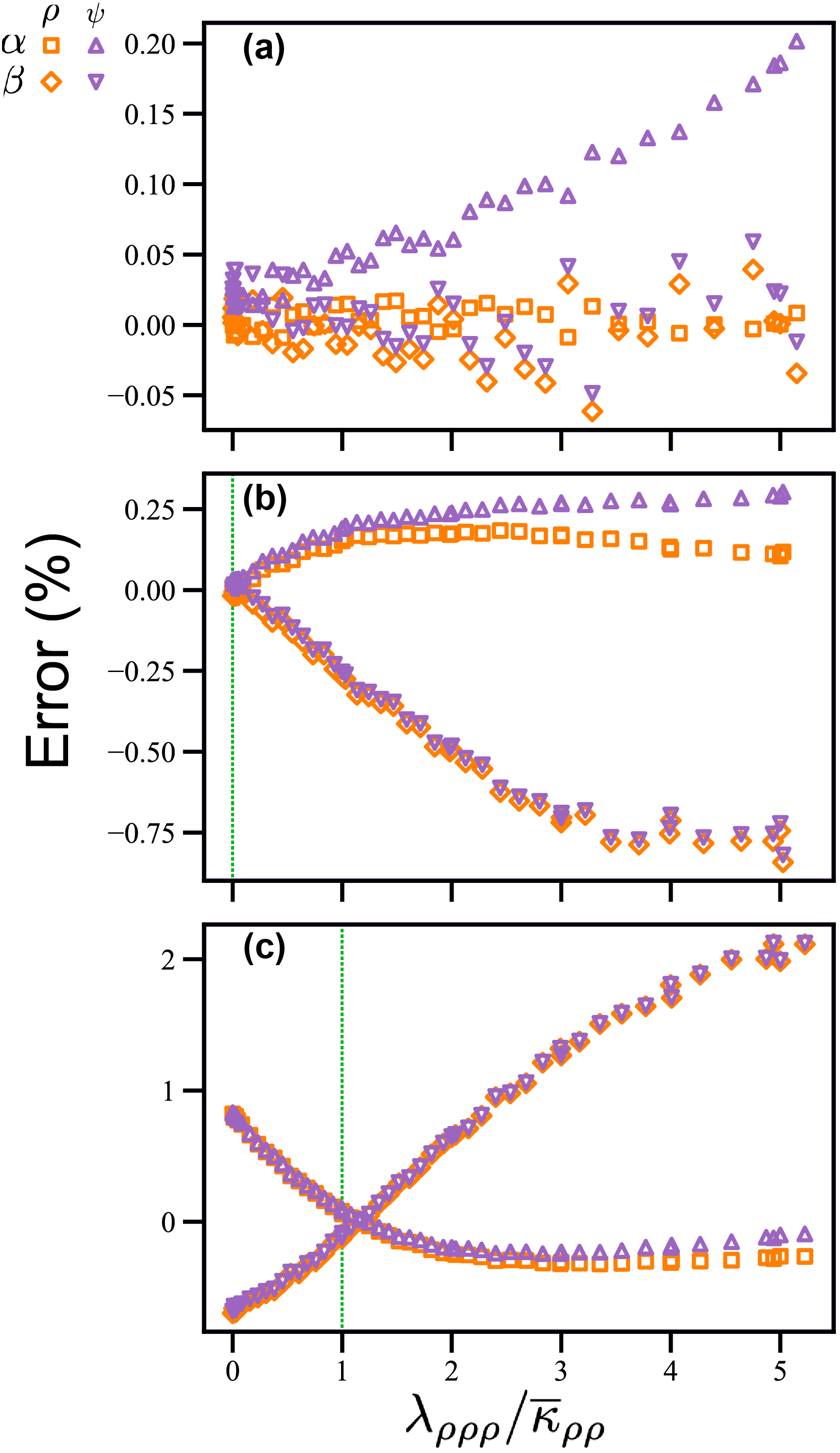}
	\caption{\protect\small{{Percent error of binodals in the $\alpha$ and $\beta$ phases predicted using the active theory in Fig.~3 in the main text compared to the steady-state numerical solutions of AMC. All parameters match those in Fig.~3 in the main text, covering systems where the final coexistence criterion is (a) exact, (b) well-defined but approximate, and (c) poorly-defined and approximate. The dotted green lines in (b) and (c) indicate the passive limit and the limit where there is a solution for $\boldsymbol{\mathcal{E}}^{\rm int}$, respectively.}}}
	\label{fig:ambAerror}
\end{figure}

In this Section, we analyze the results in Fig.~3 in the main text to understand the effectiveness of our approximate criterion.
In Fig.~\ref{fig:ambAerror}, we plot the percent error of our predictions compared to numerical results, defined as the difference in the predicted and numerical binodals divided by the numerical binodal itself.
The systems in Fig.~\ref{fig:ambAerror}(a), (b), and (c) correspond to the respective systems in Fig.~3 in the main text.
When the coexistence criteria are exact, as shown in Fig.~\ref{fig:ambAerror}(a), our predictions are within $<0.2\%$ of the numerically determined binodals at activities up to $\lambda_{\rho \rho \rho} / \overline{\kappa}_{\rho \rho} = 5$.
The magnitude of the percent error rises to $>0.75\%$ at  $\lambda_{\rho \rho \rho} / \overline{\kappa}_{\rho \rho} > 3$ when the final criterion is approximate but well-defined, as shown in Fig.~\ref{fig:ambAerror}(b).
As expected, the error decreases in the passive limit ($\lambda_{\rho \rho \rho} / \overline{\kappa}_{\rho \rho} \rightarrow 0$) where the criterion again becomes exact.
The error further increases in magnitude when the criterion used is approximate and poorly defined, as shown in Fig.~\ref{fig:ambAerror}(c), reaching a magnitude $>2\%$ at $\lambda_{\rho \rho \rho} / \overline{\kappa}_{\rho \rho} = 5$.
This demonstrates that the predictions made using the exact final coexistence criterion are more accurate than those made using the well-defined approximate criterion which are more accurate than those made using the poorly-defined approximate criterion, as anticipated.

\subsection{\label{secSI:approxsupportpath}Influence of Integration Path in Approximate Weighted-Area Construction}
In this Section, we examine how the selected integration path influences the weighted-area construction:
\begin{subequations}
\begin{equation}
    \Delta^{\alpha \beta} \mathcal{G}^{\rm int} [\mathbf{X}] \equiv \int_{\mathbf{X}^{\alpha}}^{\mathbf{X}^{\beta}} \left( \mathbf{u}^{\rm bulk} (\mathbf{X}) - \mathbf{u}^{\rm coexist} \right) \cdot \mathbf{E}^{\rm int} (\mathbf{X}) \cdot d \mathbf{X},
\end{equation}
where $\Delta^{\alpha \beta} \mathcal{G}^{\rm int} [\mathbf{X}^{\rm c}]=0$, i.e.~the weighted-area construction vanishes when evaluated along the the spatial coexistence profile, $\mathbf{X}^{\rm c}$.
We aim to avoid determining this profile and hence we can evaluate $\Delta^{\alpha \beta} \mathcal{G}^{\rm int} [\mathbf{X}]$ along another integration path and equate the result to zero as an approximate coexistence criterion.
Here, we consider the system in Fig.~3(b) in the main text where $\boldsymbol{\mathcal{E}}^{\rm int} = \begin{bmatrix} \exp \left( 2 \lambda_{\rho \rho \rho} \rho / \overline{\kappa}_{ \rho \rho} \right) & 0 \end{bmatrix}^{\rm T}$ and therefore the approximate criterion becomes:
\begin{equation}
    \Delta^{\alpha \beta} \mathcal{G}^{\rm int} [\mathbf{X}] = \int_{\rho^{\alpha}}^{\rho^{\beta}} \left( u_{\rho}^{\rm bulk} (\mathbf{X}) - u_{\rho}^{\rm coexist} \right) \exp \left( 2 \lambda_{\rho \rho \rho} \rho / \overline{\kappa}_{ \rho \rho} \right) d \rho =0.
\end{equation}
\end{subequations}

\begin{figure}
    \centering
    \includegraphics[width=0.55\linewidth]{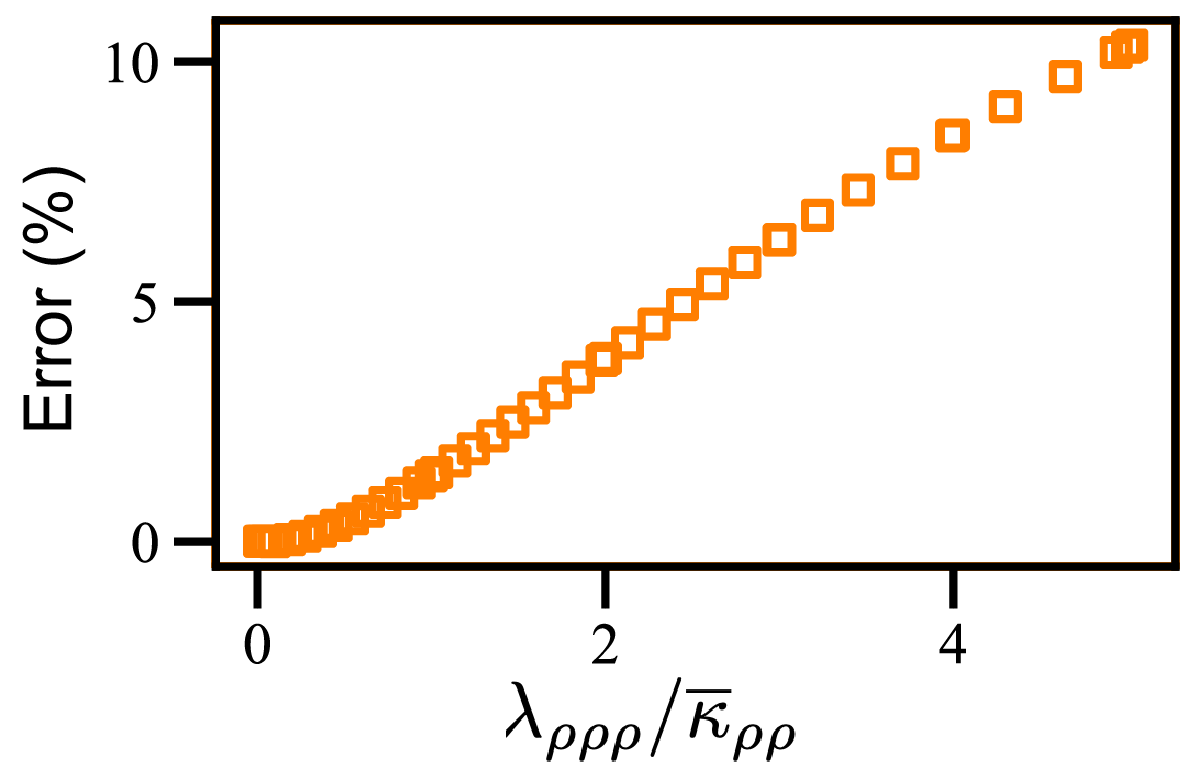}
    \caption{Percent error of $\Delta^{\alpha \beta} \mathcal{G}^{\rm int} [ \mathbf{X}^*]$ compared to $\Delta^{\alpha \beta} \mathcal{G}^{\rm int} [ \mathbf{X}^{\rm c}]$ for the system in Fig.~3(b) in the main text.}
    \label{fig:Gpath}
\end{figure}

We construct Fig.~3(b) in the main text using $\Delta^{\alpha \beta} \mathcal{G}^{\rm int} [\mathbf{X}^*]=0$.
To understand the accuracy of this approximation, we compute the percent error of $\Delta^{\alpha \beta} \mathcal{G}^{\rm int}$ evaluated along $\mathbf{X}^*$ instead of $\mathbf{X}^{\rm c}$:
\begin{equation}
    {\rm Error \ (\%)} \equiv \frac{\Delta^{\alpha \beta} \mathcal{G}^{\rm int} [\mathbf{X}^*] - \Delta^{\alpha \beta} \mathcal{G}^{\rm int} [\mathbf{X}^{\rm c}]}{\Delta^{\alpha \beta} \mathcal{G}^{\rm int} [\mathbf{X}^{\rm c}]},
\end{equation}
as shown in Fig.~\ref{fig:Gpath}.
While $\Delta^{\alpha \beta} \mathcal{G}^{\rm int} [\mathbf{X}^{\rm c}]$ is theoretically zero, in practice it takes a very small numerical value that we use here to construct the percent error.
We see that the error goes to zero in the passive limit, $\lambda_{\rho \rho \rho} / \overline{\kappa}_{\rho \rho} \rightarrow 0$, and monotonically increases with $\lambda_{\rho \rho \rho} / \overline{\kappa}_{\rho \rho}$.
Upon comparison to Fig.~\ref{fig:ambAerror}(b), it is clear that the error in the predicted binodals increases with the error in $\Delta^{\alpha \beta} \mathcal{G}^{\rm int} [\mathbf{X}^*]$.

\subsection{\label{secSI:approxsupportbulk}Accuracy of Equating $\mathcal{G}^{\rm bulk}$ as an Approximate Criterion}
We now confirm that the approximate criterion used to construct Fig.~2 in the main text (equating $\mathcal{G}^{\rm bulk} \equiv \mathbf{u}^{\rm bulk} \cdot \boldsymbol{\mathcal{E}}^{\rm bulk} - \int \mathbf{u}^{\rm bulk} \cdot d \boldsymbol{\mathcal{E}}^{\rm int}$ across phases) is indeed a good approximation.
To do so, we compute the difference in $\mathcal{G}^{\rm bulk}$ between the two phases (as detailed in the main text):
\begin{equation}
\label{sieq:delta_G}
    \Delta^{\alpha \beta} \mathcal{G}^{\rm bulk} = \left( \frac{\chi + \alpha}{\chi - \alpha} - \frac{\overline{\kappa}_{\rho \psi}}{\overline{\kappa}_{\psi \rho}} \right) \int_{z^{\alpha}}^{z^{\beta}} u^{\rm bulk}_{\psi} (\mathbf{X}^{\rm c}) \partial_z \psi^{\rm c} dz.
\end{equation}

\begin{figure}
    \centering
    \includegraphics[width=0.6\linewidth]{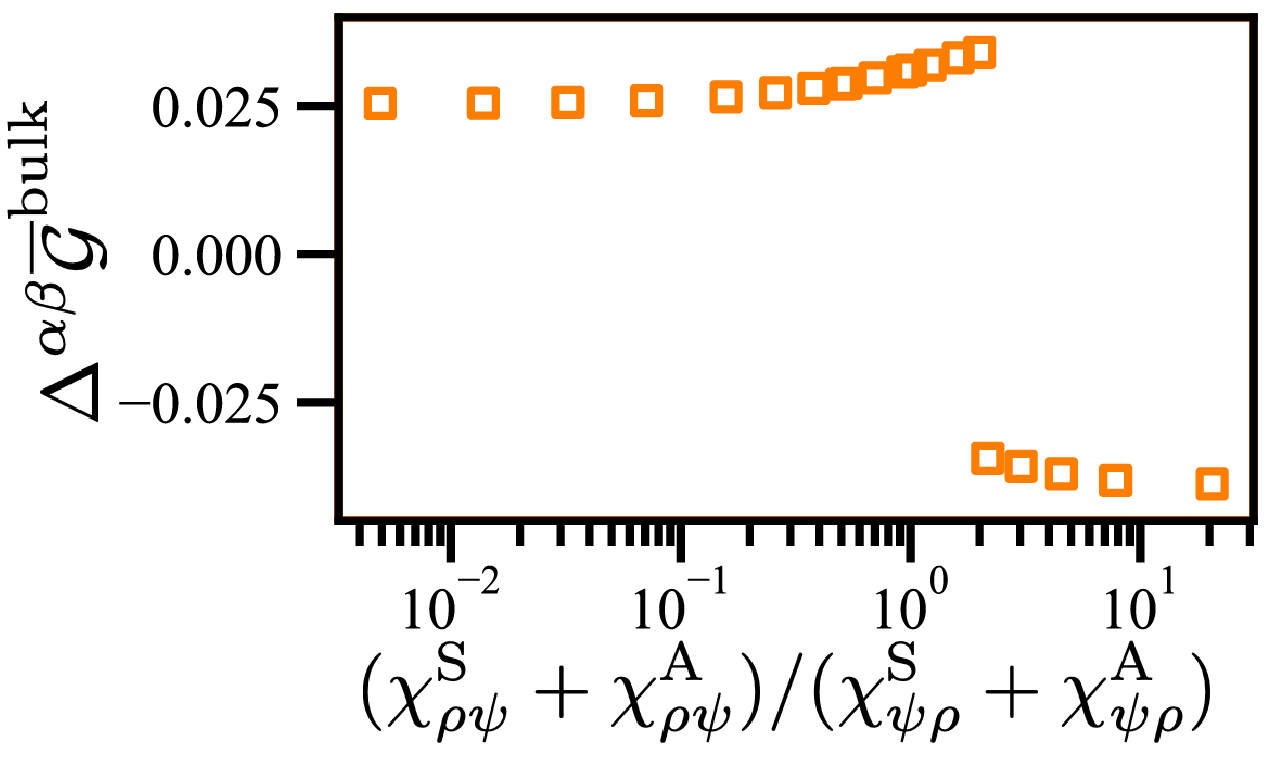}
    \caption{$\Delta^{\alpha \beta} \overline{\mathcal{G}}^{\rm bulk}$ for the system in Fig.~2 in the main text.}
    \label{fig:nrpd_supp}
\end{figure}

It is not clear from Eq.~\eqref{sieq:delta_G} what a ``small'' or ``large'' value of $\Delta^{\alpha \beta} \mathcal{G}^{\rm bulk}$ is and thus when equating it across phases is a good approximation or not, respectively.
We therefore desire to provide a scale to Eq.~\eqref{sieq:delta_G}.
One way of doing so is to treat Eq.~\eqref{sieq:delta_G} as a mean of sorts by dividing by the system length in the $z$-direction, $L_z$:
\begin{subequations}
\begin{equation}
\label{sieq:delta_G_Lz}
    \left \langle \Delta^{\alpha \beta} \mathcal{G}^{\rm bulk}_{L_z} \right \rangle \equiv \frac{\Delta^{\alpha \beta} \mathcal{G}^{\rm bulk}}{L_z} =  \left( \frac{\chi + \alpha}{\chi - \alpha} - \frac{\overline{\kappa}_{\rho \psi}}{\overline{\kappa}_{\psi \rho}} \right) \int_{z^{\alpha}}^{z^{\beta}} \frac{ u^{\rm bulk}_{\psi}(\mathbf{X}^{\rm c}) \partial_z \psi^{\rm c}}{L_z} dz,
\end{equation}
with the analogous second moment:
\begin{equation}
\label{sieq:delta_G2_Lz}
    \left \langle \left( \Delta^{\alpha \beta} \mathcal{G}^{\rm bulk}_{L_z} \right)^2 \right \rangle =  \left( \frac{\chi + \alpha}{\chi - \alpha} - \frac{\overline{\kappa}_{\rho \psi}}{\overline{\kappa}_{\psi \rho}} \right)^2 \int_{z^{\alpha}}^{z^{\beta}} \frac{\left(u^{\rm bulk}_{\psi} (\mathbf{X}^{\rm c}) \partial_z \psi^{\rm c}\right)^2}{L_z} dz.
\end{equation}
We can now define a variance as ${\rm Var}\left( \Delta^{\alpha \beta} \mathcal{G}^{\rm bulk}_{L_z} \right) \equiv \left \langle \left( \Delta^{\alpha \beta} \mathcal{G}^{\rm bulk}_{L_z} \right)^2 \right \rangle - \left \langle \Delta^{\alpha \beta} \mathcal{G}^{\rm bulk}_{L_z} \right \rangle^2$ to define the dimensionless $\Delta^{\alpha \beta} \overline{\mathcal{G}}^{\rm bulk} \equiv \left \langle \Delta^{\alpha \beta} \mathcal{G}^{\rm bulk}_{L_z} \right \rangle / \sqrt{{\rm Var}\left( \Delta^{\alpha \beta} \mathcal{G}^{\rm bulk}_{L_z} \right)}$, which we plot in Fig.~\ref{fig:nrpd_supp}.
\end{subequations}
The consistently low $\Delta^{\alpha \beta} \overline{\mathcal{G}}^{\rm bulk}$ across parameter values suggests equating $\mathcal{G}^{\rm bulk}$ across phases is a serviceable approximate coexistence criterion, as expected from the accuracy of the predictions in Fig.~2 in the main text.